\documentclass[english]{article}
\usepackage[T1]{fontenc}
\usepackage[latin1]{inputenc}
\usepackage{graphicx}

\makeatletter

\providecommand{\tabularnewline}{\\}

\newcommand{\lyxaddress}[1]{
\par {\raggedright #1
\vspace{1.4em}
\noindent\par}
}

\usepackage{babel}
\makeatother
\begin{document}

\title{}

\title{Magnetism and structure of magnetic multilayers based on the fully
spin polarized Heusler alloys Co$_{2}$MnGe and Co$_{2}$MnSn}

\author{K. Westerholt, A. Bergmann, J. Grabis, A. Nefedov and H. Zabel}

\maketitle

\lyxaddress{Institut für Experimentalphysik IV, Ruhr-Universität Bochum, D-44780
Bochum, Germany }

\begin{abstract}
Our Introduction starts with a short general review of the magnetic
and structural properties of the Heusler compounds which are under
discussion in this book. Then, more specifically, we come to the discussion
of our experimental results on multilayers composed of the Heusler
alloys Co$_{2}$MnGe and Co$_{2}$MnSn with V or Au as interlayers.
The experimental methods we apply combine magnetization and magneto-resistivity
measurements, x-ray diffraction and reflectivity, soft x-ray magnetic
circular dichroism and spin polarized neutron reflectivity. We find
that below a critical thickness of the Heusler layers at typically
$d_{cr}=1.5$ nm the ferromagnetic order is lost and spin glass order
occurs instead. For very thin ferromagnetic Heusler layers there are
peculiarities in the magnetic order which are unusual when compared
to conventional ferromagnetic transition metal multilayer systems.
In {[}Co$_{2}$MnGe/Au] multilayers there is an exchange bias shift
at the ferromagnetic hysteresis loops at low temperatures caused by
spin glass ordering at the interface. In {[}Co$_{2}$MnGe/V] multilayers
we observe an antiferromagnetic interlayer long range ordering below
a well defined N\'{e}el temperature originating from the dipolar
stray fields at the magnetically rough Heusler layer interfaces. \emph{}\textbf{\emph{Copy
of a book article edited by P.H. Dederichs and J.~Galanakis {}``Half-metallic
Alloys - Fundamentals and Applications'' Lecture Notes in Physics
Springer, p67-110 (2005)}}
\end{abstract}

\section{Introduction}

\label{sec:intro} In the main part of this article we will describe
our experimental results on thin films and multilayers based on the
Heusler phases Co$_{2}$MnGe and Co$_{2}$MnSn. However, before coming
to this special topic, we will give a short review of the structural
and magnetic properties of the Heusler compounds in general with special
emphasis on the half metallic Heusler compounds. This is intended
to give a general introduction to the experimental chapters of this
book, where the focus of the investigations is on thin films of Heusler
phases with theoretically predicted half metallic character. A knowledge
of the metallurgical, structural and magnetic properties of bulk Heusler
compounds will help the reader to estimate the problems one encounters
when trying to grow thin films of the Heusler compounds optimized
for spintronic applications.

The Heusler compounds have a long history in magnetism, starting more
than 100 years ago with the detection of the ternary metallic compound
Cu$_{2}$MnAl by A. Heusler \cite{heusler03}. That time the most
interesting point with this phase was that it represented the first
transition metal compound with ferromagnetically ordered Mn-spins.
Later it turned out that a whole class of isostructural ternary metallic
alloys with the general composition X$_{2}$YZ exists, where X denotes
a transition metal element such as Ni, Co, Fe or Pt, Y is a second
transition metal element, e.g. Mn, Cr or Ti and Z is an atom from
3$^{rd}$ , 4$^{th}$ or 5$^{th}$ group of the periodic system such
as Al, Ge, Sn or Sb. More than 1000 different Heusler compounds have
been synthesized until now, a comprehensive review of the experimental
work until the year 1987 can be found in \cite{zie86}.

\begin{figure}[ht]
 \centering \includegraphics[height=8cm]{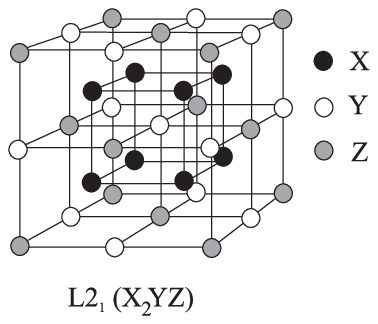}

\caption{Schematic representation of the L2$_{1}$-structure.}

\label{fig:1-1} 
\end{figure}

The original Heusler phase Cu$_{2}$MnAl, which later gave its name
to the whole class of ternary inter-metallic compounds, can be considered
as the prototype of all other Heusler compounds, including those with
half metallic properties. It has been studied intensively over the
past 100 years and we begin with describing its properties in some
detail. The crystal structure of Cu$_{2}$MnAl is cubic with three
different structural modifications: At high temperatures the crystal
structure is bcc with random occupation of the Cu, Mn and Al atoms
on the lattice sites of a simple bcc lattice with a lattice parameter
$a=0.297$~nm. At low temperatures the equilibrium crystal structure
is fcc, space group L2$_{1}$, with the lattice parameter doubled
to $a=0.595$~nm compared to the bcc high temperature phase. The
L2$_{1}$-structure is most important for the predicted half metallic
properties of the Heusler compounds discussed in this book, since
the theoretical band structure calculations usually refer to this
structure. The L2$_{1}$ unit cell is depicted in Fig.\ref{fig:1-1}
and can be imagined to be combined of four interpenetrating fcc-sublattices
occupied by Mn, Al and Cu atoms respectively and shifted along the
space diagonal with the corner of the Al-sublattice at (0,0,0), the
first Cu-sublattice at (1/4, 1/4, 1/4), Mn at (1/2,1/2,1/2) and the
second Cu-sublattice at (3/4, 3/4, 3/4) . There exists a third intermediate
crystal structure for the Cu$_{2}$MnAl compound having B2-symmetry.
It has the same lattice parameter as the L2$_{1}$ phase, but in the
unit cell depicted in Fig.1 the Mn- and Al- atoms are distributed
at random on the Mn- and Al-sublattices whereas the two Cu-sublattices
remain intact.

In Cu$_{2}$MnAl the B2 and the L2$_{1}$ phases develop from the
high temperature bcc structure by chemical ordering of the atoms on
the 4 fcc-sublattices in two steps via two second order structural
phase transitions at 1040~K (bcc-B2) and 900~K (B2-L2$_{1}$) \cite{soltys81}.
The structural order parameter of these transitions can be calculated
by measuring the intensity of the superstructure Bragg reflection
(200) for the bcc-B2 transition or the Bragg reflection (111) for
the B2-L2$_{1}$-transition. Since in these order-disorder phase transitions
diffusion of atoms is involved, long time annealing or slow cooling
is necessary to achieve complete chemical ordering. In the majority
of the Heusler compounds these or similar structural phase transitions
exist, they can be hysteretic and of first order or reversible and
of second order \cite{zie86}. For several Heusler phases there in
addition exist low temperature structural distortions of the cubic
unit cell, the most prominent example being Ni$_{2}$MnGa with a martensitic
phase transition at 175~K. Ni$_{2}$MnGa is the only ferromagnetic
transition metal alloy which exhibits shape memory properties \cite{zheludev95}.
For Cu$_{2}$MnAl and other Heusler phases one often encounters problems
with the metallurgical stability. Below $T=900$~K the Cu$_{2}$MnAl
phase tends to decompose into $\beta$-Mn, Cu$_{9}$Al$_{4}$ and
Cu$_{3}$AlMn$_{2}$ \cite{kosubski82}. Experimentally one can overcome
this problem by suppressing the nucleation of the secondary phases
by rapid quenching the Heusler alloys from high temperatures.

The magnetism of the Heusler alloys is very versatile and has been
under continuous discussion all over the past hundred years. The majority
of the Heusler alloys with a magnetic element on the Y-position order
ferromagnetically, but several antiferromagnetic compounds also exists,
Ni$_{2}$MnAl or Pd$_{2}$MnAl provide examples \cite{ziebeck75,webster68}.
The main contribution to the magnetic moments in the Heusler phases
usually stems from the atoms at the Y-position. If magnetic atoms
also occupy the X-positions, their moment is usually quite small or
even vanishing. In the above mentioned Ni$_{2}$MnAl compound e.g.
the Ni-atoms are non-magnetic. Actually there exist a few Heusler
compounds with rather large magnetic moments on both the X- and the
Y-positions. In this case the ferromagnetic state is very stable and
the ferromagnetic Curie temperatures $T_{c}$ become exceptionally
high. The best examples are provided by the Heusler phases Co$_{2}$MnSi
and Co$_{2}$FeSi with a Co-moment of about $1$~$\mu$B and Curie
temperatures of 985~K \cite{webster71} and 1100~K \cite{buschow81},
respectively, the highest $T_{c}$-values known for the Heusler alloys.
The reason for this very stable ferromagnetism is a strong next nearest
neighbor ferromagnetic exchange interaction between the spins at the
X- and the Y-position. If a non-magnetic element occupies the X-position,
the dominant exchange interaction between the Y-spins is of weaker
superexchange type, mediated by the electrons of the non-magnetic
Z- and X-atoms \cite{kuebler83}.

Heusler compounds such as Cu$_{2}$MnAl with a magnetic moment only
on the Y-position are considered as good examples of localized $3d$
metallic magnetism \cite{kuebler83}. Since in the ideal L2$_{1}$
structure there are no Mn-Mn nearest neighbours, the Mn $3d$-wave
functions overlap only weakly and the magnetic moments remain essentially
localized at the Mn-position. However, in the family of Heusler compounds
there also exist beautiful examples of weak itinerant ferromagnetism
with strongly delocalized magnetic moments. The compound Co$_{2}$TiSn
with magnetic moments only on the X-positions belongs to this category
\cite{buschow83}. As it is obvious from the crystallographic structure
(Fig.\ref{fig:1-1}), there are nearest neighbor X-atoms making the
overlap of the $3d$-wave-functions and the delocalized character
of the $d$-electrons much larger than in the case of only the atoms
at the Y-position being magnetic. Replacing the Co-atoms in Co$_{2}$TiSn
by Ni, this delocalization effect proceeds further, making the compound
Ni$_{2}$TiSn a Pauli paramagnet \cite{pierre93}. Even more interesting,
the Heusler compounds Fe$_{2}$TiSn and Fe$_{2}$VAl also do not order
magnetically, but are marginally magnetic. They belong to the rare
class of transition metal compounds exhibiting heavy Fermion like
properties in the low temperature specific heat and the electrical
resistivity, which attracted much interest in the literature in recent
years \cite{nishino09,siebarski00,okamura00}.

Now, coming to the fully spin polarized Heusler compounds which are
of main interest in the present book, the first Heusler compounds
with a gap in the electron density of states for the spin down electron
band at the Fermi level were detected in 1983 by electron energy band
structure calculations \cite{deGroot83}. The first two compounds
with this spectacular property were NiMnSb and PtMnSb, so called semi
Heusler alloys (space group C1$_{b}$) where in the Heusler unit cell
of Fig.\ref{fig:1-1} one of the X-sublattices remains empty. The
semi Heusler structure only exists if the Z-position is occupied by
atoms with large atomic radii such as Sn or Sb. The NiMnSb and PtMnSb
compounds have been dubbed half metals \cite{deGroot83}, since only
for one spin direction there is metallic conductivity, for the other
spin direction the conductivity is of semiconducting type. In a ferromagnetic
transition metal alloy this half metallicity is a very rare property,
since usually $s$- or $p$-bands with a small exchange splitting
cross the Fermi energy and contribute states of both spin directions.
For several years PtMnSb and NiMnSb remained the only ferromagnetic
alloys with half metallic character, before starting from 1990 a second
group of half metallic Heusler alloys, Co$_{2}$MnSi, Co$_{2}$MnGe
and Co$_{2}$Mn(Sb$_{x}$Sn$_{1-x}$) was detected \cite{fujii90,ishida95b,ishida95}.\\
 The experimental proof of half metallicity in these Heusler alloys
is a long and still ongoing controversial issue. The first attempts
to prove the half metallicity used electron transport measurement
to test the existence of a gap in the spin down electron band \cite{hansen90,otto89}.
Since in the half metal for temperatures small compared to the gap
in the minority spin band there is only one spin direction at the
Fermi level available, it is expected that electronic scattering processes
involving spin flips and longitudinal spin wave excitations are inhibited.
Thus one should expect an increasing electron mobility and a change
of the power law describing the temperature dependence of the resistivity
when the temperature becomes smaller than the gap for the minority
spin band. Actually for NiMnSb this type of behavior for temperatures
below 80~K has been detected, additionally in this temperature range
the Hall coefficient exhibits an anomalous temperature dependence,
strongly suggesting a thermal excitation of charge carriers across
a gap coexisting with metallic conductivity \cite{hansen90,otto89}.
More directly, positron annihilation experiments on bulk single crystals
from the NiMnSb phase were found to be consistent with 100\% spin
polarization at the Fermi level \cite{hordequin96}. Finally, analyzing
the current-voltage characteristic below the superconducting gap of
a point contact between a Nb-superconductor and a bulk PtMnSb sample,
which is dominated by Andreev reflections at the ferromagnet/superconductor
interface, the authors derived a spin polarization of 90\% at the
Fermi level \cite{soulen98}.\\
 For the Co$_{2}$Mn(Si,Ge,Sn) fully spin polarized group of Heusler
compounds spin polarized neutron diffraction measurements on bulk
samples have been employed to determine the degree of spin polarization
at the Fermi level \cite{brown00}. This methods probe the spatial
distribution of the magnetization, details of which depend sensitively
on the spin polarization. The main conclusion is that the spin polarization
is large but not 100\%. More recent superconducting/ferromagnetic
point contact measurements on a Co$_{2}$MnSi single crystal gave
a spin polarization of 55\% \cite{ritchie03}. Similarly, the degree
of spin polarization determined from the analysis of spin resolved
photoemission spectra was always found to be definitely below 100\%
\cite{ristoiu00,zhu01}.

During the first years after the discovery of the half metallic character
in the Heusler compounds they were considered as exotic ferromagnets
of mainly academic interest. This attitude has changed completely
with the development of new ideas of data storage and processing designed
to use both, the charge and the spin degree of freedom of the conduction
electrons, nowadays called spin electronics (spintronics) \cite{prinz98,awschalom99}.
In the spintronic community there is a strong belief that in the future
these new concepts have the perspective to complement or even substitute
conventional Si technology. It was rapidly realized how valuable it
would be for spintronic devices to have a ferromagnet available with
only one conduction electron spin direction at the Fermi level. With
an electrode possessing 100\% spin polarization the generation of
a fully spin polarized current for spin injection into semiconductors
would be very easy \cite{schmidt00}, spin filtering and spin accumulation
in metallic thin film systems would be most effective \cite{cacho02}
and the tunneling magnetoresistance (TMR) \cite{moo98} of a device
prepared of two half metallic electrodes should have a huge magneto-resistance,
since the tunneling probability is vanishing to first order if the
two electrodes have an antiparallel magnetization direction.

Thus the novel concepts of spintronics started an upsurge of interest
in ferromagnetic half metals in the literature. Principally there
are three different classes of ferromagnetic half metals which are
under intense discussion in the literature and regarded as possible
candidates to be used in spintronic devices. These three groups combine
semiconductors such as GaAs or ZnO doped by magnetic transition metal
ions \cite{lee04}, magnetic oxides such as CrO$_{2}$ , Fe$_{3}$O$_{4}$
or (Sr$_{1-x}$La$_{x}$)MnO$_{3}$ \cite{schwarz86,coey99} and,
last but not least, the fully spin polarized Heusler compounds under
discussion in this book.

Actually with each of these three essentially different groups of
materials one encounters serious problems when thinking of possible
applications. The question which group or which compound is the best
is completely open. The magnetically doped semiconductors all share
the problem of magnetic inhomogeneity and rather low ferromagnetic
Curie temperatures, making them unsuitable for room temperature applications.
For the ferromagnetic oxides, on the other hand, the main problem
is that the preparation and handling of these compounds is completely
incompatible with current semiconductor technology. In this respect
the fully spin polarized Heusler compounds appear to be the better
choice, since they can be prepared by conventional methods of metallic
thin film preparation. A further advantage of the Heusler compounds,
especially in comparison with the magnetically doped semiconductors,
is their high ferromagnetic Curie temperature. However, as already
stated above and frequently discussed in the following chapters of
this book, the half metallicity of the Heusler compounds is a subtle
property which is easily lost in a real sample.

Recent intense work by theoretical groups, several of them represented
in this book, increased the number of Heusler compounds with predicted
half metallic properties to more than 20, among them Rh$_{2}$MnGe,
Fe$_{2}$MnSi and Co$_{2}$CrAl, to mention a few of the new compounds
\cite{galanakis02,galanakis02b,felser03,picozzi04,raphael02}. From
the experimental side, most of these phases have not been studied
thoroughly yet, thus leaving a vast field for future experimental
investigations. The experimental work in the literature mainly concentrated
on the classical Heusler half-metallic phases NiMnSb and PtMnSb, the
alloys Co$_{2}$MnSi and Co$_{2}$MnGe and the newly discovered compound
Co$_{2}$(CrFe)Al \cite{felser03}.

When thinking of possible applications in spintronic devices, which
at the moment is the main motivation behind the experimental research
on the fully spin polarized Heusler compounds in the literature, it
is obvious that one needs thin films of the Heusler compounds, bulk
samples are not very useful. This is the reason why in the majority
of the experimental chapters in this book thin film systems are studied.
Additional detailed research on high quality bulk material, however,
is highly welcome and urgently needed for purposes of comparison with
the thin films and in order to elucidate the basic properties of the
compounds. The recent XAFS and neutron scattering study of Co$_{2}$MnSi
single crystals provides a beautiful example \cite{ravel02,hordequin98}.
It allowed a precise determination of the number antisite defects
in the crystal (Co sitting on the Mn position and vice versa), which
would be very difficult in thin films, but is important for a realistic
judgment of the perspective to reach full spin polarization in a Co$_{2}$MnSi
thin films prepared under optimum conditions.

In the experimental chapters of this book it will become clear to
the reader that using the predicted full spin polarization of the
Heusler compounds in thin film structures is difficult and the technical
development is still in its infancy. Thin film preparation in general
and especially the preparation of thin film heterostructures, often
imposes limits on the process parameters and this might severely interfere
with the needs to have of a high degree of spin polarization. For
obtaining a large spin polarization a perfect crystal structure with
a small number of grain boundaries is important. This can best be
achieved by keeping the substrate at high temperatures during the
thin film deposition. However, most Heusler phases grow in the Vollmer-Weber
mode (three-dimensional islands) at high temperatures, thus when using
high preparation temperatures there might be a strong roughening of
the surfaces which for spintronic devices is strictly prohibited.
In addition, in thin film heterostructures combining different metallic,
semiconducting or insulating layers with the Heusler compounds, high
preparation temperatures are forbidden, since excessive interdiffusion
at the interfaces must be avoided.

Interfaces of the Heusler compounds with other materials are a very
delicate problem for spintronic devices. For spin injection into semiconductors
or a tunneling magnetoresistance the spin polarization of the first
few monolayers at the interfaces is of utmost importance. A large
spin polarization in the bulk of a Heusler compound does not guarantee
that it is a good spintronic material, unless it keeps its spin polarization
down to the first few monolayers at the interfaces. Thin films and
TMR-devices have first been investigated systematically using the
semi Heusler compounds PtMnSb and NiMnSb \cite{hordequin98}. The
performance was rather disappointing e.g. the maximum value for the
TMR achieved was of the order of 12\% only. Later, TMR devices based
on Co$_{2}$MnSi was shown to work much better \cite{schmalhorst04}
(see article by J. Schmalhorst in this book), but still the calculated
spin polarization is definitely smaller than 100\%.

This experience leads to the suspicion that at least for a few monolayers
at the interfaces the full spin polarization is somehow lost. One
possible explanation for this could be modifications in the electronic
energy band structure at surfaces and interfaces, since the theoretical
predictions of a full spin polarization in a strict sense only hold
for bulk crystals. Recent band structure calculations taking the surfaces
explicitly into account revealed that for certain crystallographic
directions the full spin polarization at the surfaces is actually
lost \cite{galanakis02c,ishida98}. Even more seriously, it seems
that atomic disorder of the Heusler compound at surfaces and interfaces
has the tendency to severely affect the full spin polarization. Theoretical
band structure calculations predicting the half metallicity assume
L2$_{1}$-symmetry with perfect site symmetry of all four X, Y and
Z sublattices. As mentioned above, the existence of some antisite
disorder, however, cannot be completely avoided even in well annealed
single crystals \cite{ravel02,hordequin98} .

An essential question is, which type of disorder is most detrimental
for the spin polarization of the Heusler compounds. Model calculations
taking different types of point defects into account are very illustrative
in this respect \cite{orgassa99,picozzi04} (see also the article
by S. Picocci in this book). These calculations show that for the
Co$_{2}$MnGe and the Co$_{2}$MnSi Heusler phase the electronic states
of Co-antisite atoms sitting on regular Mn-sites fill the gap in the
minority spin band. The magnetic moment of these Co antisite defect
remains virtually unchanged and couples ferromagnetically to the surrounding
spins. A Mn-antisite atom sitting on a regular Co-position, on the
other hand, does not introduce electronic states in the minority spin
gap, however has its magnetic moment coupled antiferromagnetically
to the surrounding spins. This leads to a drastic reduction of the
saturation magnetization, which frequently has been observed for ferromagnetic
Heusler phases not prepared under optimum conditions.

Introducing now the present article which reviews our work during
the last four years on thin films and multilayers based on the Heusler
alloys Co$_{2}$MnGe and Co$_{2}$MnSn, our main interest was not
on spintronic devices for the moment, although we will present examples
of GMR devices in chapter IV. We, instead, go one step backwards and
study first the basic structural and magnetic properties of very thin
films of these Heusler compounds. The magnetism of a ternary metallic
compound is complex and little is known of what happens with the magnetic
order in the limit of very thin films or at interfaces. Then we present
our results on the magnetic order in multilayers of the Heusler alloys
using V and Au as interlayers, which can be grown with high quality.
Neutron and x-ray scattering provide well established methods to get
insight into the structural and magnetic properties of the interfaces.
And, by the way, we follow the interesting and until now unsolved
question, whether an oscillating interlayer exchange coupling (IEC)
between the Heusler layers can be observed when varying the thickness
of the V- and Au-layers systematically.

In detail, our article below is organized as follows: In Sec. \ref{sec:single}
we discuss the magnetic and structural properties of single films
of the Co$_{2}$MnGe and Co$_{2}$MnSn phase for a thickness range
down to about $3$~nm. In Sec. \ref{sec:multi} we discuss the properties
of multilayers prepared by combining these Heusler layers with Au
and V interlayers. In this section we also review in some detail a
novel type of antiferromagnetic interlayer long range order which
we have observed in {[}Co$_{2}$MnGe/V]$_{n}$ multilayers and an
unidirectioal exchange anisotropy originating from the interface magnetic
order in the {[}Co$_{2}$MnGe/Au]$_{n}$ multilayers. In Sec. \ref{sec:transport}
we present results of our resistivity and magnetoresistivity measurements
on the single films, multilayers and GMR devices, before we summarize
and draw conclusions in Sec. \ref{sec:summary}.

\section{Single Co$_{2}$MnGe and Co$_{2}$MnSn thin films}

\label{sec:single}

\subsection{Structural properties}

\label{sec:single-1}

All films of the present study were grown by rf-sputtering with a
dual source HV sputtering equipment on single crystalline Al$_{2}$O$_{3}$
(11-20) surfaces (sapphire a-plane). The base pressure of the system
was $5\times10^{-8}$~mbar after cooling of the liquid nitrogen cold
trap. We used pure Ar at a pressure of $5\times10^{-3}$~mbar as
sputter gas, the targets were prepared from stoichiometric alloys
of the Heusler phases. The sputtering rates during the thin film preparation
were 0.06 nm/s for the Co$_{2}$MnSn phase and 0.04~nm/s for the
Co$_{2}$MnGe phase, the Au or V seed layers were deposited at a sputtering
rate of 0.05 and 0.03~nm/s respectively. The sapphire substrates
were cleaned chemically and ultrasonically after cutting and immediately
before the deposition they were additionally etched by an ion beam
for 300~s in order to remove any residual surface contaminations.
There are two conditions to achieve good structural and magnetic quality
of the Heusler layers. First, the substrate temperature must be high,
in our optimized procedure the substrate temperature was 500$^{\circ}$C.
Second, seed layers of a simple metal with a good lattice matching
to the Heusler compounds are important to induce epitaxial or textured
growth with a flat surface. Growing the films directly on sapphire
results in polycrystalline films and surface roughening. In our previous
investigations we tested different possible seed layers such as Cr,
Nb and Ag, later we concentrated on Au and V which gave the best structural
results \cite{geiersbach02,geiersbach03}. The surfaces of the Heusler
compounds oxidize rapidly and must be covered by a protection layer
before handling in air. If nothing else is stated we use an amorphous
Al$_{2}$O$_{3}$ film of 2-nm thickness as a cap layer.

\begin{figure}
\centering \includegraphics[height=8cm,keepaspectratio]{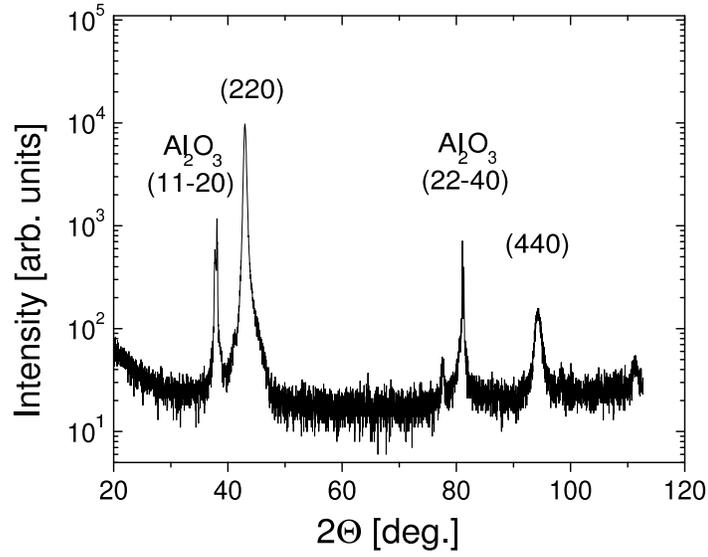}
\includegraphics[height=8cm,keepaspectratio]{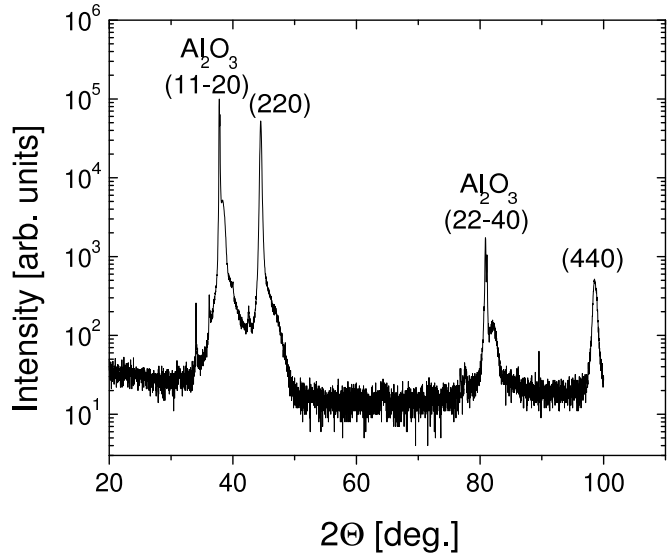}

\caption{Upper panel (a): Out of plane Bragg scans of a Au(3nm)/Co$_{2}$MnSn(100nm)
film measured with Cu K$_{\alpha}$ radiation. Bottom panel (b): same
for a V(3nm)/Co$_{2}$MnGe(100nm) film.}

\label{fig:2-1} 
\end{figure}

\begin{table}[ht]
 \centering \begin{tabular}{|c|c|c|c|c|c|}
\hline 
phase &
\multicolumn{2}{c|}{~lattice parameter {[}nm]~}&
\multicolumn{2}{c|}{~sat. magnetization {[}$\frac{emu}{g}$]~}&
~sat. moment {[}$\mu_{B}$]~ \rule{0mm}{5mm} \tabularnewline
\hline 
&
~bulk~ &
film &
~bulk~ &
film &
film \rule{0mm}{5mm} \tabularnewline
\hline 
Co$_{2}$MnGe &
~0.5743~ &
0.575~ &
111 &
114 &
5.02 \rule{0mm}{5mm} \tabularnewline
Co$_{2}$MnSn &
0.600 &
0.6001 &
91 &
92 &
4.95 \rule{0mm}{5mm} \tabularnewline
\hline
\end{tabular}

\caption{Lattice parameters, saturation magnetization and saturation magnetic
moments (per formula unit) of the pure, thick Heusler films in comparison
to the bulk values. The bulk values have been taken from reference
\protect{\cite{zie86}}.}

\label{tab1} 
\end{table}

\begin{figure}
\centering \includegraphics[height=8cm,keepaspectratio]{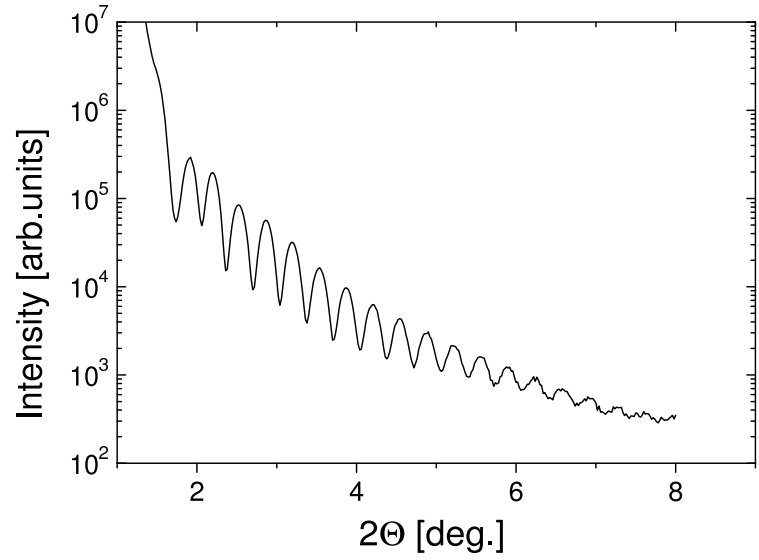}
\includegraphics[height=8cm,keepaspectratio]{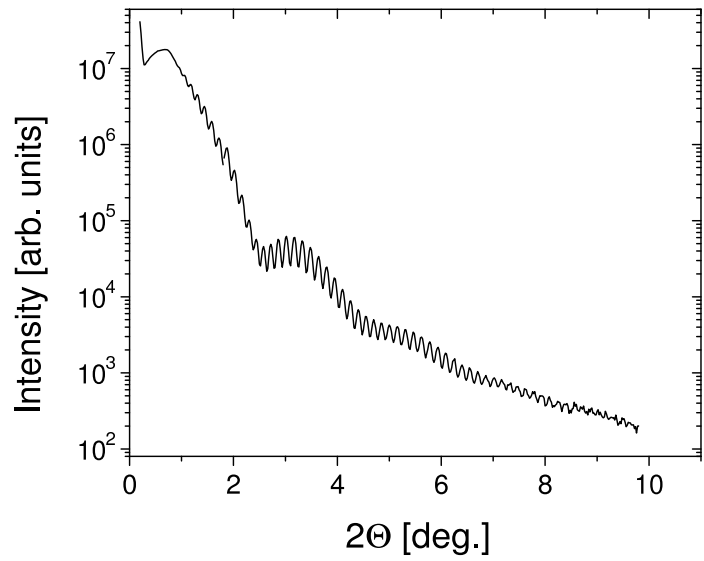}

\caption{X-ray reflectivity scans of Au(3nm)/Co$_{2}$MnSn(30nm) in the upper
panel (a) and V(4nm)/Co$_{2}$MnGe(60nm) in the bottom panel (b).}

\label{fig:2-2} 
\end{figure}

The structural quality of the samples was studied by small angle x-ray
reflectivity and high angle out-of-plane Bragg scattering using Cu
$K_{\alpha}$ radiation. Figure \ref{fig:2-1} show an out of plane
x-ray Bragg scan of a Co$_{2}$MnSn film with a 2-nm thick Au seed
layer and Co$_{2}$MnGe film with a 5-nm V seed layer. Only the (220)
and the (440) Bragg peaks of the Heusler phases are observed, proving
perfect (110) texture out-of-plane. In plane the films are polycrystalline.
Table \ref{tab1} summarizes the relevant structural data derived
from the x-ray scans. The lattice parameters virtually coincide with
those of the bulk phases. The half width of the rocking curve (width
in 2$\Theta$ at half maximum) characterizing the mosaicity of the
crystallite is 1.5$^{\circ}$ for the Co$_{2}$MnSn phase and 0.8$^{\circ}$
for the Co$_{2}$MnGe phase. The half width of the radial Bragg scans
corresponds to the experimental resolution of the spectrometer implying
that the total thickness of the Heusler layer is structurally coherent
in the growth direction. Examples of a small angle x-ray reflectivity
scan for films of the same phases are shown in figure \ref{fig:2-2}.
Total thickness oscillations are observed up to 2$\theta=10^{\circ}$,
indicatiove of a very flat surface morphology. additional oscillations
from the buffer layer are also observed. A simulation of the curves
with the Parratt formalism gives a typical rms roughness for the film
surface of about 0.3 nm. This is corroborated by atomic force microscopic
(AFM) pictures of the surface which reveal a very smooth surface morphology.
We also have grown films at lower preparation temperatures down to
a substrate temperature $T_{sub}=100^{\circ}$C. The structure of
films prepared below $T_{s}=500^{\circ}$C is still perfect (110)
texture, however with a rocking width slightly increasing with decreasing
preparation temperature.

\begin{figure}[ht]
 \centering \includegraphics[height=8cm]{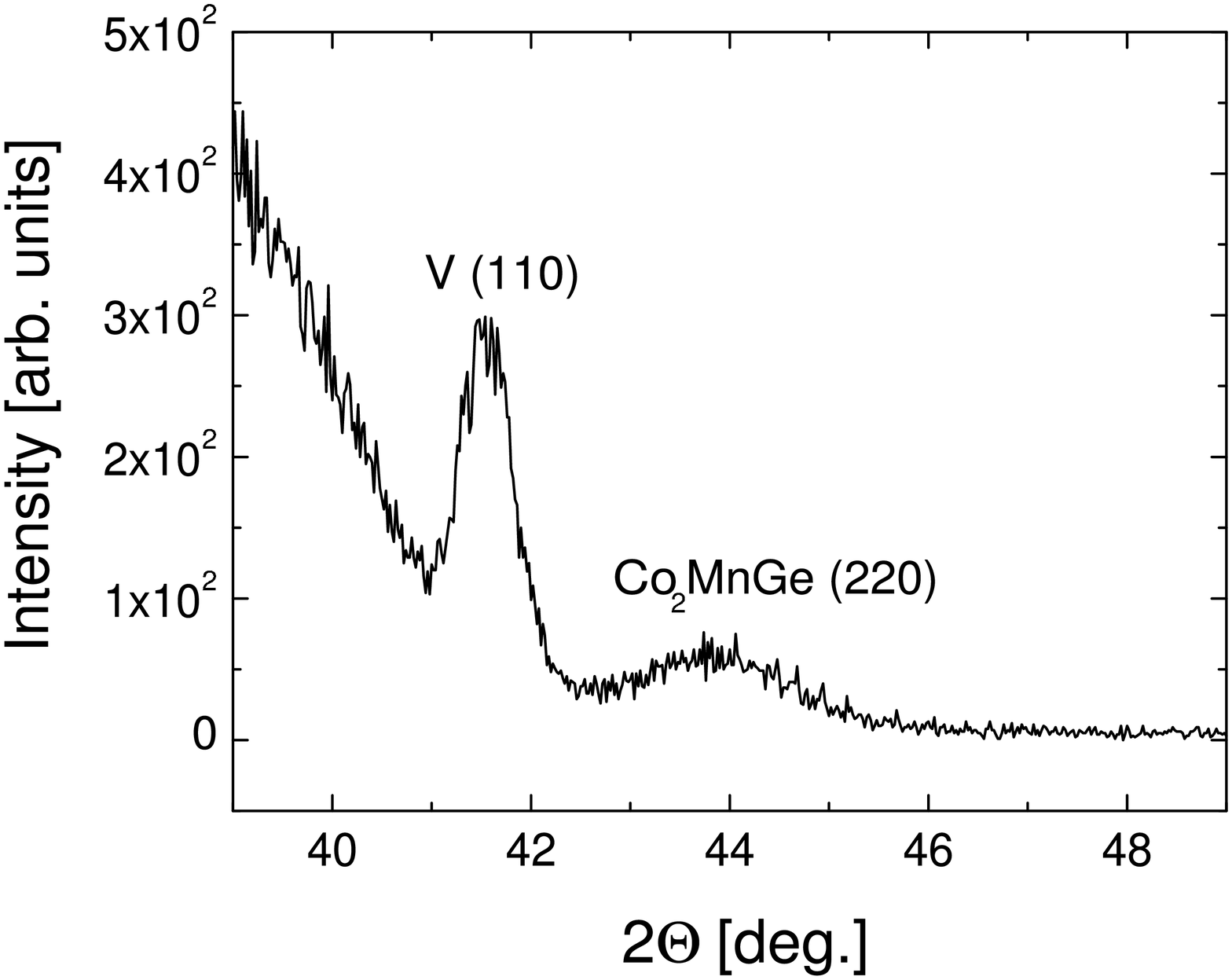}

\caption{Out of plane Bragg scan of a V(5nm)/Co$_{2}$MnGe(4nm)/V(5nm) trilayer.}

\label{fig:2-3} 
\end{figure}

Since for the multilayers discussed in the next chapter the typical
layer thickness of the Heusler layers is of the order of 3 nm only
and $T_{s}$ is limited to $300^{\circ}$C in order to avoid strong
interdiffusion at the interfaces, it seems worthwhile to study the
structural and magnetic properties of single very thin Heusler layers
prepared under the same conditions. Figure \ref{fig:2-3} shows an
out-of- plane Bragg scan of a trilayer V(5nm)/Co$_{2}$MnGe(4nm)/V(5nm).
One observes a very broad (220) Bragg peak with a half width of $\Delta2\Theta=2^{\circ}$
at 2$\Theta=43^{\circ}$ from the Co$_{2}$MnGe phase. Using the Scherrer
formula for the Bragg reflection of small particles, which correlates
the coherence length $L_{c}$ of the particles with the half width
of the Bragg peaks

\begin{eqnarray}
L_{c}=0.89\lambda/[\Delta(2\Theta)\cdot\cos(\Theta)]\end{eqnarray}

\noindent we derive $L_{c}=3$ nm i.e. a value slightly smaller than
the layer thickness. This is an indication that the mean size of the
crystalline grains in the film is slightly smaller than the film thickness.
For Co$_{2}$MnGe layers of 4-nm thickness between two Au layers,
which we also have studied, the grain size was found to be definitely
larger than the film thickness, since the correlation length is only
limited by the finite film thickness.

\begin{figure}
\centering \includegraphics[height=8cm,keepaspectratio]{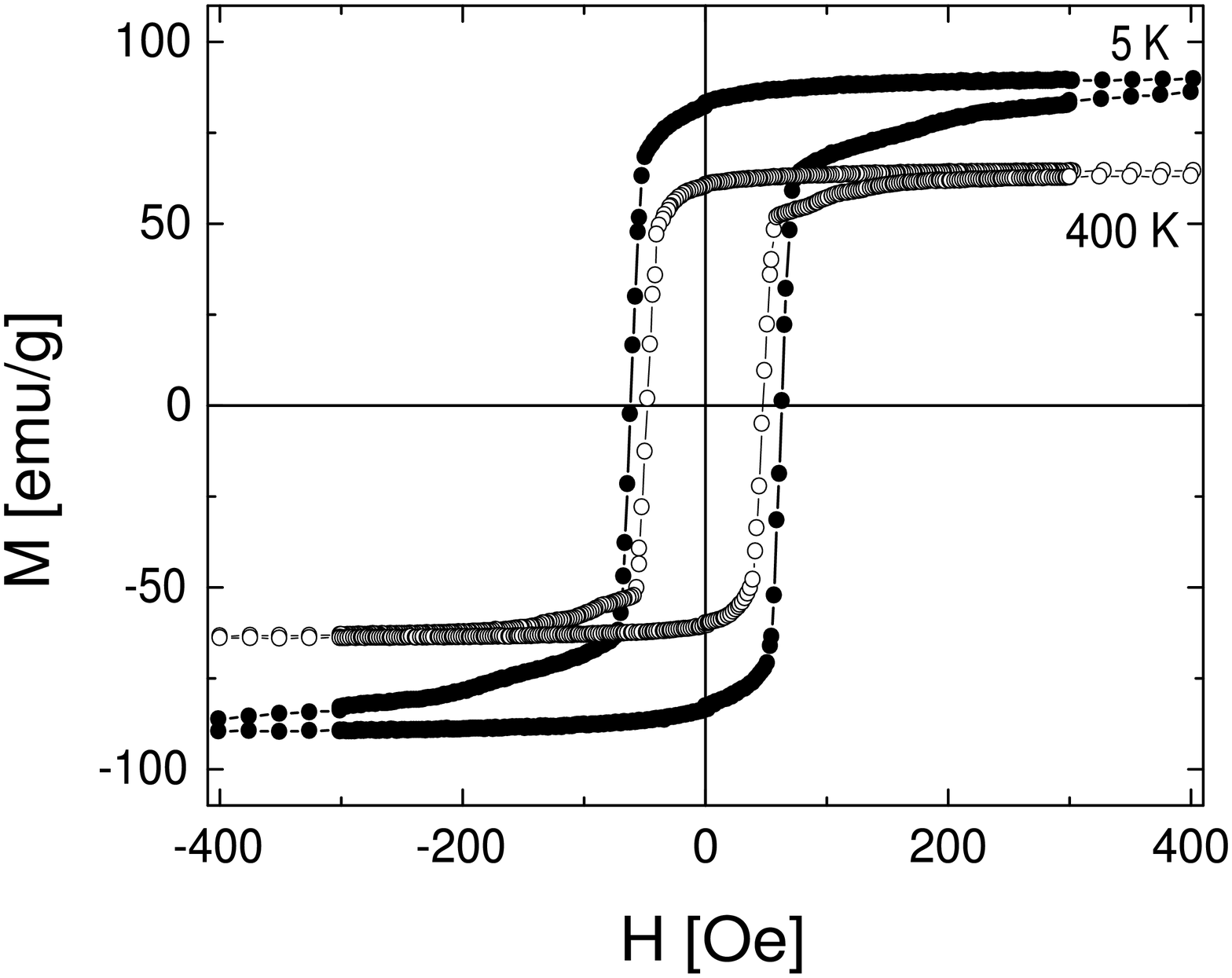}
\includegraphics[height=8cm,keepaspectratio]{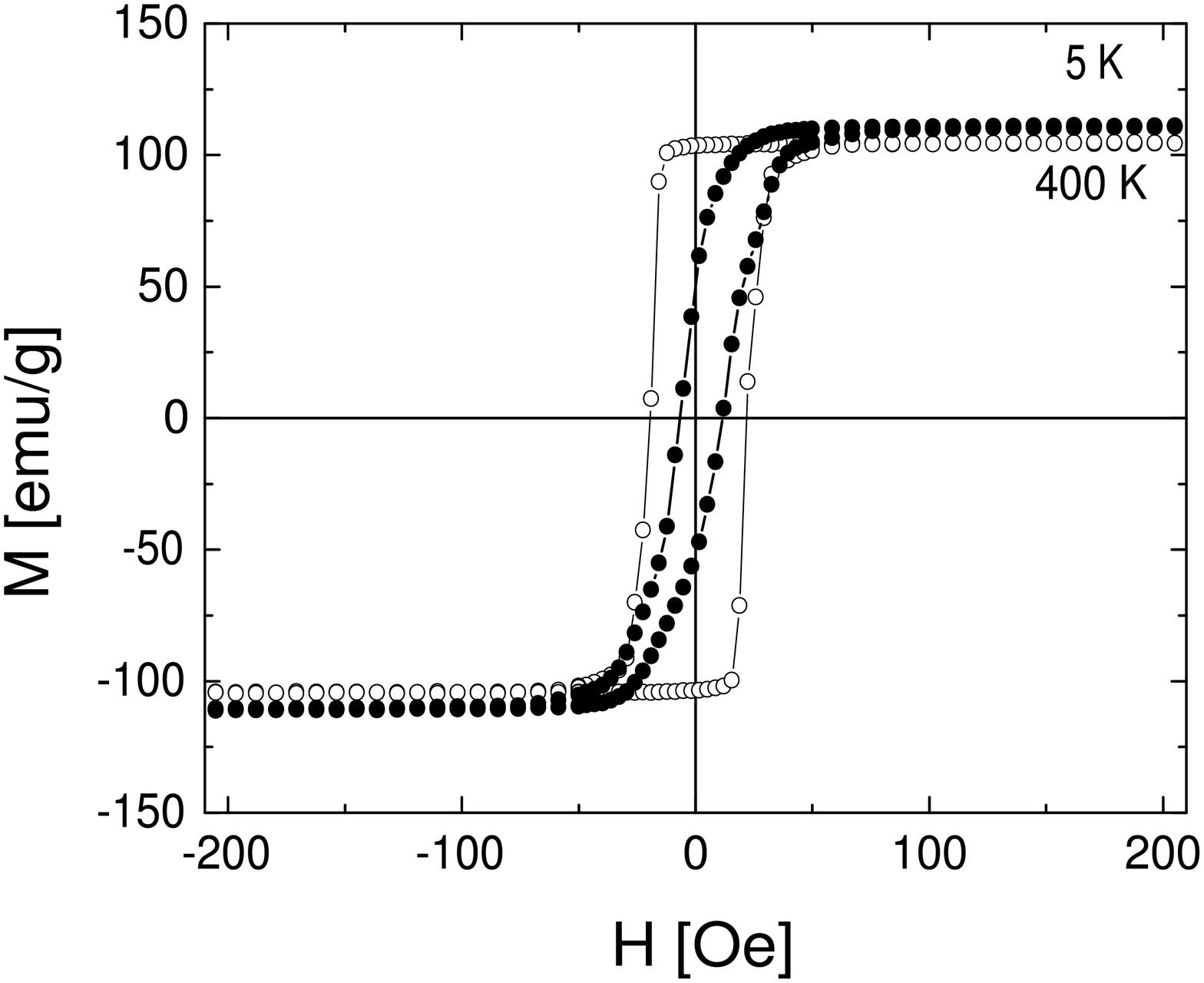}

\caption{Hysteresis loops of a Co$_{2}$MnSn in the upper panel (a) and a
Co$_{2}$MnGe film in the bottom panel (b) from figure \ref{fig:2-1}
measured at 5~K and 400~K.}

\label{fig:2-4} 
\end{figure}

\subsection{Magnetic properties}

\label{sec:single-2}

\begin{figure}
\centering \includegraphics[height=8cm]{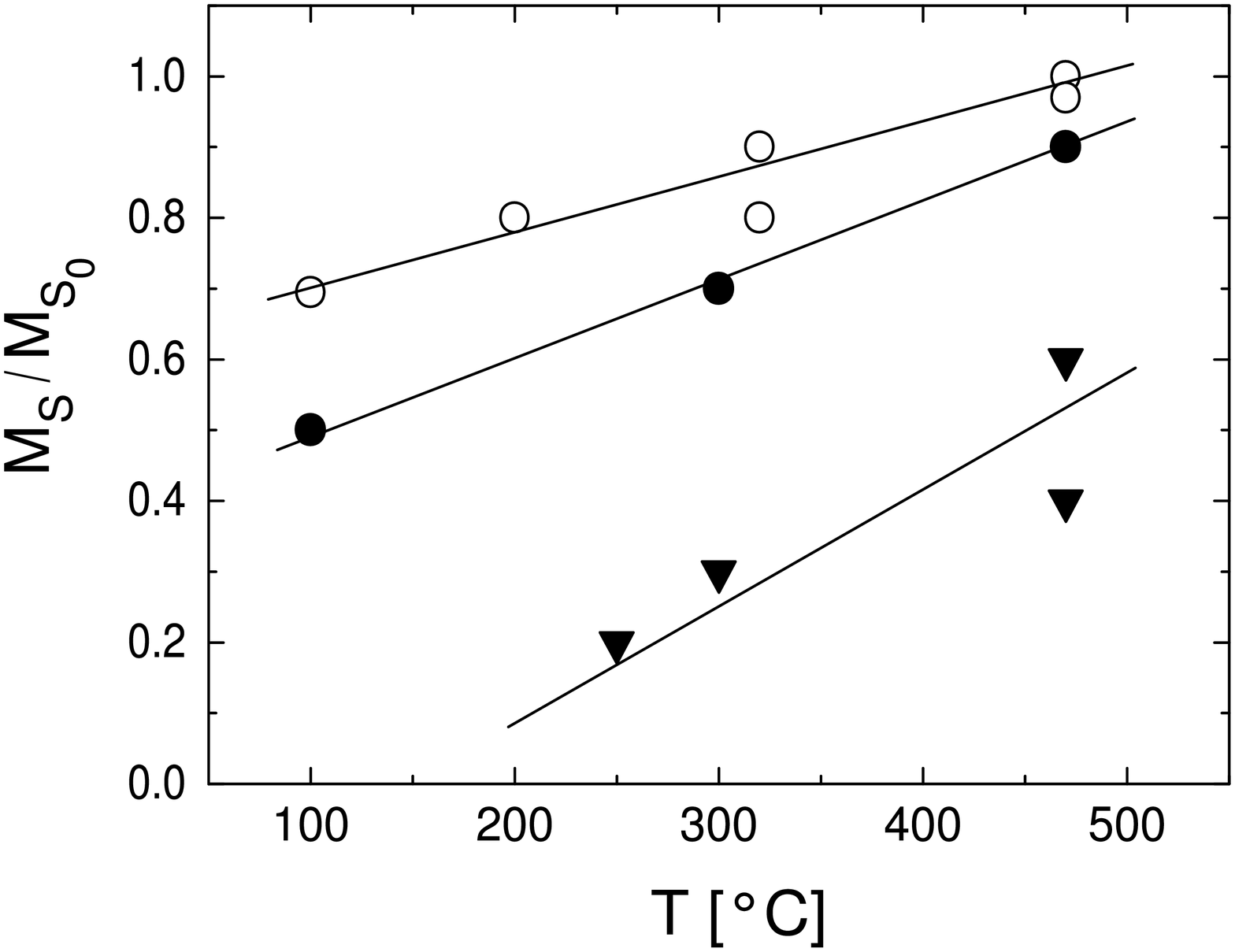}

\caption{Saturation magnetization of Co$_{2}$MnGe (open circles), Co$_{2}$MnSn
(solid circles) and Cu$_{2}$MnAl (triangles) versus the substrate
temperature during preparation.}

\label{fig:2-5} 
\end{figure}

The dc magnetization of our films was studied by a commercial superconducting
quantum interference device (SQUID) based magnetometer (Quantum Design
MPMS system). Examples of magnetic hysteresis loops of the Co$_{2}$MnGe
and the Co$_{2}$MnSn films are presented in figure \ref{fig:2-4}.
The films possess a growth induced weak, uniaxial anisotropy with
an anisotropy field $H_{K}$ of about 50~Oe. For the measurements
in figures \ref{fig:2-4} and \ref{fig:2-5} the external field axis
was directed parallel to the magnetic easy axis, thus the hysteresis
loops are rectangular. The coercive field for the Co$_{2}$MnGe phase
is $H_{c}=20$~Oe, for the Co$_{2}$MnSn film we get $H_{c}=50$~Oe
at room temperature. The saturation magnetization at 5~K is in good
agreement with the values measured in bulk samples within the experimental
error bars (see table \ref{tab1}). The saturation magnetic moment
per formula unit calculated from the saturation magnetization is 4.95~$\mu_{B}$
for Co$_{2}$MnSn and 5.02~$\mu_{B}$ for Co$_{2}$MnGe. These values
agree with those derived from the theoretical band structure calculations
for perfect L2$_{1}$ type of order \cite{galanakis02}, indicating
that the films have a high degree of metallurgical order. Thus we
can conclude that thick films of the Co based Heusler phases can be
grown at high temperatures with a quality comparable to bulk samples.

\begin{figure}
\centering \includegraphics[height=10cm]{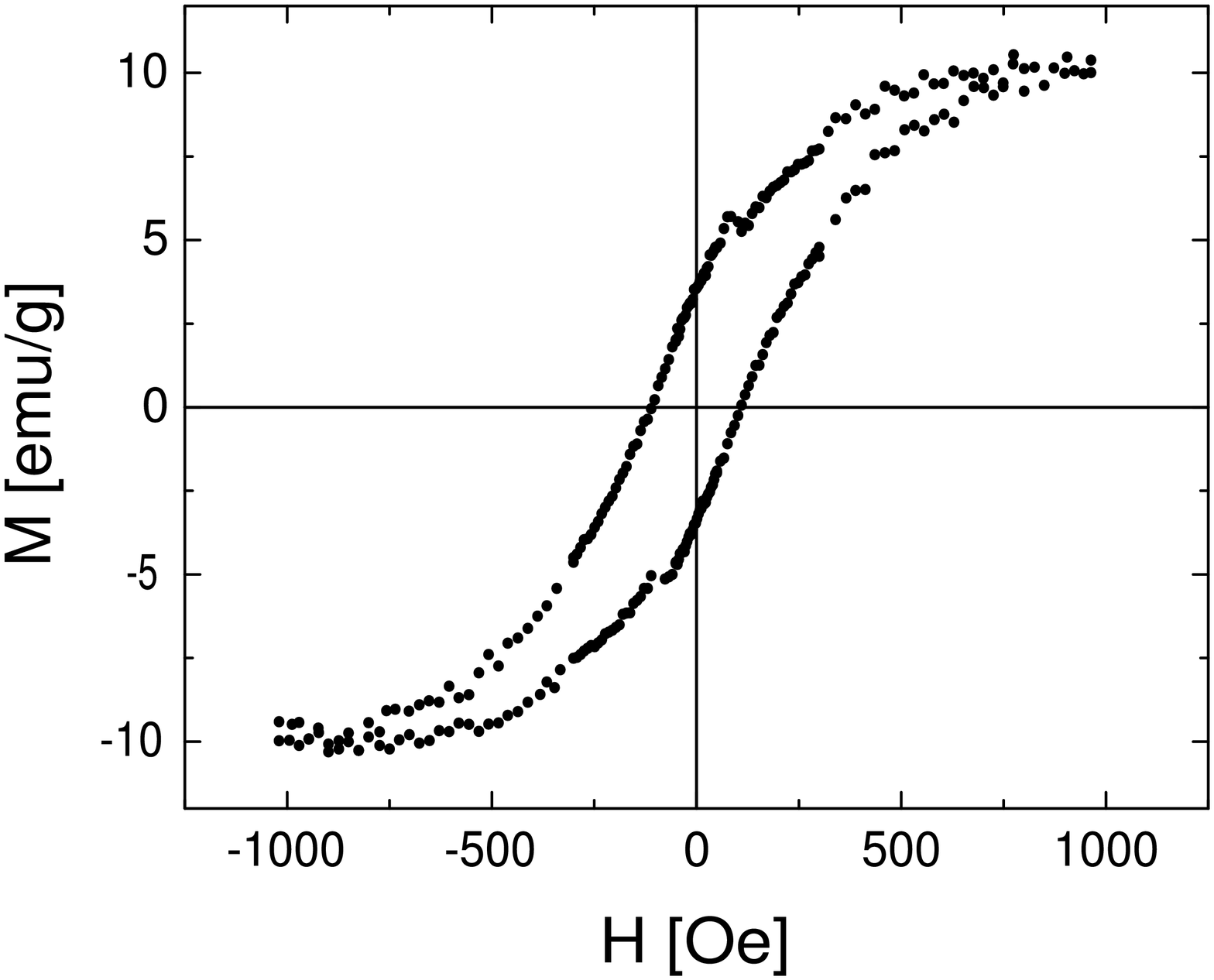}

\caption{Hysteresis loop of a 4-nm thick Co$_{2}$MnGe film grown directly
on sapphire a-plane at a substrate temperature of 300$^{\circ}$C
measured at 5~K.}

\label{fig:2-6} 
\end{figure}

\begin{figure}
\includegraphics[height=8cm,keepaspectratio]{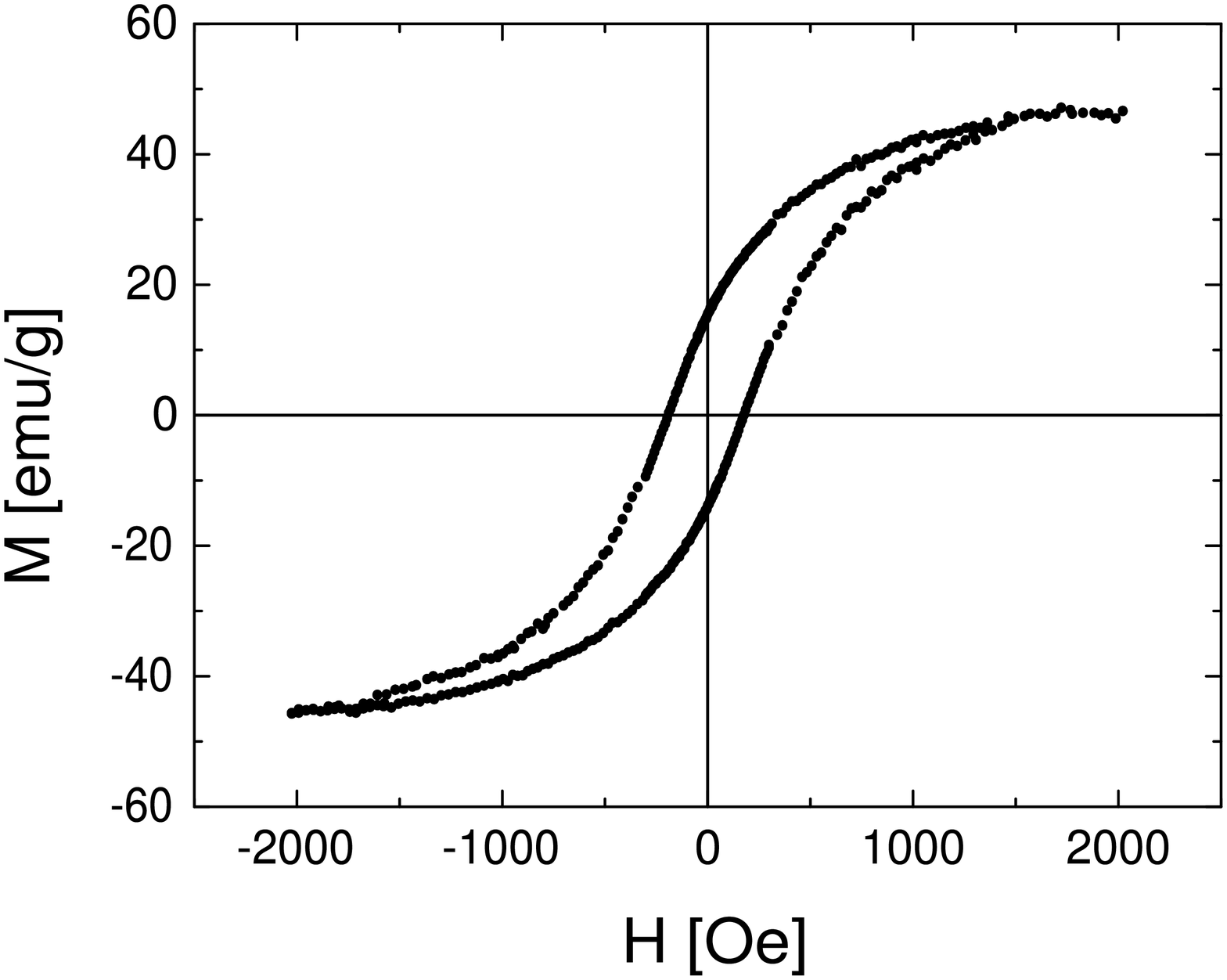} \includegraphics[height=8cm,keepaspectratio]{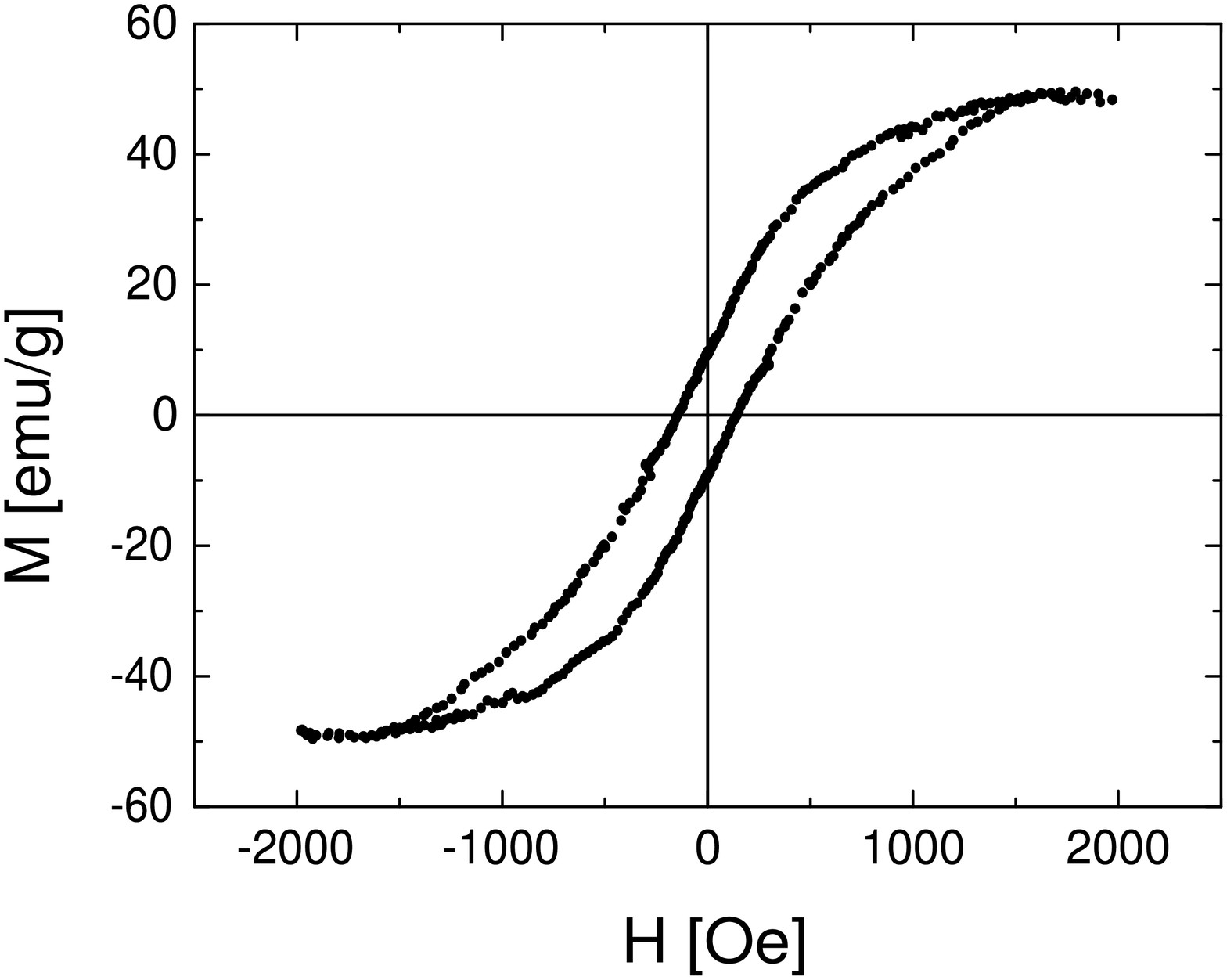}

\caption{Hysteresis loops measured at 5~K of V(3nm)/Co$_{2}$MnGe(4nm)/V(3nm)
with the field applied parallel to the sample plane in the upper panel
(a) and perpendicular to the sample plane in the bottom panel (b).}

\label{fig:2-7} 
\end{figure}

However, when decreasing the substrate temperature while keeping the
thickness of the Co$_{2}$MnGe layer and all other parameters constant,
we observe a continuous decrease of the saturation magnetization down
to about 70\% at $T_{s}=100^{\circ}$C (figure \ref{fig:2-5}). This
decrease of the magnetization is accompanied by an increase of the
lattice parameter of about 1\%. For the Co$_{2}$MnSn phase and Cu$_{2}$MnAl
the degradation of the ferromagnetic saturation magnetization with
decreasing $T_{s}$ is even faster (figure \ref{fig:2-5}). It seems
plausible to attribute the decrease of the saturation magnetization
to an increasing number of antisite defects in the L2$_{1}$ structure.
This effect is well known in Cu$_{2}$MnAl, where the disordered B2
phase which can be prepared by quenching from high temperatures has
a very low saturation magnetization \cite{goto95}. For films with
a small thickness of the Heusler phase of the order of a few nm the
situation becomes even worse. In this case a preparation at $500^{\circ}$C
is prohibited, since then interdiffusion at the seed layer/Heusler
interface is too strong. A practical limit for the substrate temperature
for avoiding excessive interdiffusion is $300^{\circ}$C. As a first
example of a very thin film, figure \ref{fig:2-6} depicts the hysteresis
loop measured at 5~K of a 4-nm thick Co$_{2}$MnGe film grown directly
on sapphire a-plane at $T_{s}=300^{\circ}$C. The x-ray structural
analysis showed no resolvable Bragg peak, thus the crystalline structure
seems to be polycrystalline with very small grains. The film has a
very low saturation magnetization of only about 10\% of the bulk value,
showing that the ferromagnetic properties are completely different
from those of the Co$_{2}$MnGe phase in the L2$_{1}$ structure.
Growing the same film thickness for Co$_{2}$MnGe on a V seedlayer,
a (220) Bragg peak can be observed (see figure \ref{fig:2-3}) and
about 50\% of the ferromagnetic saturation magnetization of the Co$_{2}$MnGe
phase is recovered (figure \ref{fig:2-7}). The magnetization is isotropic
in the film plane with a magnetic remanence of only about 30\% of
the saturation magnetization.

It is interesting to note that for a field direction perpendicular
to the film plane (figure \ref{fig:2-7}b) the hysteresis curves are
very similar to those observed in a parallel field, thus the strong
anisotropy of the demagnetising field characterizing a homogeneous
magnetic thin film is absent. This clearly shows that the magnetization
of the film in figure \ref{fig:2-7} is not homogeneous, but breaks
up into weakly coupled small magnetic particles pointing in their
own magnetically easy direction given by the geometric shape and the
crystal magnetic anisotropy. For a 4-nm thick Co$_{2}$MnGe film grown
on Au the small particle character of the hysteresis loop also exists,
but the magnetic anisotropy of a homogenous thin film is partly recovered.
For this film the ferromagnetic saturation magnetization is also strongly
reduced compared to the bulk value of Co$_{2}$MnGe. We would attribute
the different magnetic behaviour of Co$_{2}$MnGe grown on V and Au
to much smaller crystalline grains in the case of a V seed layer,
where the single grains seem to be nearly decoupled magnetically.

\subsection{XMCD on Co$_{2}$MnGe}

\label{sec:single-3}

In order to elucidate the microscopic origin of the moment reduction
in the Co$_{2}$MnGe Heusler alloy when prepared at low substrate
temperatures (see figure \ref{fig:2-5}) x-ray magnetic circular dichroism
(XMCD) is a very suitable experimental method as it allows an element
specific study of the magnetism \cite{stoehr99}. We therefore prepared
a Co$_{2}$MnGe film specially designed for an XMCD study with 16-nm
thickness grown on a V seed layer and with a 2-nm Au cap layer, all
layers prepared at $T_{s}=300^{\circ}$C. The saturation magnetization
of the film as measured by SQUID magnetometry was found to be slightly
smaller than the value given in Fig.5, it corresponds to a magnetic
moment per formula unit of 2.3~$\mu_{B}$ at room temperature and
2.98~$\mu_{B}$ at 4~K.

\begin{figure}
\centering \includegraphics[height=8cm]{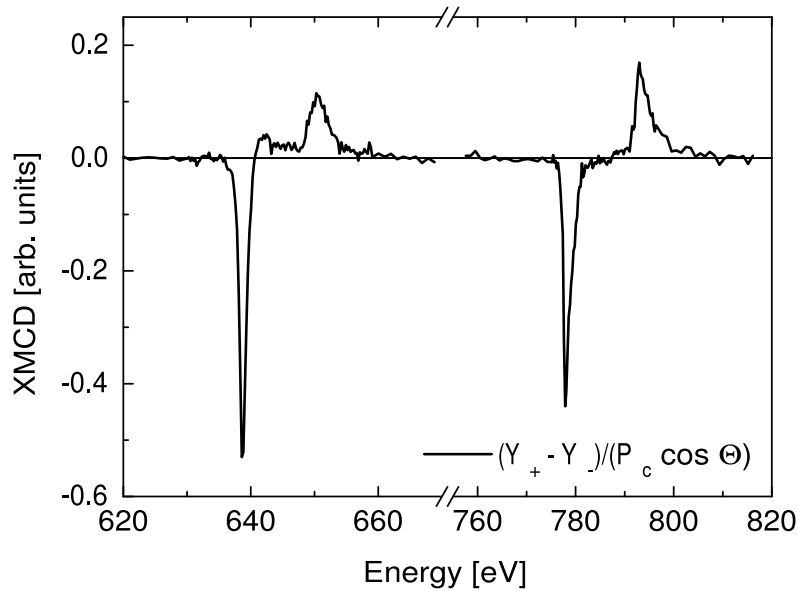}

\caption{XMCD spectra of a 11-nm thick Co$_{2}$MnGe film at the Mn L$_{3,2}$
and at the Co L$_{3,2}$ edge.}

\label{fig:2-8} 
\end{figure}

The XMCD measurements were performed at the bending magnet beamline
PM3 at BESSY~II (Berlin Germany) using the new ALICE chamber for
spectroscopy and resonant scattering. \cite{grabis03}. The measurements
were taken by the total electron yield (TEY) method. At an angle of
incidence of 40$^{\circ}$ saturation effects are small and the TEY
is proportional to the absorption coefficient to a good approximation.
During the experiment the helicity of the photons was fixed whereas
the magnetization of the sample was switched by a magnetic field of
$\pm0.1$~T thus providing the electron yield with the magnetization
parallel (Y$_{+}$) and antiparallel (Y$_{-}$) to the photon helicity.
The Y$_{+}$ and Y$_{-}$ scans measured at the $L_{3}$ edge of Mn
and Co are normalized to the flux of the incoming photon beam. The
XMCD spectrum (Y$_{+}$-Y$_{-}$) at the $L_{2,3}$ edges of Mn and
Co measured at room temperature is plotted in figure \ref{fig:2-8}.
The XMCD spectra contain quantitative information on the spin and
orbital magnetic moments which can be extracted via the sum rule analysis
\cite{grabis04}. There are several sources of systematic errors in
this analysis which might affect the absolute values for the magnetic
moment questionable. This is first the number of holes in the $d-$band,
which we can precisely take from electronic band structure calculations.
Second, it is the neglecting of magnetic dipolar interactions in the
model, which in our case seems justified because of the cubic symmetry
of the Co$_{2}$MnGe phase. Third, for the Mn atom there might be
a mixing of the $L_{3}$ and the $L_{2}$ levels by the relatively
strong $2p-3d$ Coulomb interactions. The correction factor $x$ taking
this effect into consideration has been calculated ranging from $x=1$
for negligible $jj$-mixing to $x=1.5$ for strong $jj$-mixing \cite{duerr97}.
Keeping these reservation in mind, the sum rule analysis yields for
the case of the Co atom $m_{spin}=0.55~\mu_{B}$ for the spin magnetic
moment and $m_{orb}=0.028~\mu_{B}$ for the orbital magnetic moment.
For the case of the Mn atom the analysis yields $m_{spin}=0.98~\mu_{B}$
(1.47 $\mu_{B}$) and $m_{orb}=0.056~\mu_{B}$, where for the first
value it is assumed that $x=1.0$ holds, for the value given in brackets
$x=1.5$ is assumed. Summing up all values for the atomic magnetic
moments and extrapolating to 4~K we determine a saturation magnetic
moment of $m=0.75~\mu_{B}$ for Co and $m=1.36~\mu_{B}$ (1.97~$\mu_{B}$)
for Mn \cite{grabis04}. The moment for Co agrees reasonable with
the theoretical value from band structure calculations, for the Mn
atom the theoretical calculations give $m=3.6~\mu_{B}$ i.e. a much
higher value than the experiment, irrespective of the exact value
of the correction factor $x$. The magnetization data yielded a saturation
magnetic moment of 2.98 $\mu_{B}$ per Co$_{2}$MnGe formula unit,
the XMCD results yield 2.83 $\mu_{B}$ (3.46 $\mu_{B}$), i.e. within
the uncertainty range of the XMCD result the agreement is satisfactory.

As mentioned in the introduction, the theoretical model calculations
\cite{orgassa99,picozzi04} show that antisite disorder in the L2$_{1}$
structure severely affects the Mn magnetic moments, since a Mn spin
sitting on a regular Co position has a spin direction antiparallel
to the other Mn and Co spins. This will strongly reduce the experimentally
determined mean Mn moment. The value of the Co moment when sitting
on a Mn position remains essentially unaffected. Thus the plausible
hypothesis formulated above that a lower preparation temperatures
causes site disorder in the L2$_{1}$ structure and concomitantly
a lowering of the saturation magnetization finds strong support from
the XMCD results and the theoretical model calculations.

Summarizing this chapter, we have shown that with optimized preparation
conditions high quality thick films of the Co$_{2}$MnGe and the Co$_{2}$MnSn
phase can be grown. But if experimental constraints are imposed when
preparing devices such as limits for the substrate temperature, non
applicability of seed layers or if in devices very thin Heusler layers
are needed, one faces problems. Site disorder in the interior of the
films and mixing and disorder at interfaces have the tendency to lower
the ferromagnetic magnetization. The magnetic behaviour of very small
Co$_{2}$MnGe grains as e.g. the nearly complete loss of ferromagnetism
for Co$_{2}$MnGe deposited on bare sapphire or the typical small
particle magnetic behaviour in very thin Co$_{2}$MnGe grown on V
suggest that the grain boundaries are only weakly ferromagnetic or
even non ferromagnetic.

\section{Multilayers {[}Co$_{2}$MnSn(Ge)/Au(V)]$_{n}$}

\label{sec:multi}

\subsection{Structural and magnetic properties}

\label{sec:multi-1}

Multilayers of the two Heusler phases Co$_{2}$MnGe and Co$_{2}$MnSn
with V and Au as interlayers have been prepared by the same dual source
rf-sputtering equipment described in section \ref{sec:single}. For
the multilayer preparation the substrate holder is swept between the
two targets, if nothing else is stated, the number of bilayers prepared
is $n=30$ starting with either V or Au, and the films are protected
by a Al$_{2}$O$_{3}$ cap layer. The substrate temperature was held
fixed at $T_{S}=300^{\circ}$C for all multilayers, the deposition
rates of the materials were the same as given in section \ref{sec:single}
for the single layers. Although the structural quality of the Heusler
layers improve at higher substrate temperatures, $T_{S}=300^{\circ}$C
turned out to be the upper limit when strong interdiffusion at the
interfaces must be avoided.

Using the natural gradient of the sputtering rate, the simultaneous
preparation of up to 10 samples with the thickness of either the magnetic
layer or the non magnetic layer altered is possible. The thickness
can be varied by a factor of three and we exploit this feature e.g.
for the preparation of several series of multilayers for the investigation
of the thickness dependence of the magnetic interlayer coupling. In
our previous investigations we have also tested the growth of Heusler
multilayers with several other combinations of materials, including
the Heusler compounds Co$_{2}$MnSi and Cu$_{2}$MnAl and Cr as interlayers
\cite{geiersbach02,westerholt03}. Here we concentrate on Co$_{2}$MnGe
and Co$_{2}$MnSn with Au and V as interlayers which can be grown
with best structural quality. The x-ray characterization of the multilayers
was done by a standard thin film x-ray spectrometer or using synchrotron
radiation at the Hasylab in Hamburg, Germany.

\begin{figure}
\centering \includegraphics[height=8cm,keepaspectratio]{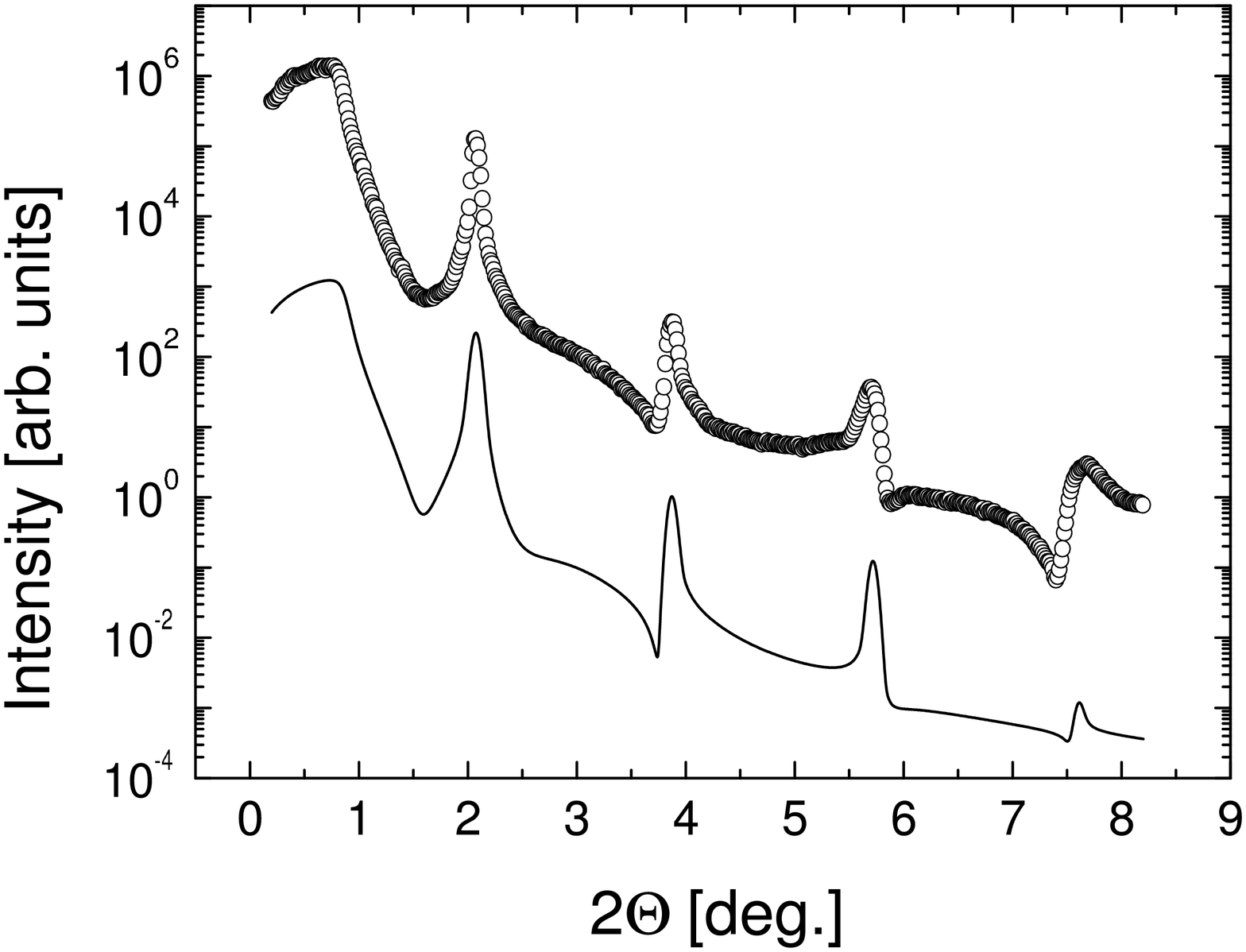}
\includegraphics[height=8cm,keepaspectratio]{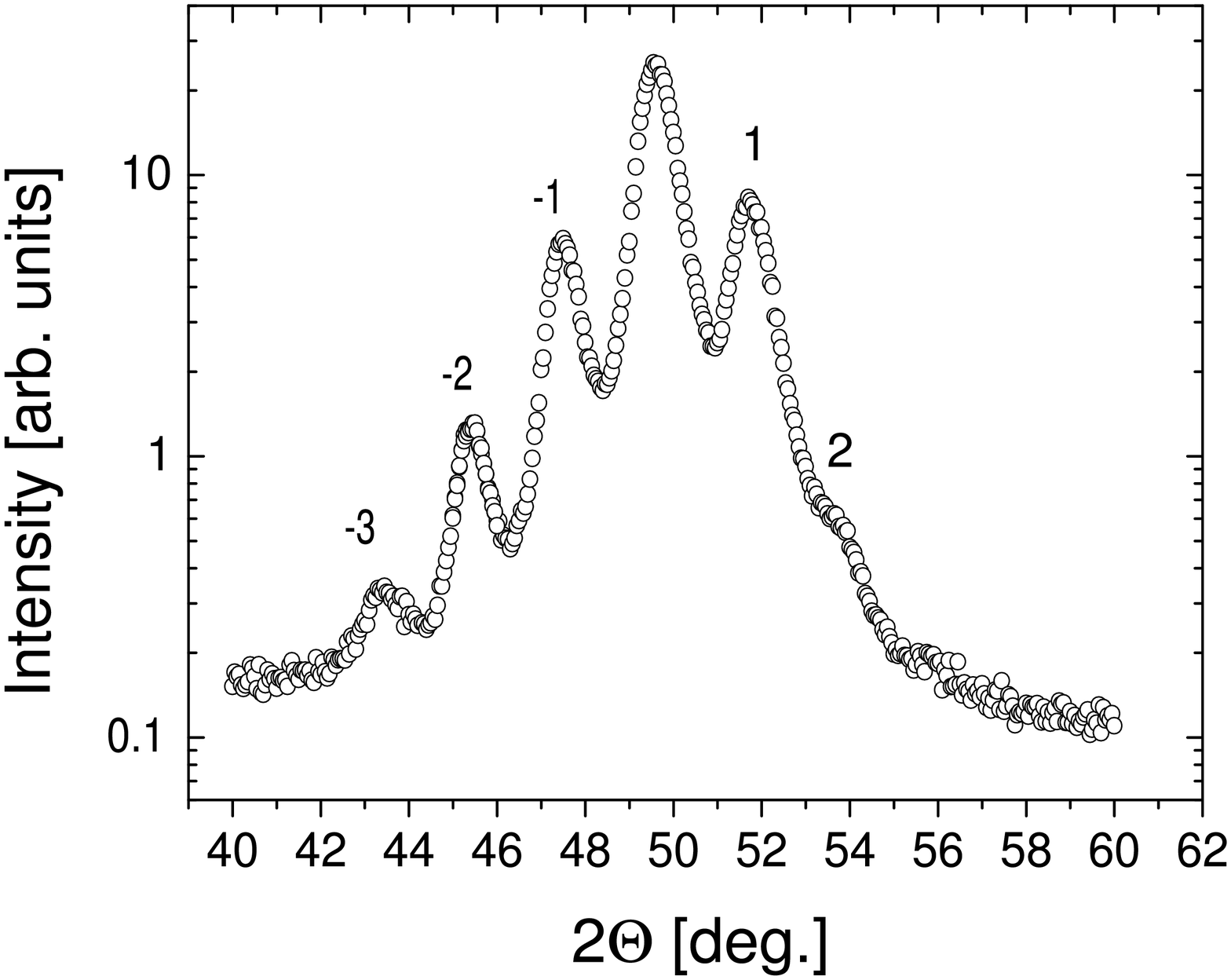}

\caption{Upper panel (a): X-ray reflectivity scan of a {[}Co$_{2}$MnGe(3nm)/V(2nm)]$_{50}$
multilayer measured at a wavelength of the synchroton radiation $\lambda=1,77$
nm. The open symbols are the measured intensity, the straight line
shows a simulation. Bottom panel: (b) Out of plane Bragg scan of the
same multilayer at $\lambda=1,77$ nm. The numbers in the figure denote
the order of the superlattice reflections and the order of the satellites.}

\label{fig:3-1} 
\end{figure}

Figure \ref{fig:3-1}a shows a low angle x-ray reflectivity scan measured
on a {[}Co$_{2}$MnGe(3nm)/V(2nm)]$_{50}$ multilayer (the number
in round brackets denotes the nominal thickness of the single layers,
$n=50$ is the number bilayers) by synchrotron radiation at a wavelength
of 0.177 nm. One observes superlattice reflections up to the 4th order
revealing a smooth layered structure. From a fit with the Parratt
formalism shown by the displaces thin line one derives a thickness
of $d_{V}=2.2$~nm and $d_{Heusler}=2.8$~nm and a roughness parameter
of $\sigma=0.5$~nm for V and $\sigma=0.7$~nm for Co$_{2}$MnGe.
Thus at the interfaces there is interdiffusion and/or topological
roughness on a scale of about 0.6~nm.

\begin{figure}
\centering \includegraphics[height=8cm,keepaspectratio]{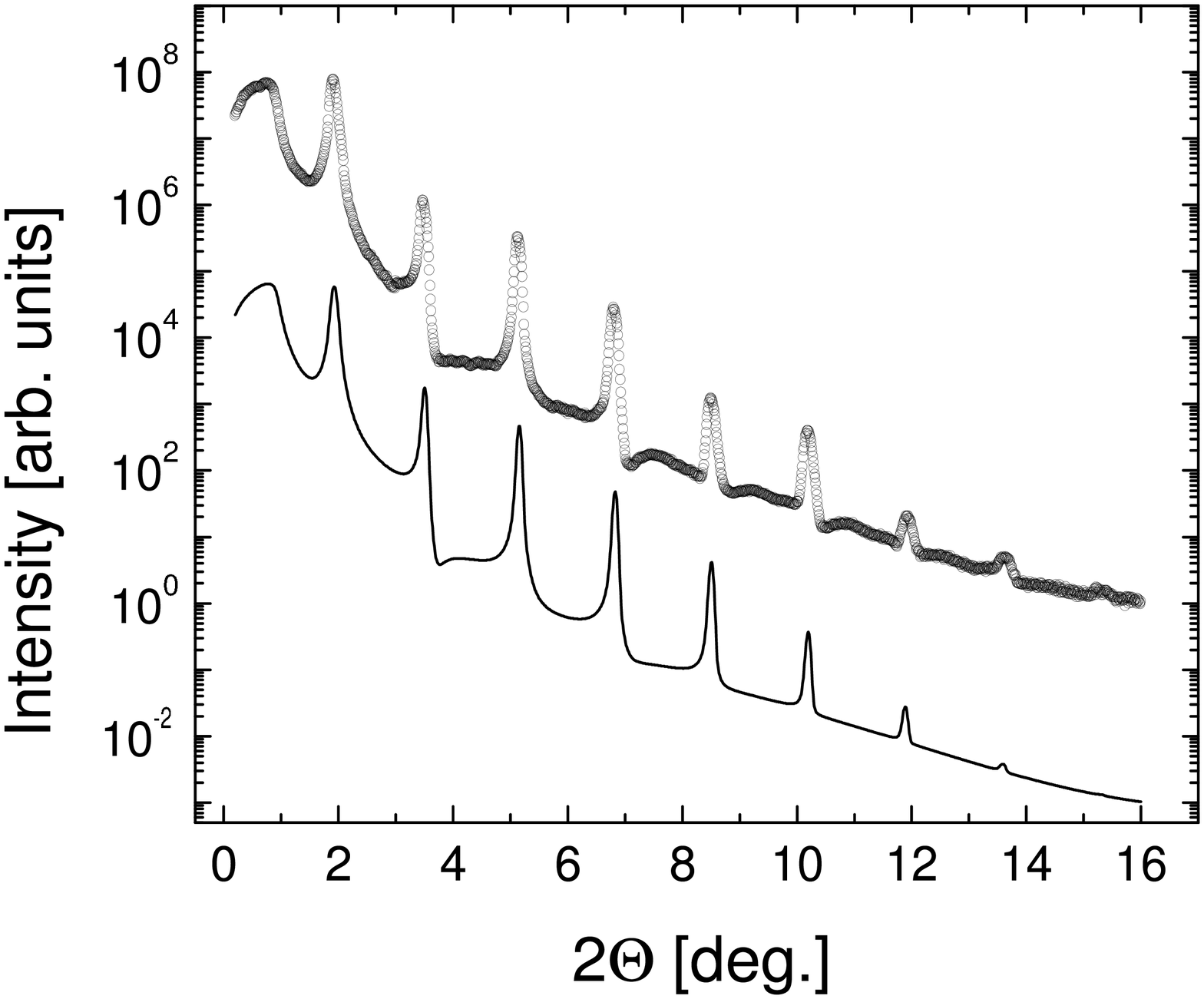}
\includegraphics[height=8cm,keepaspectratio]{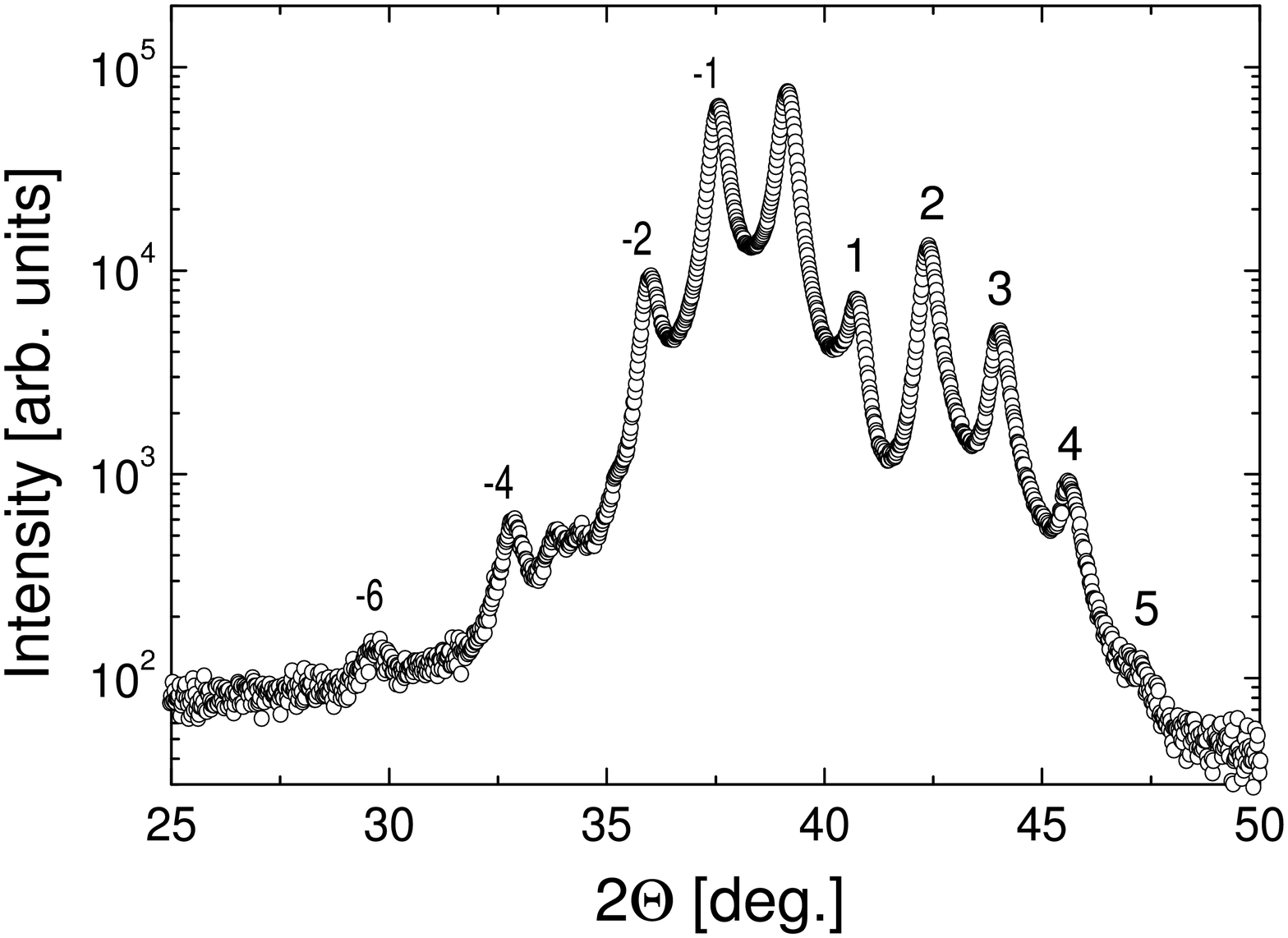}

\caption{Upper panel (a): X-ray reflectivity scan of a {[}Co$_{2}$MnGe(3nm)/Au(2.5nm)]$_{50}$
multilayer measured at $\lambda=1,54$~nm. The open symbols are the
measured intensity, the straight line shows a simulation. Bottom panel
(b): Out of plane Bragg scan of the same multilayer at $\lambda=1,54$~nm.}

\label{fig:3-2} 
\end{figure}

Figure \ref{fig:3-1}b depicts a scan across the (220)/(110) Bragg
reflection of the same multilayer. There is a rich structure with
satellites up to 3rd order proving the growth of coherent V and Co$_{2}$MnGe
layers. The out of plane coherence length derived from the width of
the Bragg peak is about 13~nm i.e. comprises about 3 bilayers. Figure
\ref{fig:3-2}a shows the small angle reflectivity scan for a nominal
{[}Co$_{2}$MnGe(3nm)/Au(2.5nm)]$_{50}$ multilayer. The fit with
the Parratt formalism gives a thickness of 3.0~nm and 2.3~nm for
the Heusler and Au layer respectively. As already evident from the
narrower superlattice reflection peaks, the quality of the layered
structure is even better than for the {[}Co$_{2}$MnGe/V] multilayer
of figure \ref{fig:3-1}a. This is corroborated by the simulation
which gives a roughness parameter $\sigma_{Au}=0.5$~nm and $\sigma_{Heusler}=0.3$~nm.
The Bragg scan (figure \ref{fig:3-2}b) at the (111)/(220) Bragg position
also reveals an out of plane coherent crystalline lattice with a coherence
length of 14~nm.

We have also grown successfully {[}Co$_{2}$MnSn/V] and {[}Co$_{2}$MnSn/Au]
multilayers with comparable quality \cite{geiersbach03} i.e. with
an out of plane coherence length and a roughness parameter similar
to the {[}Co$_{2}$MnGe/V] and {[}Co$_{2}$MnGe/Au] multilayers. Multilayers
of similar quality can be grown in the thickness range of down to
1~nm for the Heusler compounds and the interlayers, for a smaller
thickness the layered structure and the crystalline coherence is gradually
lost.

\begin{figure}
\centering \includegraphics[height=8cm,keepaspectratio]{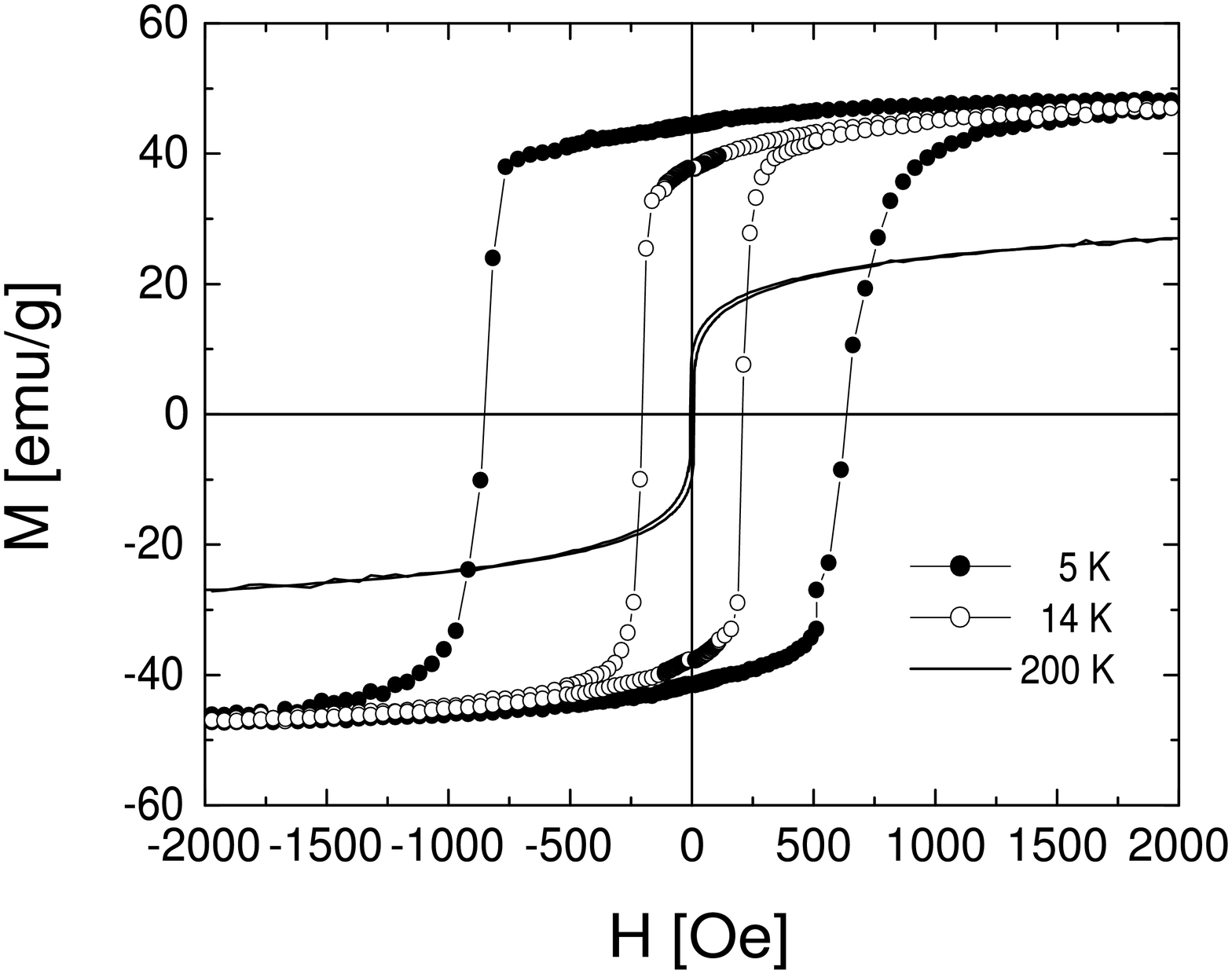}
\includegraphics[height=8cm,keepaspectratio]{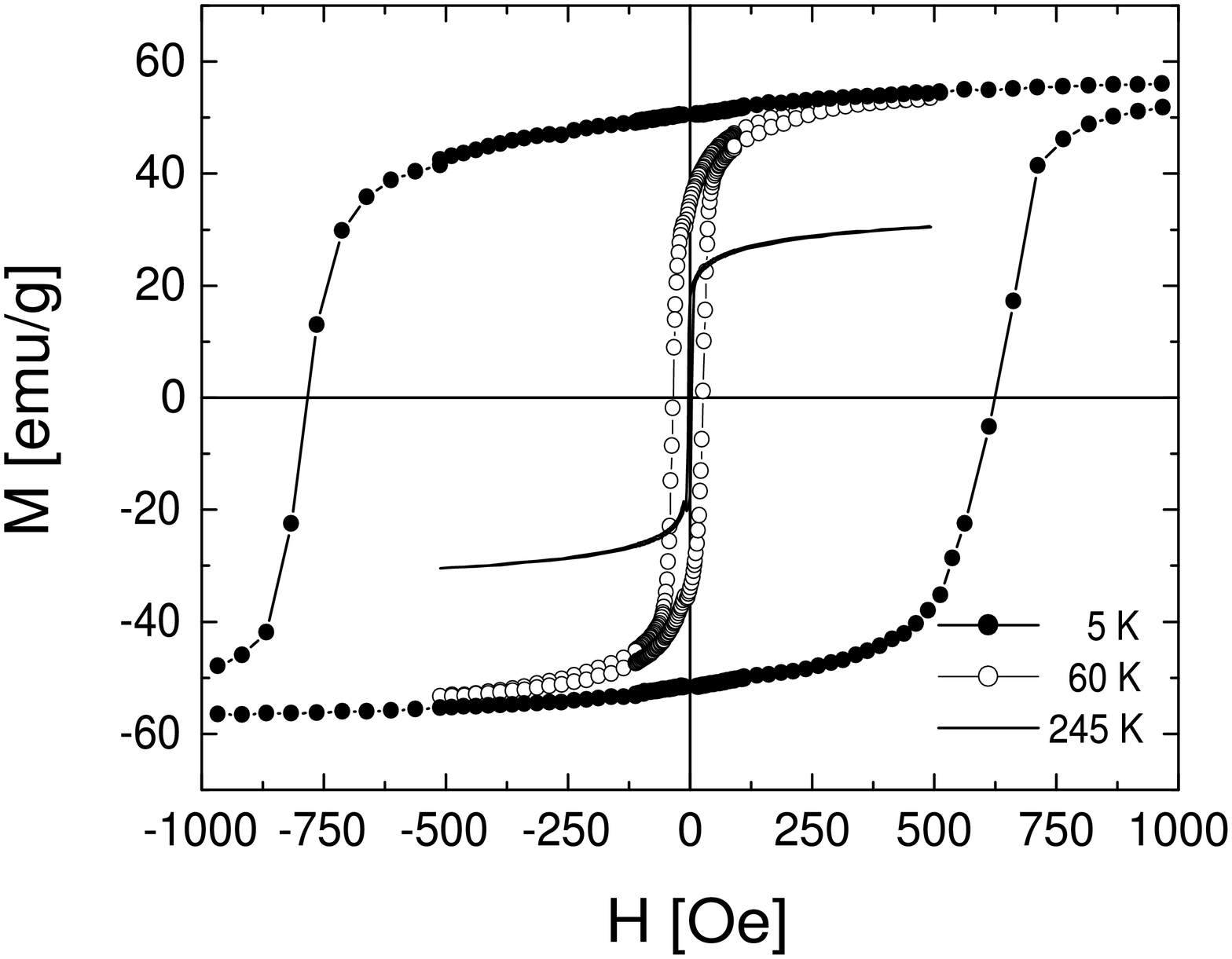}

\caption{Hysteresis loops of the multilayers {[}Co$_{2}$MnGe(2.2nm)/Au(3nm)]$_{30}$
in the upper panel (a) and {[}Co$_{2}$MnSn(3nm)/Au(1.5nm)]$_{30}$
in the bottom panel (b) measured at different temperatures given in
the figure.}

\label{fig:3-3} 
\end{figure}

\begin{figure}
\centering \includegraphics[height=8cm,keepaspectratio]{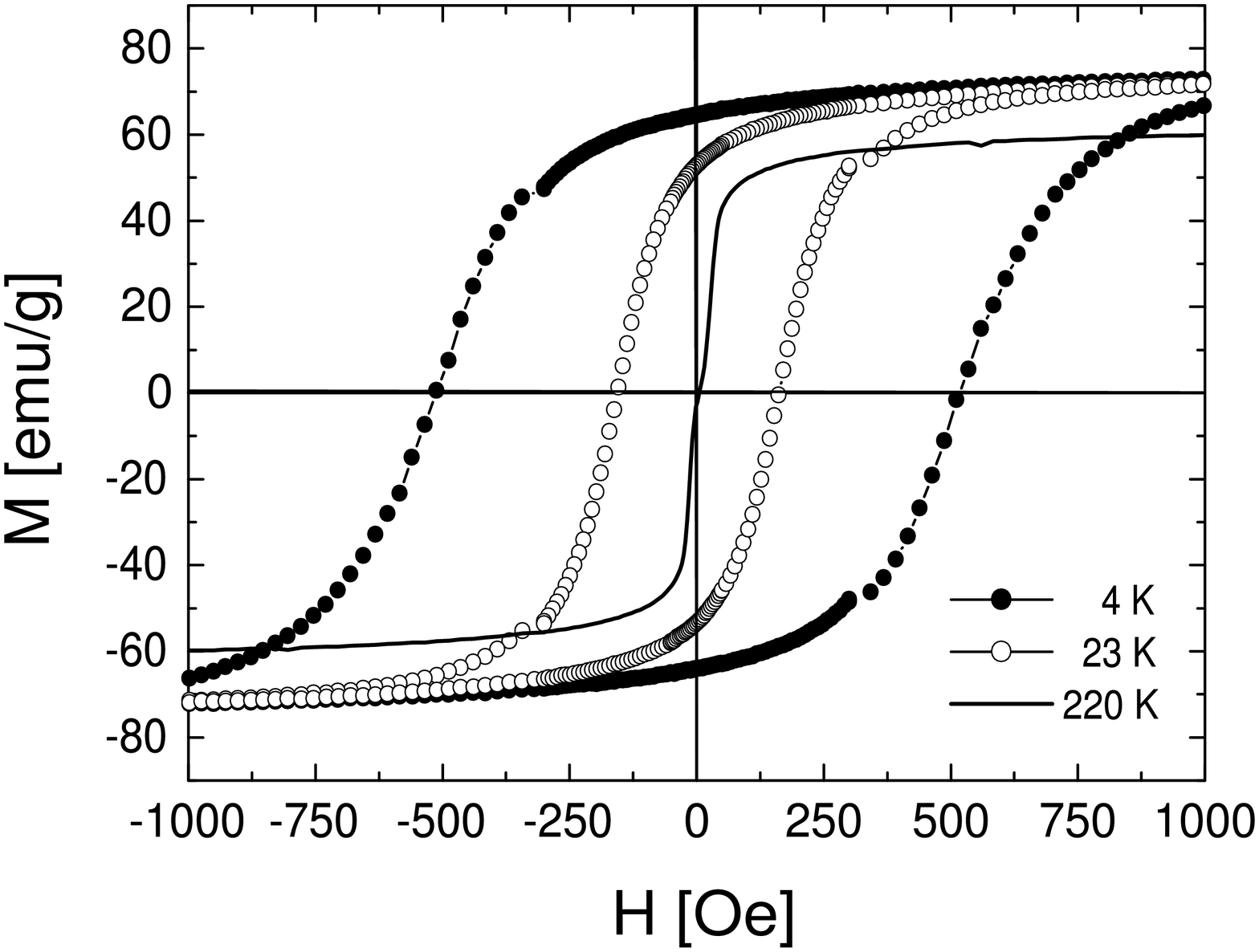}
\includegraphics[height=8cm,keepaspectratio]{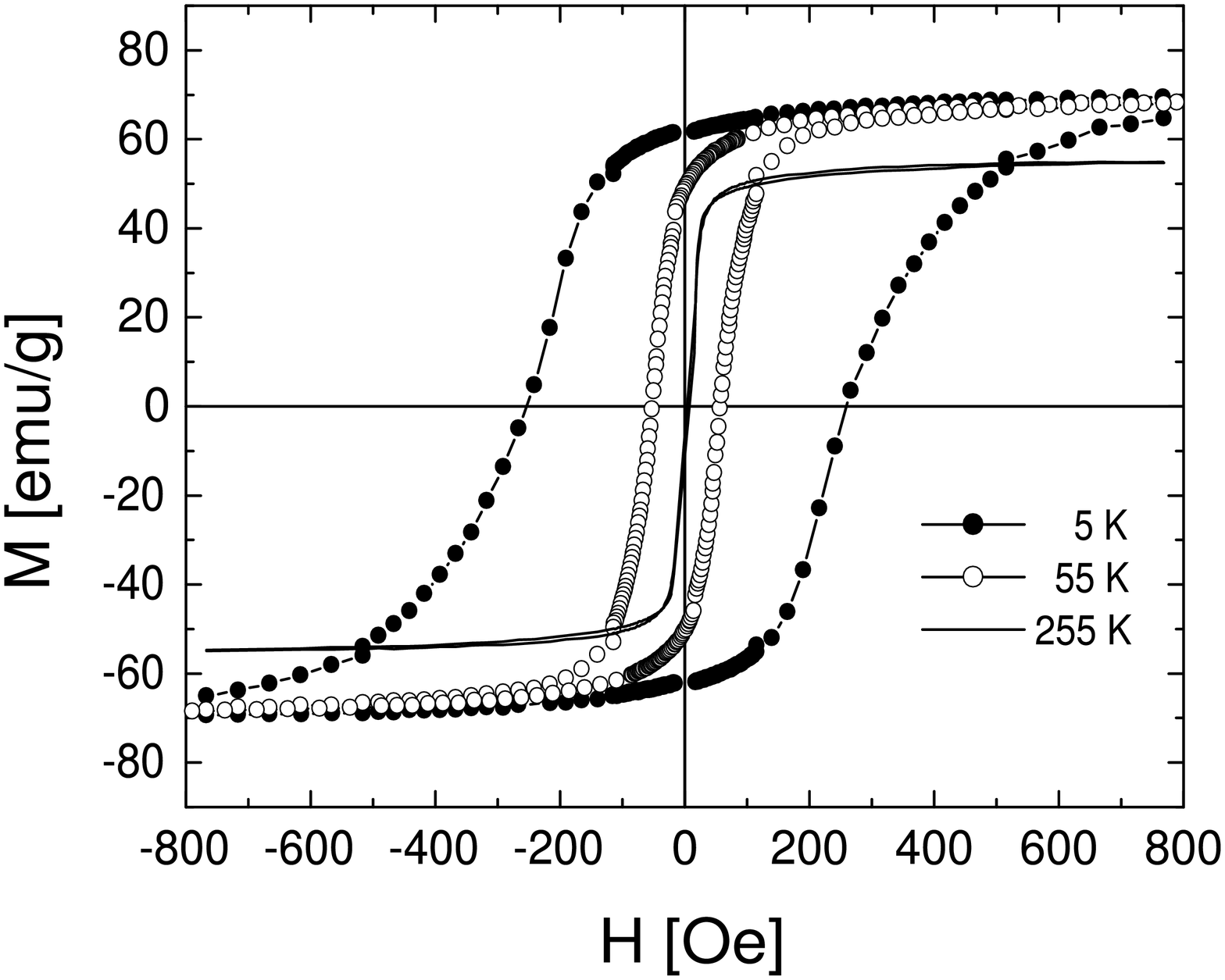}

\caption{Hysteresis loops measured at different temperatures for the samples
{[}Co$_{2}$MnGe(3nm)/V(3nm)]$_{30}$ in the upper panel (a) and {[}Co$_{2}$MnSn(3nm)/V(3nm)]$_{30}$
in the bottom panel (b).}

\label{fig:3-4} 
\end{figure}

The magnetic properties of the multilayers were studied by measurements
of the magnetic hysteresis loops. Figures \ref{fig:3-3} and \ref{fig:3-4}
show examples of hysteresis loops from the {[}Co$_{2}$MnGe/Au], the
{[}Co$_{2}$MnSn/Au], the{[}Co$_{2}$MnGe/V] and the {[}Co$_{2}$MnSn/V]
multilayer systems at different temperatures. Several important features
should be noted.

\begin{figure}
\centering \includegraphics[height=8cm]{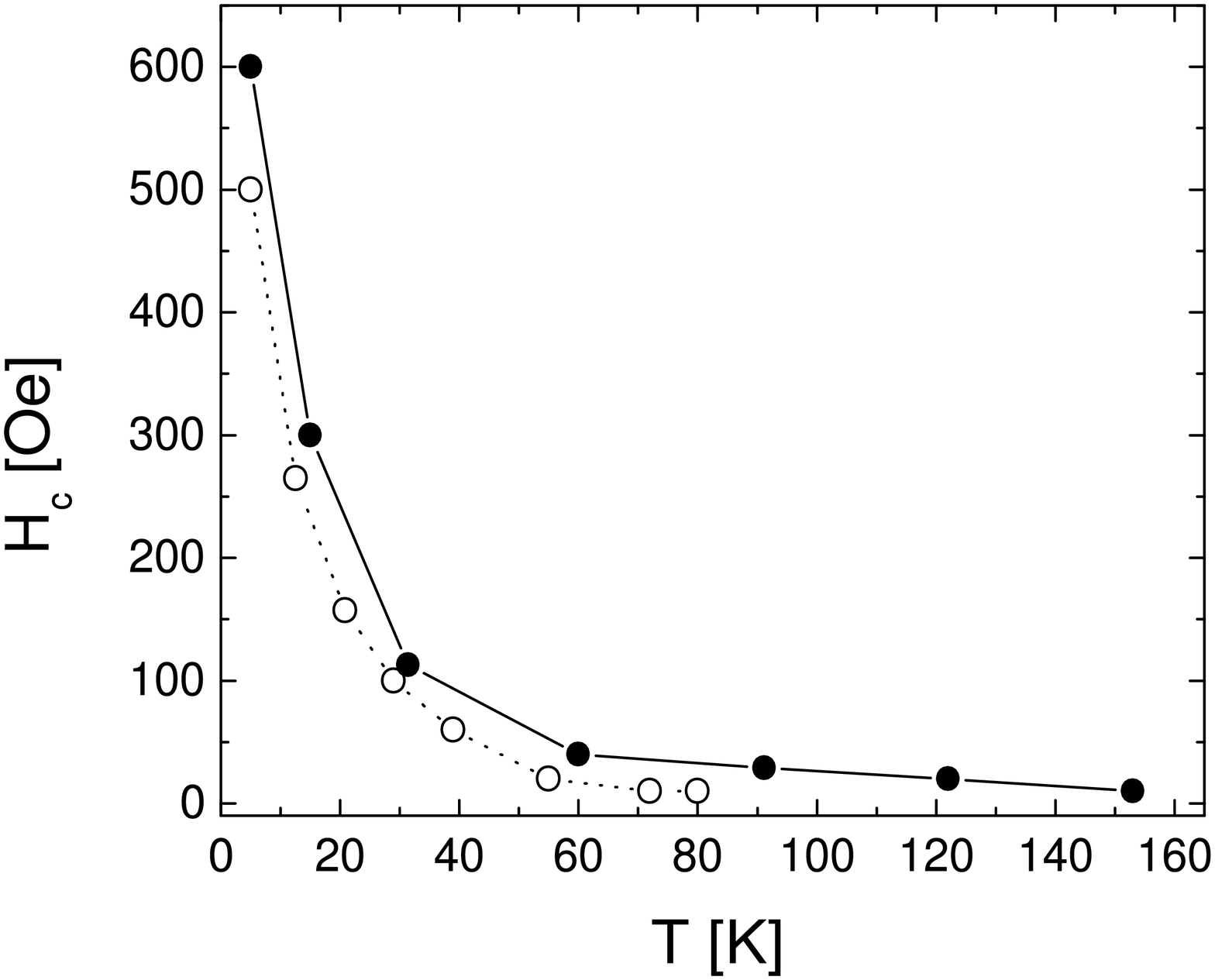}

\caption{Coercive field versus temperature for the multilayers {[}Co$_{2}$MnSn(3nm)/Au(1.5nm)]$_{30}$
(solid circles) and {[}Co$_{2}$MnGe(3nm)/V(4nm)]$_{50}$ (open circles).}

\label{fig:3-5} 
\end{figure}

The saturation magnetization measured at 5~K is smaller than the
value for the bulk compound, the reduction is comparable to that we
observed in single films of the same systems in section \ref{sec:single}.
The coercive field is strongly increasing at low temperatures from
values of typically $H_{c}=50$~Oe at 60~K to several hundred Oe
at 4~K. This effect is shown in more detail in figure \ref{fig:3-5}
where the coercive force drastically increases below 50~K. This feature
is often observed in magnetically inhomogeneous films and indicates
that thermal activation plays an important role in the remagnetization
processes at higher temperatures. Interestingly for the multilayers
with V in figure \ref{fig:3-4} there is no observable magnetic remanence
at higher temperature, the magnetization curve is completely reversible
for temperatures above about 150~K. On the other hand for the multilayers
with Au (figure \ref{fig:3-3}) there is a hysteresis with a finite
remanent magnetization up to the ferromagnetic Curie temperature,
as it should be for a normal ferromagnetic compound. Most of our multilayers
possess a growth induced uniaxial magnetic anisotropy similar to the
thick films discussed in section \ref{sec:single} , but with a definitely
smaller amplitude of the order of 20~Oe for the anisotropy fields.
For the hysteresis curve measurements in figure \ref{fig:3-3} the
field is applied along the magnetic easy axis, thus the vanishing
remanent magnetization cannot simply be explained by a magnetic anisotropy
field perpendicular to the field direction.

\begin{figure}
\centering \includegraphics[height=9cm]{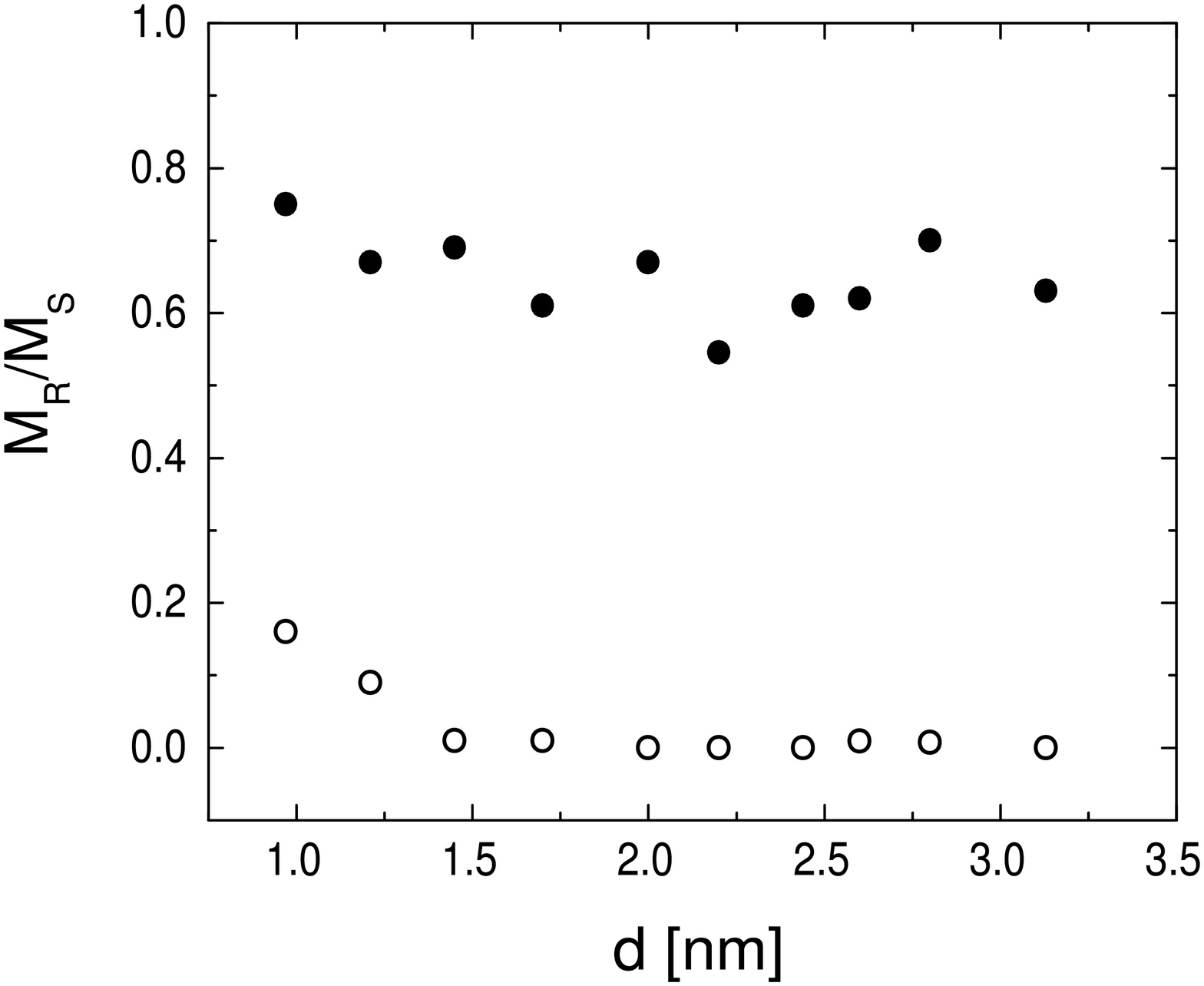}

\caption{Relative remanent magnetization for {[}Co$_{2}$MnSn(3nm)/Au($d$)]$_{30}$
(solid circles) and {[}Co$_{2}$MnGe(3nm)/V($d$)]$_{30}$ (open circles)
multilayers determined at $T=200$~K versus thickness of the non
magnetic interlayer $d$.}

\label{fig:3-6} 
\end{figure}

\begin{figure}[ht]
 \centering \includegraphics[height=9cm]{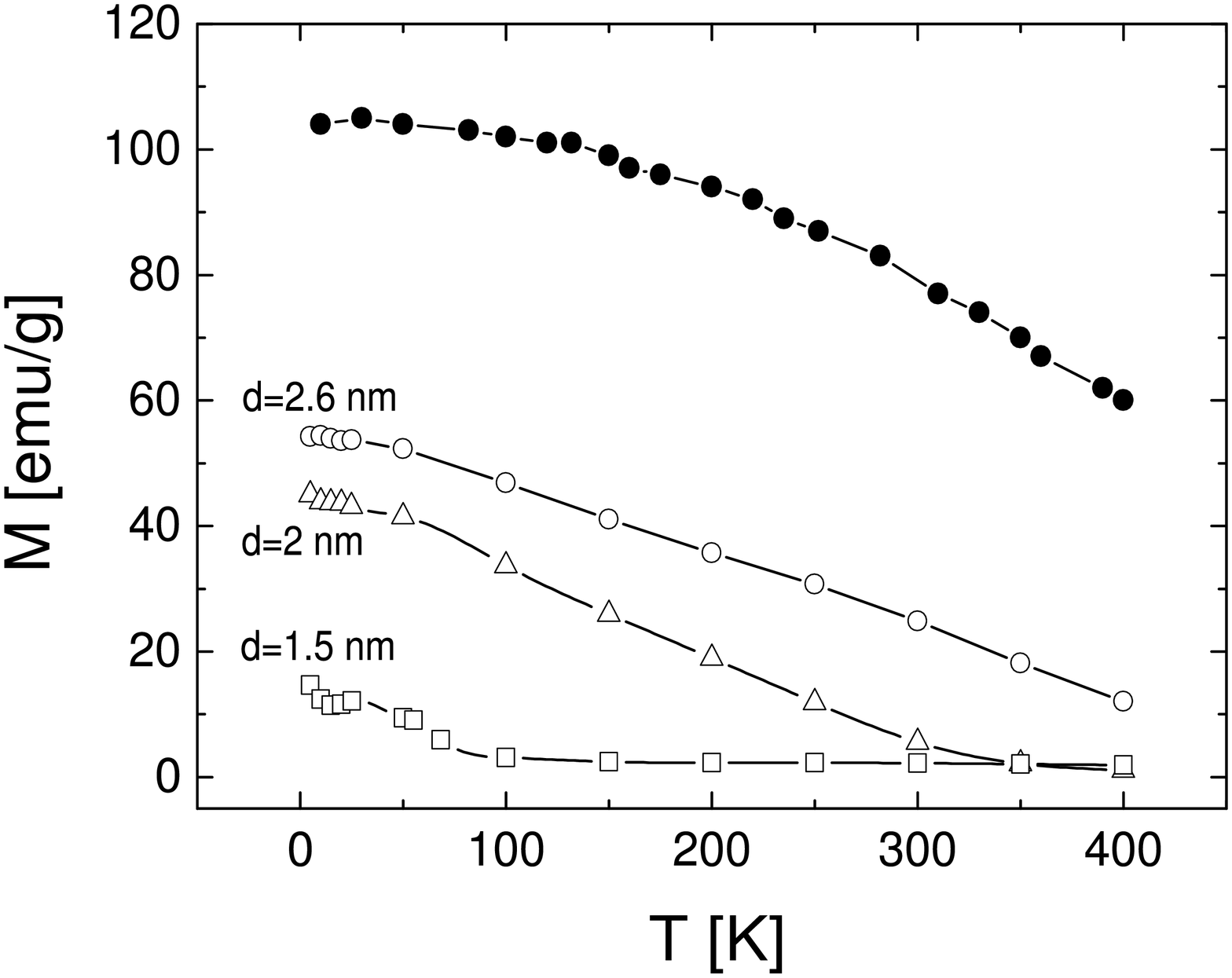}

\caption{Ferromagnetic saturation magnetization of bulk Co$_{2}$MnGe (solid
circles) and {[}Co$_{2}$MnGe($d$)/Au(3nm)]$_{30}$ multilayers (open
symbols) with varied Co$_{2}$MnGe layer thickness $d$ (given in
the figure) versus temperature.}

\label{fig:3-7} 
\end{figure}

\begin{figure}
\centering \includegraphics[height=8cm]{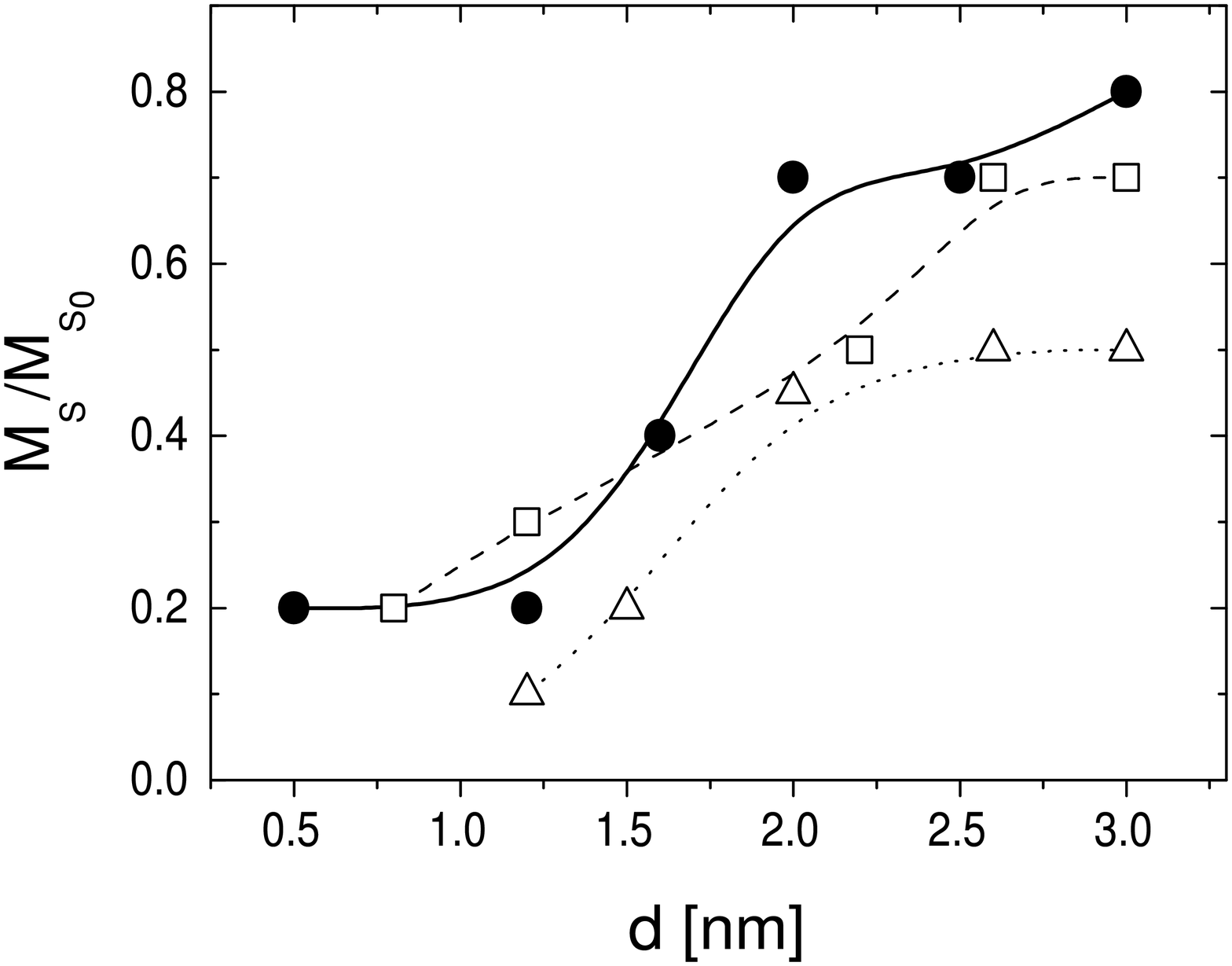}

\caption{Relative saturation magnetization versus the thickness of the Co$_{2}$MnGe
and Co$_{2}$MnSn layer for multilayers {[}Co$_{2}$MnSn($d$)/V(3nm)]$_{30}$
(circles), {[}Co$_{2}$MnGe($d$)/V(3nm)]$_{30}$ (squares) and {[}Co$_{2}$MnSn($d$)/Au(3nm)]$_{30}$
(triangles).}

\label{fig:3-8} 
\end{figure}

We prepared series of 10 multilayer samples with either the thickness
of the Heusler compound or the thickness of the V or Au layer varied
between 1~nm and 3~nm and systematically studied the hysteresis
curves. These results are summarized in the next three figures. In
figure \ref{fig:3-6} we have plotted the relative remanent magnetization
for {[}Co$_{2}$MnGe/V] and {[}Co$_{2}$MnSn/Au] multilayers determined
at a temperature of 200~K as a function of the thickness of the non
magnetic interlayers. We find no systematic variation with the thickness,
for the multilayers with Au the remanent magnetization is about 60\%
of the saturation magnetization independent of the thickness of the
Au layer, for {[}Co$_{2}$MnGe/V]$_{n}$ the remanent magnetization
vanishes for all thicknesses above $d=1.5$~nm. Thus there is no
indication of an oscillation of the interlayer exchange interaction
in {[}Co$_{2}$MnSn/Au]$_{n}$ which in the case of an antiferomagnetic
coupling should give a lowering of the remanent magnetization. Similarly
in {[}Co$_{2}$MnGe/V]$_{n}$ the vanishing remanent magnetization
could indicate an af coupling, but then the af coupling would be independent
of the interlayer thickness, which does not at all fit into the scheme
of an IEC mechanism \cite{bruno95}. We will come back to this question
in section \ref{sec:multi-3} below. In figure \ref{fig:3-7} we have
plotted the ferromagnetic saturation magnetization measured in a field
of 2~kOe when varying the thickness of the Co$_{2}$MnGe layers while
keeping the thickness of the Au layers constant. It is apparent that
the saturation magnetization and the ferromagnetic Curie temperatures
are much lower than for the bulk Co$_{2}$MnGe phase and continuously
decrease further with decreasing film thickness. For a layer thickness
of Co$_{2}$MnGe of $d=2.6$~nm we estimate a ferromagnetic Curie
temperature $T_{c}$ of about 500~K, for $d=2$~nm we get $T_{c}\sim340$~K
and for $d=1.5$~nm $T_{c}\sim90$~K. For the latter sample the
saturation magnetization is very low and the magnetic ground state
is a spin glass state rather than a ferromagnetic state (see next
section). Finally in figure \ref{fig:3-8} we have plotted the ferromagnetic
saturation magnetization measured at $T=5$~K and $H=2$~kOe as
a function of the magnetic layer thickness for the different multilayer
systems under study here. One finds that for a layer thickness of
$d=3$~nm the saturation magnetization reaches only 50 to 80\% of
the bulk saturation magnetization, depending on the combination. Typically
below $d=1.5$~nm the ferromagnetic moment breaks down completely,
suggesting that at this thickness mixing and disorder from both sides
of the ferromagnetic layer destroys the ferromagnetic ground state
completely.

Summarizing this section, we have shown that high quality multilayers
of the Heusler compounds Co$_{2}$MnGe and Co$_{2}$MnSn can be grown
on sapphire a-plane. The ferromagnetic saturation magnetization is
definitely lower than the bulk value but similar to what we have observed
for the very thin single films. The reduction of the magnetization
has two different sources. First there are many antisite defects since
the multilayers are prepared at $300^{\circ}$C and not at the much
higher optimum temperature. Second, there is an additional drastic
breakdown of the saturation magnetization at the interfaces due to
strong disorder and intermixing here. From the investigations of the
multilayers with very thin magnetic films we may conclude that typically
0.7~nm of the Heusler layers at the interfaces contribute only a
small magnetization and are not ferromagnetic. This is supported by
the results of the small angle x-ray reflectivity, from which we derived
an interface roughness of the same order of magnitude. When varying
the thickness of the non magnetic interlayers Au and V we find no
indication of an oscillating interlayer exchange coupling, but for
the case of the V interlayers there is an interesting anomalous behaviour,
namely a vanishing remanent magnetization over a broad thickness range
which will be the subject of the section \ref{sec:multi-3}.

\subsection{Exchange bias in {[}Co$_{2}$MnGe/Au] multilayers}

\label{sec:multi-2}

Apart from the more general magnetic properties of the Co-based Heusler
multilayers presented in the previous section, there are intriguing
peculiarities in the magnetic order of these multilayers which we
want to discuss in this and the next section.

After cooling in a magnetic field most of the Co-based Heusler multilayers
which we have studied exhibit asymmetric magnetic hysteresis loops
shifted along the magnetic field axis, evidencing the existence of
an unidirectional exchange anisotropy, nowadays dubbed \char`\"{}exchange
bias\char`\"{} \cite{nogues99}. This is an effect known to occur
at interfaces between an antiferromagnet (af) and a ferromagnet (f),
the classical example being the Co/CoO interface \cite{nogues99}
. The exchange bias phenomenon attracted considerable new interest
in recent years and there is a much deeper understanding of it now
than it was ten years before, although a quantitative theoretical
model is still lacking. The exchange bias sets in at a blocking temperature
$T_{B}$ slightly below the N\'{e}el temperature $T_{N}$ of the
antiferromagnet and there is general agreement in the literature that
the spins at the f/af interface which are coupled to both, the ferromagnet
and the antiferromagnet, cause the exchange bias effect. The coupling
to the ferromagnet in saturation defines the direction of the exchange
bias field $H_{EB}$. Upon cooling through $T_{B}$ the interface
spins are blocked and keep the direction of $H_{EB}$ fixed. Theoretical
models developed for the exchange bias effect in recent years focus
on different aspects of the complex problem. The domain state model
\cite{miltenyi00} assumes that the domain formation in the antiferromagnet
is the most essential point, other models start from the frustrated,
weakly coupled spins at the interface or at domain walls of the antiferromagnet
\cite{mauri87,malozemoff87}.

\begin{figure}[ht]
 \centering \includegraphics[height=8cm]{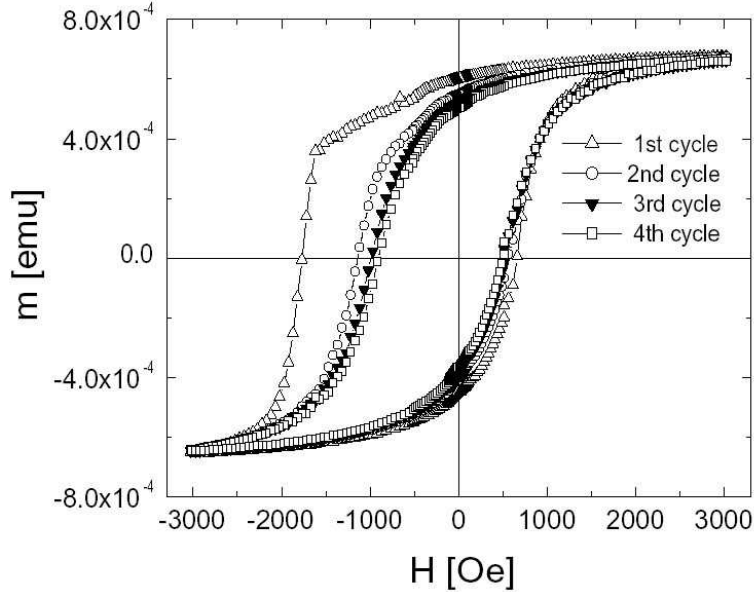}

\caption{Sequence of magnetic hysteresis loops of the multilayer {[}Co$_{2}$MnGe(2.6nm)/Au(3nm)]$_{30}$
measured at 2~K after field cooling in $H=2$~kOe.}

\label{fig:3-9} 
\end{figure}

We have observed an exchange bias shift of the hysteresis loop in
multilayers of Co$_{2}$MnSn and Co$_{2}$MnGe when combined with
Au, Cr or Cu$_{2}$MnAl \cite{westerholt03}. Only with V interlayers
it does not occur. Since certainly there is no antiferromagnetism
in these multilayers, the existence of $H_{EB}$ is puzzling at the
first glance.

\begin{figure}[ht]
 \centering \includegraphics[height=9cm]{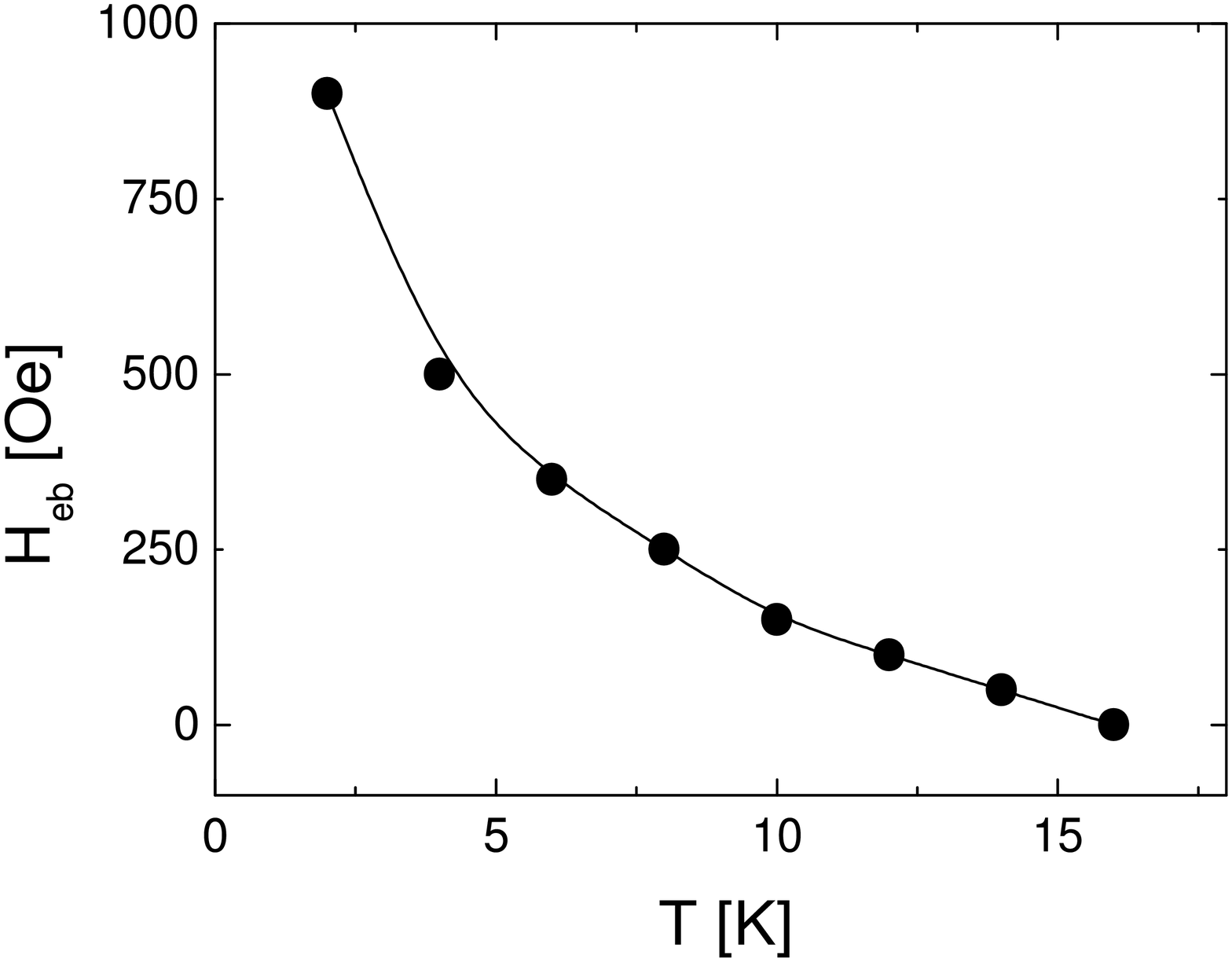}

\caption{Exchange bias field versus temperature for the multilayer {[}Co$_{2}$MnGe(2.6nm)/Au(3nm)]$_{30}$
for a cooling field $H=200$~Oe.}

\label{fig:3-10} 
\end{figure}

Before coming back to this question we shortly introduce the main
experimental results for the example of the {[}Co$_{2}$MnGe/Au] multilayer
system. In figure \ref{fig:3-9} we show the hysteresis loop measured
at 2~K after field cooling in an applied field of 2~kOe. The loop
is found to be shifted by a rather large exchange bias field $H_{EB}=500$~Oe.
Similar to many other exchange bias systems \cite{nogues99} there
is a strong relaxation of $H_{EB}$ when sweeping the magnetic field
several times (see sweep nb. 1 to 4 in figure \ref{fig:3-9}). The
exchange bias effect sets in rather sharply at a blocking temperature
of 15~K (figure \ref{fig:3-10}), showing that the magnetic exchange
interactions defining $T_{B}$ are rather weak. We have also tested
the dependence of $H_{EB}$ on the thickness of the magnetic layer
$d_{m}$ and found $H_{EB}\sim1/d_{m}$ as usual for exchange bias
systems and proving that the exchange anisotropy originates from interface
spins.

\begin{figure}[ht]
 \centering \includegraphics[height=9cm]{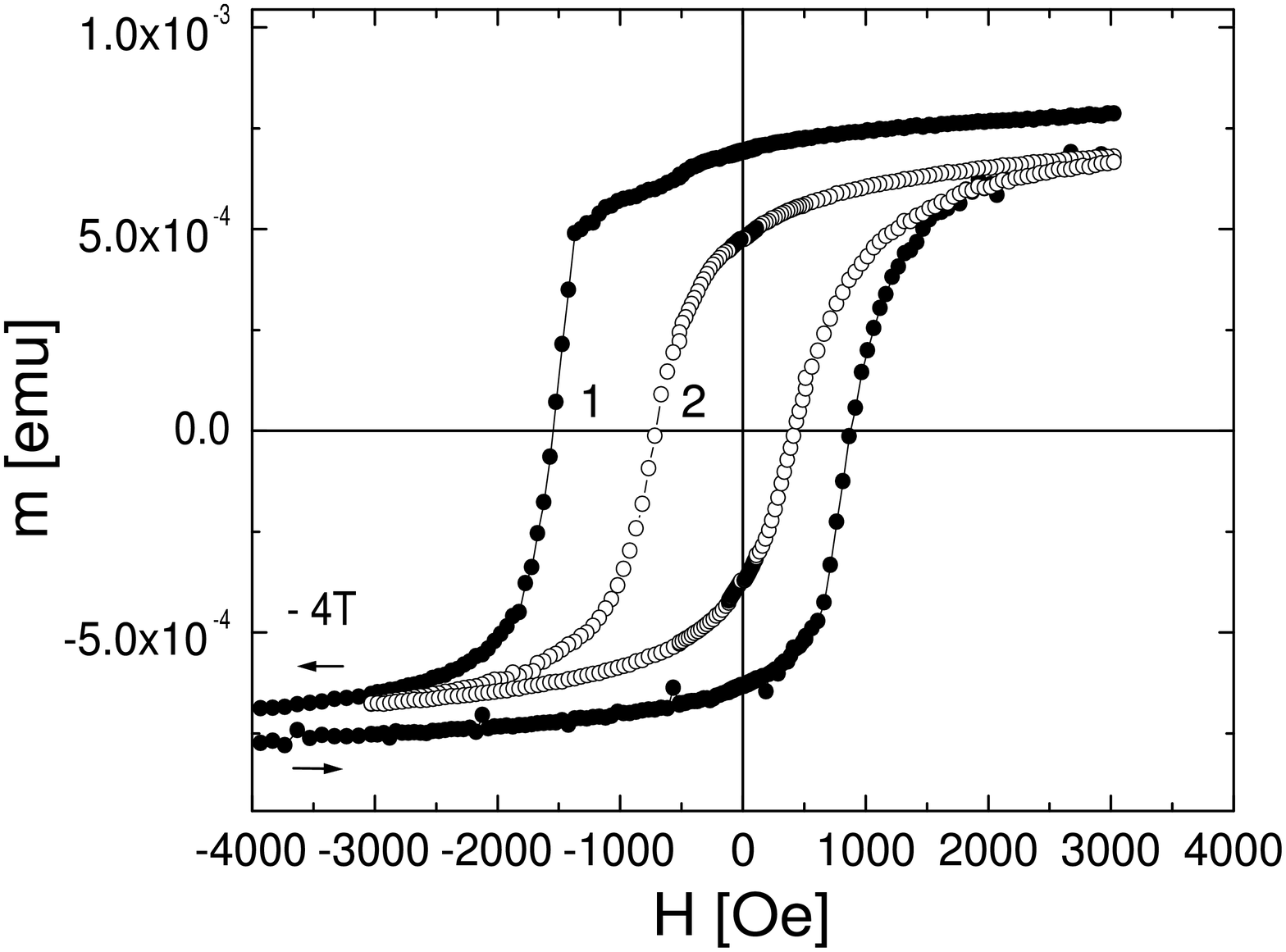}

\caption{Magnetic hysteresis loops of the multiayer {[}Co$_{2}$MnGe(2.6nm)/Au(3nm)]$_{30}$
measured at 2~K after field cooling in $H=40$~kOe. The labels in
the figure denote the number of the field cycles. The field has been
driven to -40~kOe during the first field cycle.}

\label{fig:3-11} 
\end{figure}

\begin{figure}
\centering \includegraphics[height=8cm]{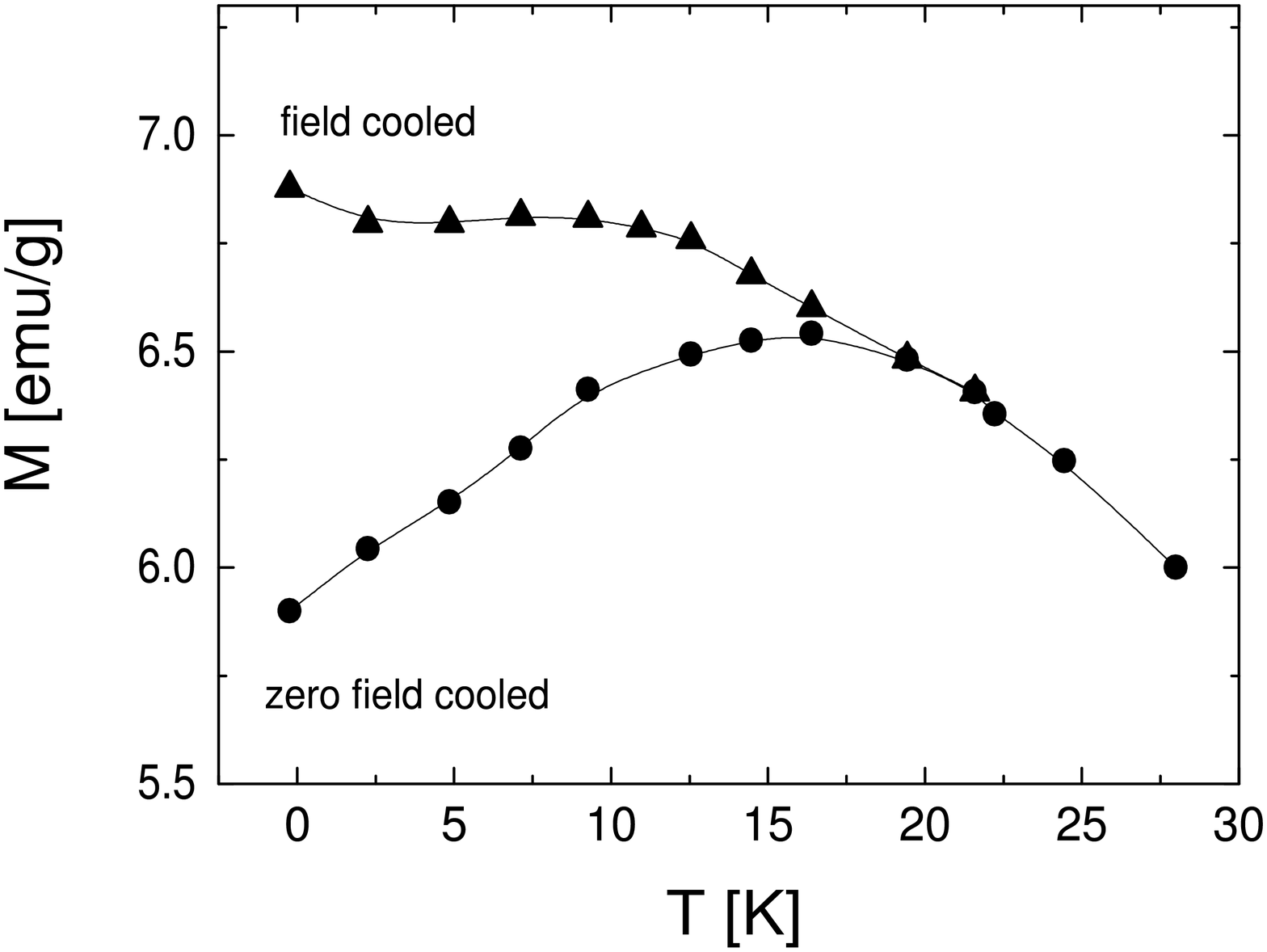}

\caption{Field cooled and zero field cooled magnetization for the sample {[}Co$_{2}$MnGe(1.2nm)/Au(3nm)]$_{30}$
in a magnetic field of 500~Oe.}

\label{fig:3-12} 
\end{figure}

There is one enlightening experiment which is difficult to realize
in conventional f/af exchange bias systems and which we show in figure
\ref{fig:3-11}. When sweeping the external field to -4~T at 5~K,
one can nearly eliminate $H_{EB}$. Simultaneously the excess magnetization
(asymmetry of the loop along the magnetization axis) vanishes, directly
proving the intimate correlation of the excess positive magnetization
and $H_{EB}$. It seems quite natural to conclude that interface spins
determining $H_{EB}$ are rotated out of their metastable orientation
when sweeping the field up to 4~T. Now coming back to the question
concerning the origin of the exchange bias shift in our multilayers,
the mere existence of $H_{EB}$ shows that at the Co$_{2}$MnGe/Au
interface antiferromagnetic exchange interactions must be present.
The excess positive magnetization which vanishes after magnetizing
to high negative fields shows that there are spins at the interface
with a component antiparallel to the ferromagnet. Adopting the point
of view of localized models for magnetic exchange interactions, there
seems to be an antiferromagnetic superexchange type of interaction
at the interface e.g. an af Co-Au-Co superexchange or an af Co-Au-Mn
superexchange. This interaction competing with the ferromagnetic exchange
of the pure Co$_{2}$MnGe phase creates a few monolayers with frustrated
spins and spin glass type of order at the interface. Actually the
magnetic order of very thin Co$_{2}$MnGe layers with a thickness
of the order of 1~nm exhibit all ingredients of spin glass order
(figure \ref{fig:3-12}). This spin glass phase takes over the role
of the antiferromagnet in conventional f/af exchange bias systems.
The spin glass provides a source of weakly coupled spins which are
blocked at the spin glass freezing temperature $T_{f}$, which replaces
the blocking temperature $T_{B}$ of the f/af interface.

The experimental fact that the exchange bias effect occurs in many
Co based Heusler multilayers shows that the existence of antiferromagnetic
interactions and spin glass order are the rule rather than the exception
in these multilayers.

\subsection{Antiferromagnetic interlayer coupling in {[}Co$_{2}$MnGe/V] multilayers}

\label{sec:multi-3}

It is an interesting principal question whether in the Co Heusler
based multilayers an oscillatory interlayer exchange coupling (IEC)
of the type existing in multilayers combined of the $3d$ transition
element ferromagnets and normal metals \cite{bruno95} can be observed,
too. Thus we carefully inspected the magnetic hysteresis loops of
the Co-based Heusler multilayers from section \ref{sec:multi-1} in
order to find indications for an IEC mechanism. We were successful
in the {[}Co$_{2}$MnGe/V] multilayer system, where a vanishing remanence
at room temperature (see fiogure \ref{fig:3-4}a) is seen, giving
a first indication of an antiferromagnetic IEC.

We then started a systematic investigation of the magnetic order in
this system by magnetic neutron reflectometry. The experiments were
carried out at the ADAM reflectometer at the ILL in Grenoble. This
instrument is equipped with neutron spin analysers so that spin polarized
neutron reflectivity (PNR) measurements can be performed. PNR gives
additional and important information about the magnetization direction
in the film \cite{blundell92,zabel94,zabel03}, since a magnetization
direction parallel (or antiparallel) to the neutron spin polarization
causes scattering with the polarization direction unchanged (up-up
(+,+) and down-down (-,-) channels) only, whereas a magnetization
component perpendicular to the spin quantization axis causes spin
flip scattering appearing in the down-up (-,+) and up-down (+,-) channels.

\begin{figure}
\centering \includegraphics[height=8cm]{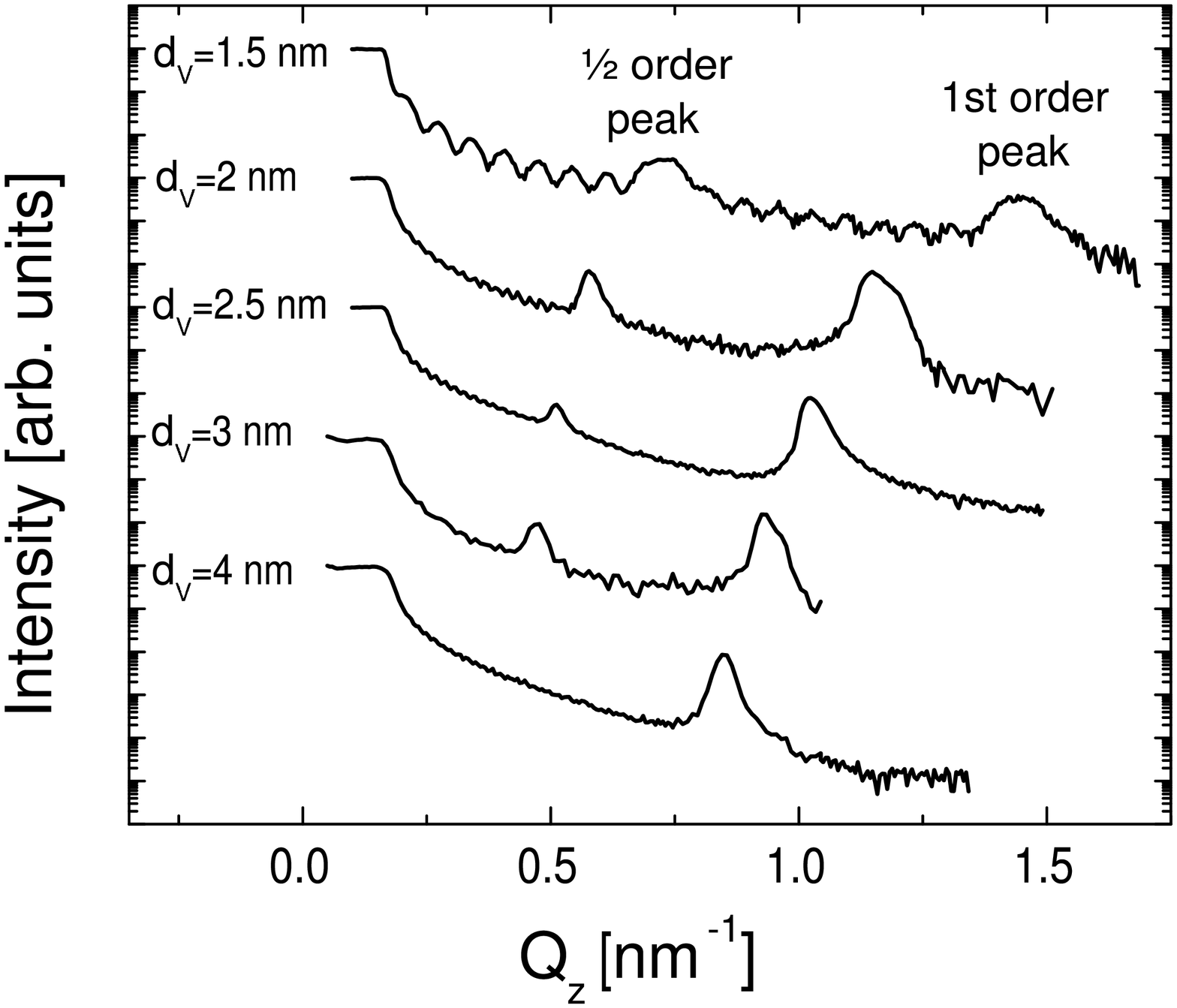}

\caption{Specular unpolarized neutron reflectivity scans of {[}Co$_{2}$MnGe(3nm)/V($d$)]$_{n}$
multilayers. The V layer thickness $d_{V}$ is given in the figure.
For $d_{V}=1.5$~nm the number of bilayers is $n=20$; in all other
cases $n=50$.}

\label{fig:3-13} 
\end{figure}

Figure \ref{fig:3-13} displays (unpolarized) neutron reflectivity
scans for a series of {[}Co$_{2}$MnGe/V] multilayers with different
thickness of the V layer $d_{V}$ ranging from 1.5~nm to 4~nm. The
observation of the half order peak clearly proves the existence of
an antiferromagnetic (af) interlayer coupling. The af coupling is
observed for all multilayers with an interlayer thickness d$_{V}$
smaller than 3 nm, we find no indication of an oscillatory IEC in
this thickness range. We also studied other, intermediate V thicknesses
by magnetization measurements and found that the coupling is always
antiferromagnetic. At this point serious doubts arise whether actually
the IEC mechanism is responsible for the af interlayer coupling, since
the oscillatory character of this coupling is an intrinsic property
of the geometry of the Fermi surface of the interlayers \cite{bruno95}.

\begin{figure}[ht]
 \centering \includegraphics[height=8cm]{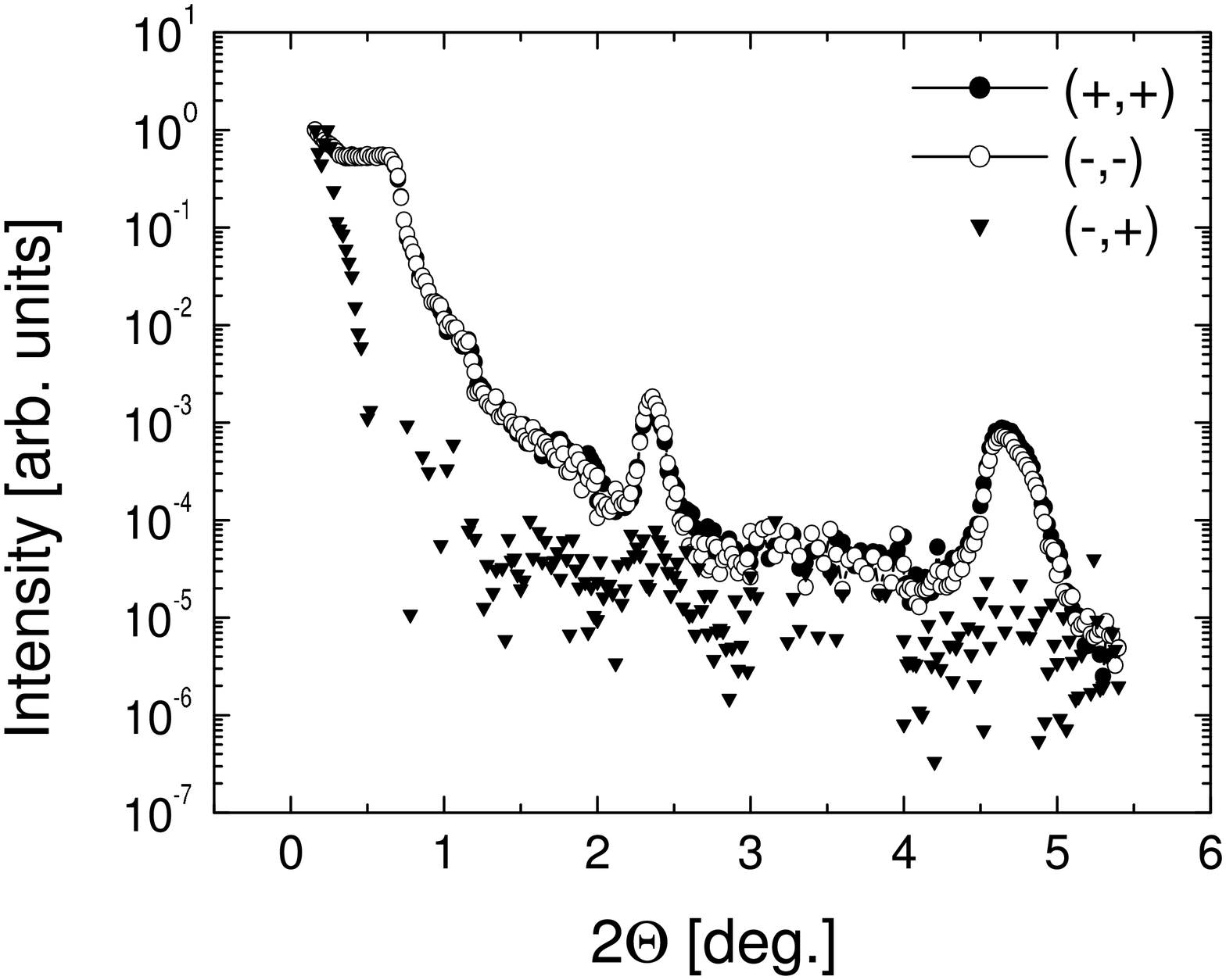}

\caption{Specular polarized neutron reflectivity scans for non spin flip ((+,+)
and (-,-)) and for spin flip (-,+) channels of the multilayer {[}Co$_{2}$MnGe(3nm)/V(2nm)]$_{50}$.}

\label{fig:3-14} 
\end{figure}

In order to characterize the af interlayer ordering further, PNR measurements
of the {[}Co$_{2}$MnGe/V] multilayers with the growth induced magnetic
easy axis of the multilayers parallel to the small neutron guiding
field were taken (figure \ref{fig:3-14}). Figure \ref{fig:3-14}
presents the intensity measured for the spin channels ((+,+), (-,-)
and (+,-)), only the non spin flip channels show a finite intensity.
Thus the sublattice magnetization of the af lattice is pointing parallel
or antiparallel to the magnetic field. The first order structural
peak in figure \ref{fig:3-14} has the same intensity for the (+,+)
and the (-,-) channel. This proves that the magnetic interlayer superstructure
has no ferromagnetic component which could occur e.g. by some canting
of the af coupled layers. From this we can conclude that the af interlayer
structure of the multilayer is well defined.

\begin{figure}[ht]
 \centering \includegraphics[height=8cm]{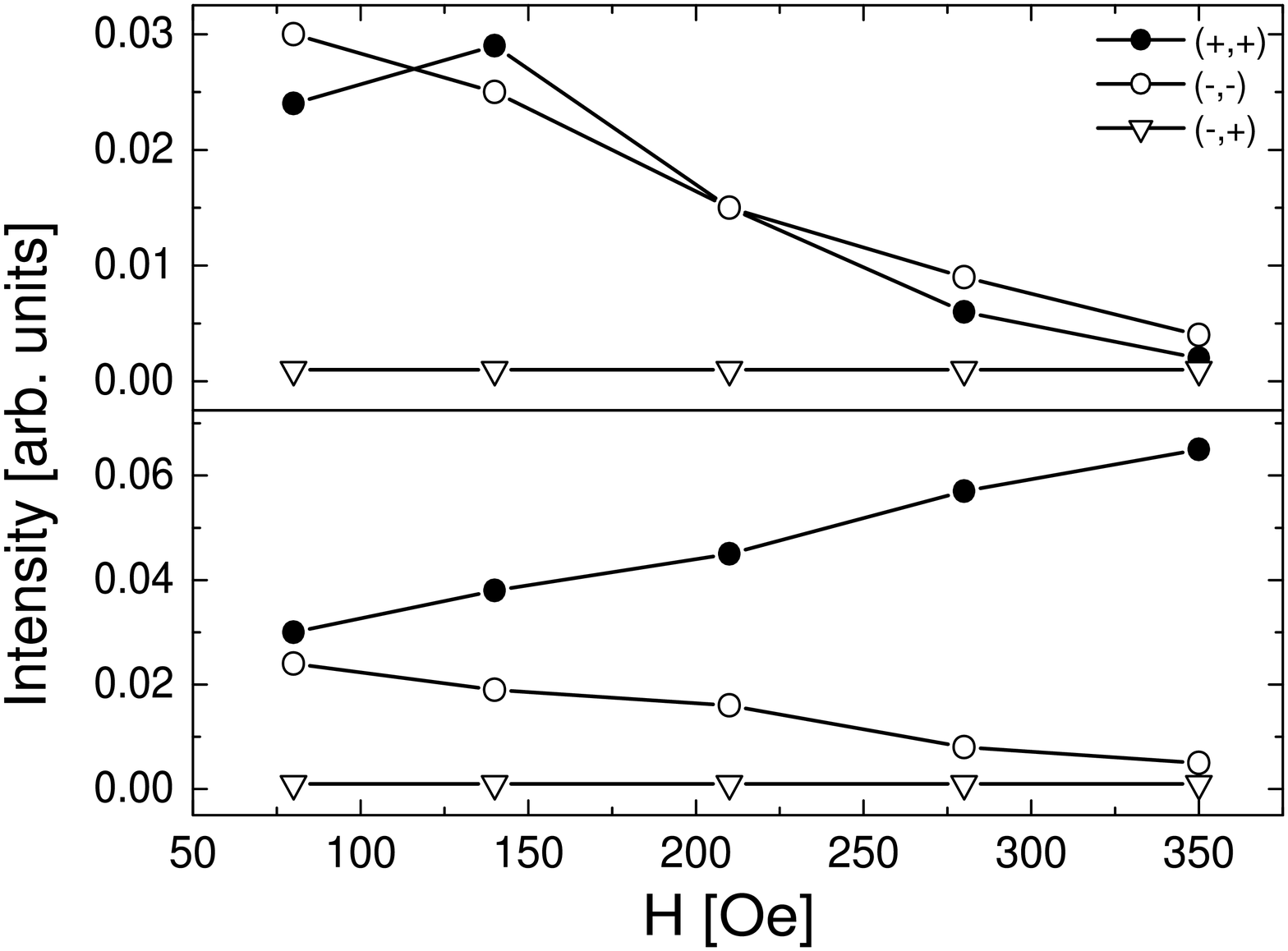}

\caption{Field dependence of the af peak intensity (upper panel) and 1st order
structural peak (lower panel) for the spin flip (-,+) and non spin
flip ((+,+) and (-,-)) channels for the multilayer {[}Co$_{2}$MnGe(3nm)/V(2nm)]$_{50}$.}

\label{fig:3-15} 
\end{figure}

The measurement of PNR scans as a function of an applied field is
shown in figure \ref{fig:3-15}. The af peak intensity is rapidly
suppressed at higher magnetic fields, simultaneously the splitting
of the (-,-) and the (+,+) scattering intensities increases until
at the ferromagnetic saturation field there is only a finite scattering
intensity in the (+,+) channel. The ferromagnetic saturation field
derived from these measurements is 50~Oe at 200~K and 300~Oe at
4~K. Thus the interlayer coupling interaction in this system is very
weak. This holds for all af ordered multilayers of the {[}Co$_{2}$MnGe/V]
system.

There are three additional features in figure \ref{fig:3-15} worth
mentioning: First, at intermediate fields there is no observable spin
flip scattering i.e. the magnetization reversal of the af state takes
place by domain wall movements within the single ferromagnetic layers
and not by coherent rotations \cite{theis03}. Second, the af structure
is irreversibly lost when magnetizing the sample to the ferromagnetic
saturation field at low temperatures. Only at high temperatures above
about 100~K the af order is partly restored. This is also clearly
observed in the series of magnetic hysteresis curves measured for
$T<T_{N}$ shown in figure \ref{fig:3-4}a. At low temperatures one
observes typical ferromagnetic hysteresis loops with a remanent magnetization
of up to 90\% of the saturation magnetization. Third, we note that
in ferromagnetic saturation in figure \ref{fig:3-15} the intensity
in the (+,+) channel vanishes completely i.e the neutron reflected
from the multilayer are completely spin polarized in the up direction.
This originates from the fact that in the Co$_{2}$MnGe layer the
magnetic and structural neutron scattering cross section have the
same magnitude so that for one spin direction in ferromagnetic saturation
the magnetic and the structural scattering length just cancel.

\begin{figure}[ht]
 \centering \includegraphics[height=8cm]{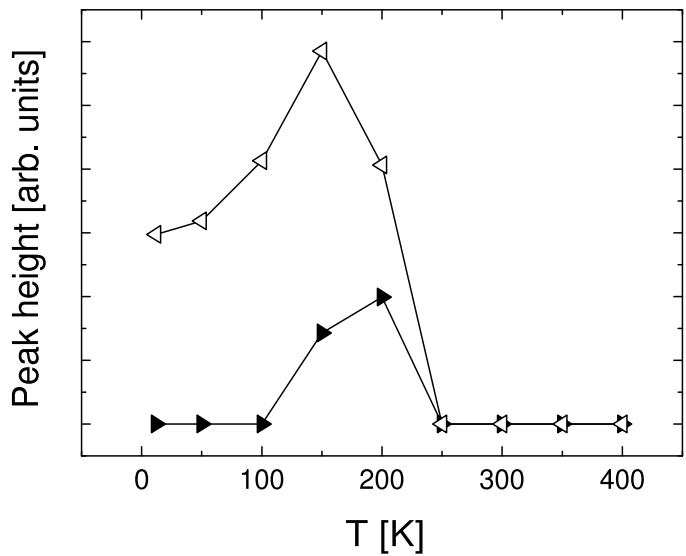}

\caption{Temperature dependence of the af peak intensity of the multilayer
{[}Co$_{2}$MnGe(3nm)/V(3nm)]$_{30}$ measured after field cooling
(solid triangles) and zero field cooling (open triangles).}

\label{fig:3-16} 
\end{figure}

The temperature dependence of the af peak intensity as measured after
cooling in zero-field and after cooling in 1000~Oe and then switching
off the field at the measuring temperature is displayed in figure
\ref{fig:3-16}. After zero-field cooling the af peak intensity develops
below 270~K in a phase transition like fashion, reaches a maximum
at about 200~K and decreases slightly towards lower temperatures.
Cooling in a high field there is no detectable intensity below 100~K,
but approaching the phase transition at 270~K the af order recovers
after switching off the field and close to the transition temperature
the peak intensity coincides with that measured after zero-field cooling.
Since the half order peak intensity is proportional to the squared
sublattice magnetization in an antiferromagnet, this behaviour clearly
reveals that there is a reversible antiferromagnetic phase transition
at 270~K.

\begin{figure}
\centering \includegraphics[height=8cm,keepaspectratio]{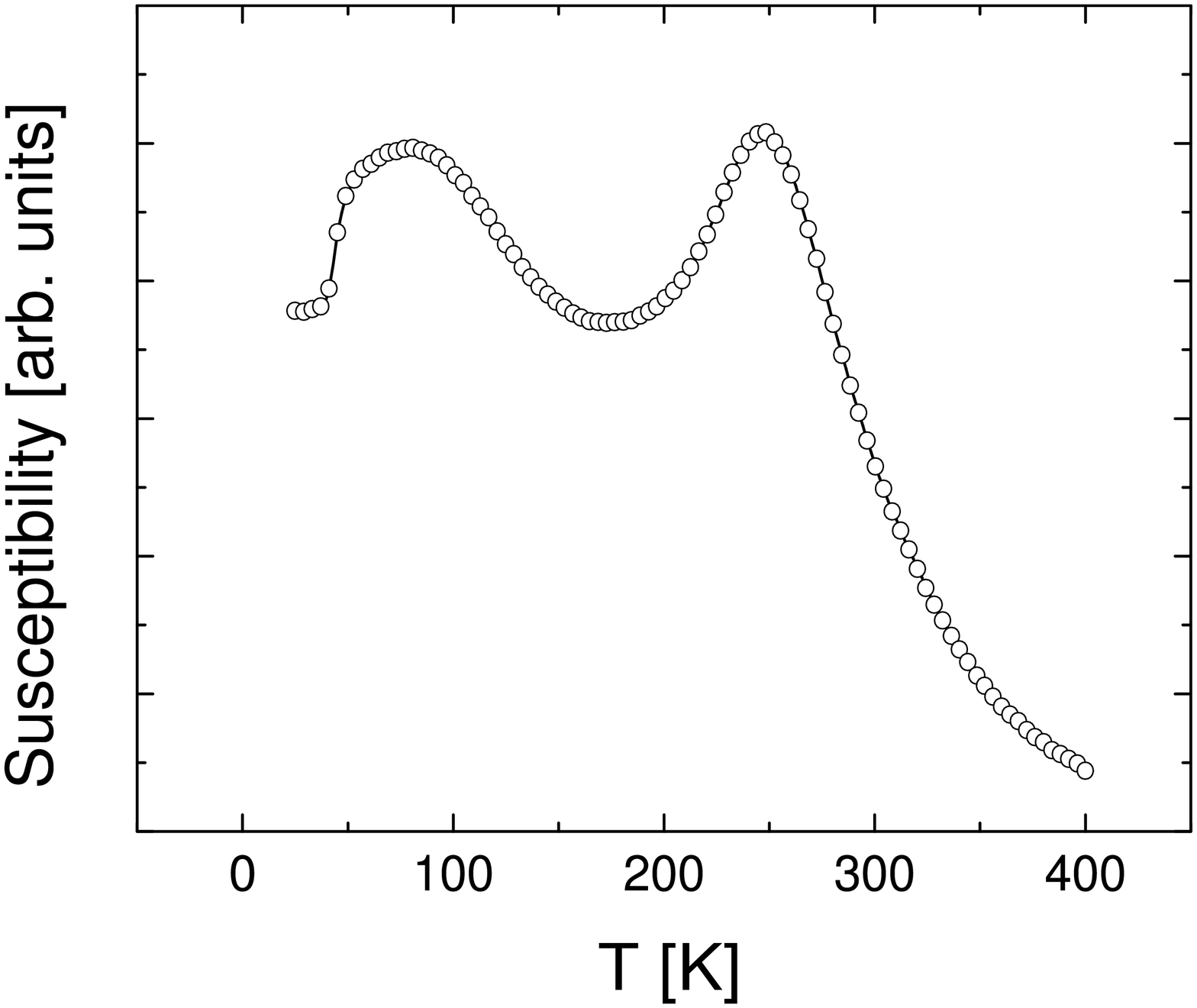}
\includegraphics[height=8cm,keepaspectratio]{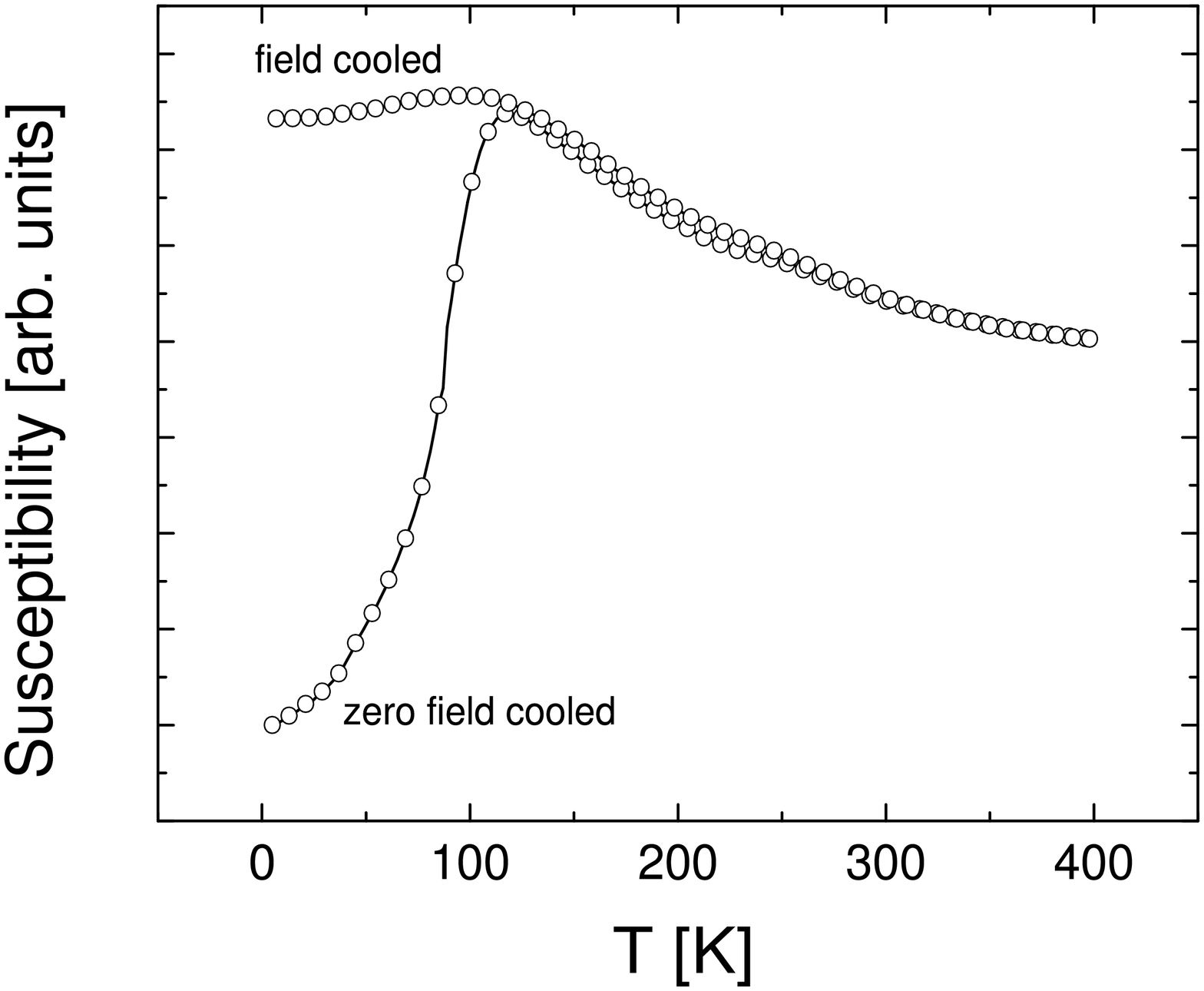}

\caption{The dc magnetic susceptibility of the {[}Co$_{2}$MnGe(3nm)/V(3nm)]$_{30}$
multilayer (upper panel) and the {[}Co$_{2}$MnGe(3nm)/V(4nm)]$_{30}$
multilayer (lower panel).}

\label{fig:3-17} 
\end{figure}

The transition can also easily be detected in the dc magnetic susceptibility
(figure \ref{fig:3-17}a) which exhibits a peak at 270~K, similar
to a conventional bulk antiferromagnet. An antiferromagnetic phase
transition in the {[}Co$_{2}$MnGe/V] system at 270~K is rather surprising,
since from the study of the single V/Co$_{2}$MnGe/V trilayers with
the same thickness of the Co$_{2}$MnGe layer we estimated a ferromagnetic
ordering temperature $T_{c}$ of 600~K (see section \ref{sec:single}).
In multilayers coupled by antiferromagnetic IEC the N\'{e}el temperature
of the af ordering between the layers and the ferromagnetic Curie
temperature of the single layers should coincide. Thus the low value
for $T_{N}$ provides further evidence that the af interlayer ordering
in {[}Co$_{2}$MnGe/V] is not explicable by the conventional IEC mechanism.

The key for the understanding of the af interlayer magnetic ordering
in {[}Co$_{2}$MnGe/V] lies in the magnetic granular behaviour of
very thin Co$_{2}$MnGe layers as discussed in chapter I. By mixing
and disorder at interfaces and grain boundaries the films break up
into small weakly interacting ferromagnetic clusters and exhibit the
typical behaviour of a small particle magnet. If the clusters are
small and the cluster interactions weak, it is expected that the clusters
are superparamagnetic at high temperatures \cite{culity76}. In this
case there is no ferromagnetic phase transition at a ferromagnetic
Curie temperature $T_{c}$ but a freezing of the clusters at a superparamagnetic
blocking temperature $T_{B}<T_{c}$. Actually this has not been observed
in our single Co based Heusler layers from section \ref{sec:single}
at least up to the maximum experimental temperature of 400~K. But
it may well be that at even higher temperatures a corresponding phenomenon
might exist. Now, in multilayers composed of the Co$_{2}$MnGe layers
with the same thickness superparamagnetic behaviour and cluster blocking
clearly appears below 400~K, probably due to a change of the microstructure
and the additional influence of the dipolar fields from the neighbouring
layers.

\begin{figure}[ht]
 \centering \includegraphics[height=8cm]{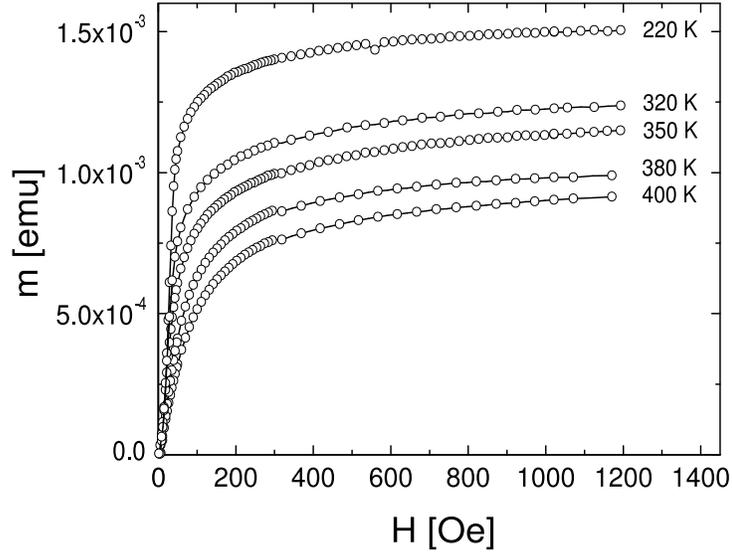}

\caption{Magnetization curves of the af coupled multilayer {[}Co$_{2}$MnGe(3nm)/V(3nm)]$_{50}$
measured at different temperatures $T>T_{N}$.}

\label{fig:3-18} 
\end{figure}

The superparamagnetic behavior is shown in figure \ref{fig:3-18},
where magnetization curves for the af ordered sample from figure \ref{fig:3-14}
are plotted for temperatures above $T_{N}$. The $M(H)$ curves are
completely reversible and saturate at fields of the order of 1~kOe
which is characteristic for a superparamagnet with a huge magnetic
moment. We applied the classical formula for the superparamagnetic
magnetization as a first approximation \begin{equation}
M(H,T)=N_{c}\mu_{c}\mathcal{L}(\frac{\mu_{c}H}{k_{b}T})\end{equation}
 with the number of clusters $N_{c}$ and the cluster moment $\mu_{c}=N\mu_{B}$.
$\mathcal{L}$ is the Langevin function. Fitting this formula to our
results in figure \ref{fig:3-18} at 400~K we find a cluster moment
of $1.6\times10^{5}$~$\mu_{B}$ corresponding to a cluster with
a lateral diameter of about 70~nm.

These experimental findings suggest that at $T_{N}$ of the {[}Co$_{2}$MnGe/V]
multilayers thermally rotating ferromagnetic clusters within each
Co$_{2}$MnGe layer undergo a second order phase transition with simultaneous
intrta-plane ferromagnetic order and inter-plane antiferromagnetic
order. This is a rather spectacular phenomenon which to the best of
our knowledge has never been observed in multilayer systems before.
An interesting question now arises concerning the type of magnetic
order for the multilayer in figure \ref{fig:3-13} with a thickness
of the V layer of 4~nm, which obviously does not exhibit af long
range order. The answer is given in figure \ref{fig:3-17}b, where
the zero field cooled and the field cooled magnetization of this sample
is compared. There is an onset of a strong magnetic irreversibility
at about 150~K, reminiscent of a cluster blocking or a cluster glass
magnetic ordering \cite{culity76}. Actually in the hysteresis loops
measured at low temperatures for this sample one finds features consistent
with this type of transition, namely a temperature dependence of the
coercive force and a magnetic remanence following a power law behaviour
$H_{c}\sim T^{-a}$.

\begin{figure}[ht]
 \centering \includegraphics[height=8cm]{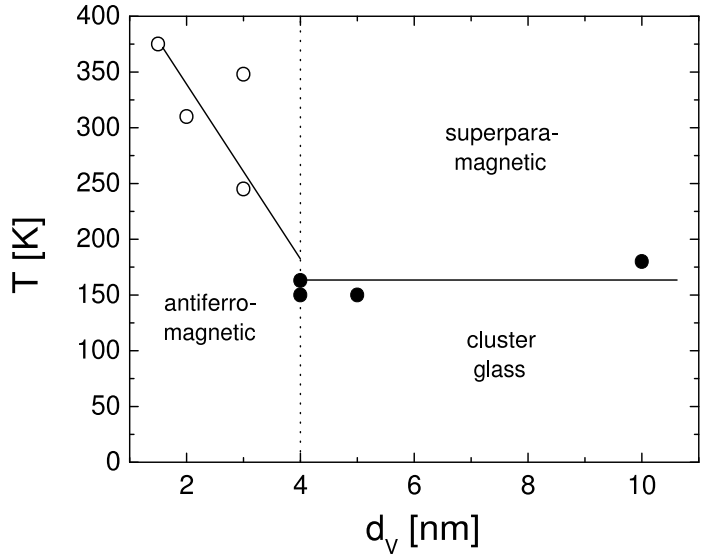}

\caption{Magnetic phase diagram for multilayers {[}Co$_{2}$MnGe(3nm)/V($d$)]$_{50}$
with different vanadium thickness $d_{V}$. There is an antiferromagnetic
range for $d_{V}\leq3$~nm and a range with cluster blocking for
$d_{V}>4$~nm.}

\label{fig:3-19} 
\end{figure}

The magnetic order in the {[}Co$_{2}$MnGe/V] multilayer system for
different thickness of the V layer is summarized in a magnetic phase
diagram in figure \ref{fig:3-19}. There is an antiferromagnetic range
for $d_{V}\leq3$~nm with the N\'{e}el temperature increasing with
decreasing $d_{V}$ up to $T_{N}=380$~K for $d_{V}=1.5$~nm. Then
a range with cluster blocking or cluster glass freezing follows with
the magnetic transition temperature nearly independent of $d_{V}$
and extending up to the maximum thickness we have studied $d_{V}=10$~nm.

Concerning the microscopic origin of the ordering phenomenon observed
in the {[}Co$_{2}$MnGe/V] multilayers here, we would like to emphasize
that the IEC coupling plays no essential role. It is either too weak
or cancels out because of thickness fluctuations. The remaining other
candidate which can produce an interlayer coupling and af long range
order are interlayer magnetic dipolar fields. Finite dipolar stray
fields at the surface of each single Co$_{2}$MnGe layer are inevitably
present, since the layers possess an internal granular magnetic structure.
Topologically the magnetic surface is rough on the length scale of
the dimension of the clusters i.e. on a scale of about several 10~nm.

Theoretical model calculations show that the dipolar coupling field
seen by the neighbouring layers depends on the topological structure
of the surface and can be ferromagnetic or antiferromagnetic, depending
on details of the roughness \cite{altbir95}. Quantitative calculations
with a surface structure mimicking typical thin film systems lead
to an estimation of a coupling field strength of the order of 100~Oe
i.e. the order of magnitude which we observed here. Experimentally
there is the well known N\'{e}el type af coupling \cite{neel62}
between thin films separated by a very thin nonmagnetic interlayer,
which is a disturbing factor in some spin valve devices \cite{wang93}.
Originally this coupling, which is of dipolar origin, has been proposed
by N\'{e}el \cite{neel62}. Dipolar fields can also produce antiferromagnetic
long range order in multilayers, as has been shown for {[}Co/Cu] multilayers
with a thickness of the Cu layer of 6~nm \cite{borchers99}. But
in this system the af long range order is only observed in the virgin
state and is lost irreversibly after once magnetizing the sample.
In {[}Nb/Fe] multilayers the existence of af interlayer dipolar interaction
has also been shown experimentally \cite{maletta97}.

In summary of this section, we find an interesting novel magnetic
ordering phenomenon in the {[}Co$_{2}$MnGe/V] multilayers which is
directly related to the granular magnetic microstructure of the Co
based Heusler thin films in the limit of very small thickness. The
antiferromagnetic order is very subtle and we have only observed it
in {[}Co$_{2}$MnGe/V] multilayers with a thickness of the Heusler
layer of 3~nm, although magnetic granularity is a common feature
of all Co based Heusler phases with a comparable thickness. However,
when combined into multilayers all other systems we have studied remain
ferromagnetic without the cluster blocking features and the af long
range order characteristic for the {[}Co$_{2}$MnGe/V] multilayers.
Thus in this sense the phenomenon is rather unique for {[}Co$_{2}$MnGe/V]
multilayers.

\begin{figure}[ht]
 \centering \includegraphics[height=8cm]{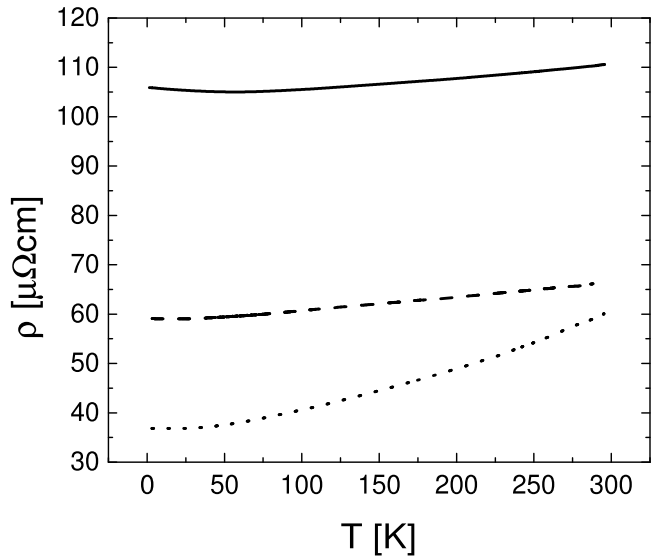}

\caption{Electrical resistivity versus temperature for V(3nm)/Co$_{2}$MnSi(100nm)
(solid line), Au(3nm)/Co$_{2}$MnSn(100nm) (dashed line) and V(3nm)/Co$_{2}$MnGe(100nm)
(dotted line).}

\label{fig:4-1} 
\end{figure}

\section{Magnetotransport Properties}

\label{sec:transport}

For possible technical applications such as GMR devices based on the
fully spin polarized Heusler compounds, the electrical transport properties
are of utmost importance. Thus we summarize here our results on the
electrical conductivity, the Hall conductivity and the magnetoresistance
of the pure Heusler films and the Heusler multilayers. The electrical
conductivity was measured by 4 point technique with silver painted
electrical contacts, for the measurement of the Hall conductivity
the films were patterned into the conventional Hall bar geometry with
five contacts produced by ion beam etching. Figure \ref{fig:4-1}
displays the electrical resistivity versus temperature for the optimized
100-nm thick pure films of the Co based Heusler alloys. The residual
resistivity is rather high, especially for the Co$_{2}$MnSi film
(taken from reference \cite{geiersbach02}) with a residual resistivity
ratio ($RRR$) (defined as the ratio of the room temperature resistivity
and the resistivity at 4~K) of $RRR=1.07$. The Co$_{2}$MnSn and
the Co$_{2}$MnGe films have a higher electrical conductivity and
a higher $RRR$ ($RRR=1.15$ for the Co$_{2}$MnSn phase and $RRR=1.55$
for the Co$_{2}$MnGe phase). Actually the $RRR$ of 1.55 for Co$_{2}$MnGe
is the highest value reported for thin films of the Co based fully
spin polarized Heusler compounds in the literature until now.

\begin{figure}
\centering \includegraphics[height=8cm]{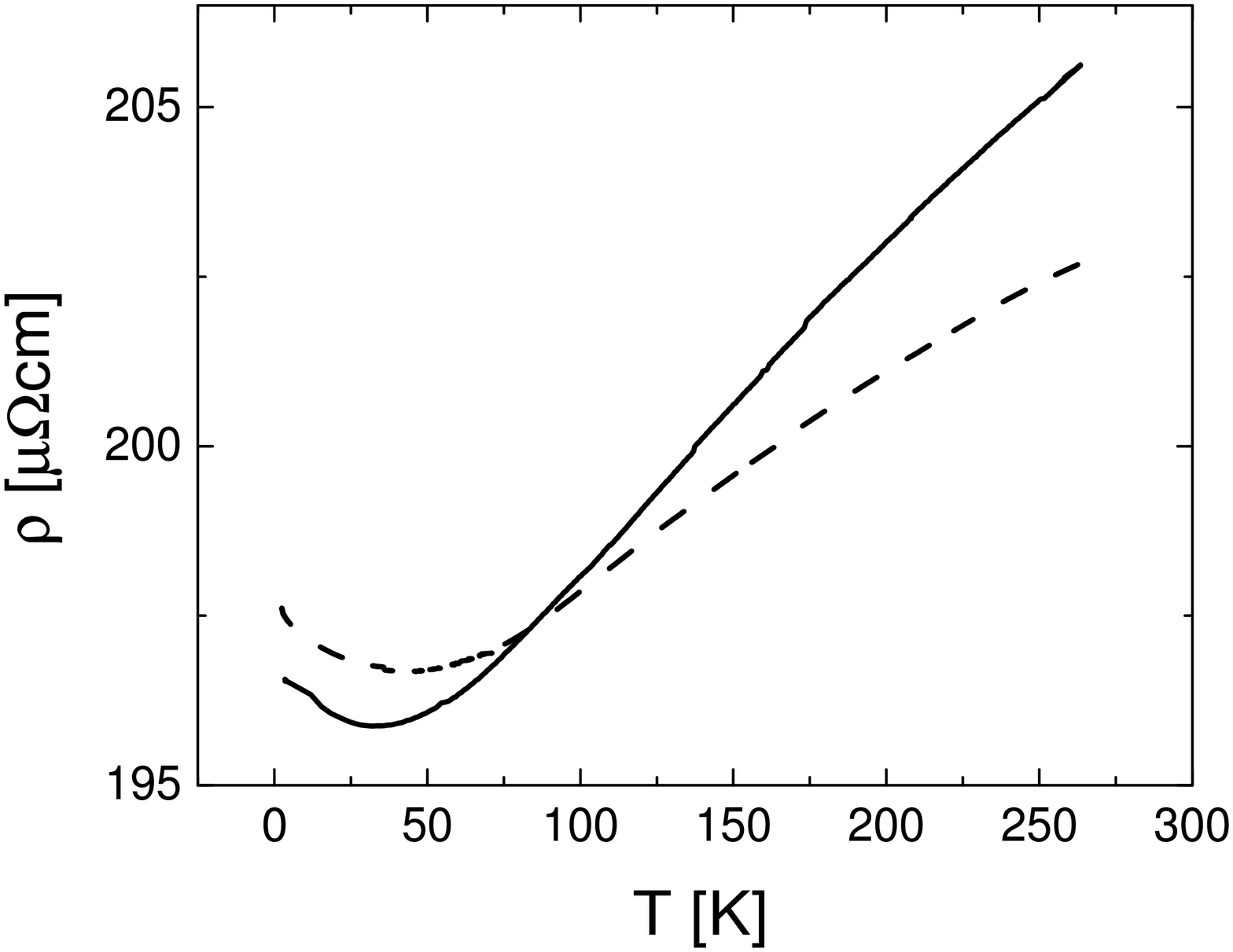}

\caption{Resistivity versus temperature for the multilayers {[}Co$_{2}$MnSn(3nm)/V(3nm)]$_{30}$
(dashed line) and {[}Co$_{2}$MnGe(3nm)/V(3nm)]$_{50}$ (straight
line).}

\label{fig:4-2} 
\end{figure}

It holds generally true for thin films of the Co based Heusler compounds
that there is strong electron scattering at defects such as grain
boundaries, antisite defects, interstitials and voids. We determined
the Hall coefficient $R_{0}$ by fitting the Hall voltage measured
up to a magnetic field $\mu_{0}H=3$~T by the formula $U_{H}=R_{0}H+R_{H}4\pi M$
with the magnetization $M$ and the anomalous Hall coefficient $R_{0}$
and derived $R_{0}=2\times10^{-10}$~m$^{3}$/As for the Co$_{2}$MnSi
phase and $R_{0}=2.2\times10^{-10}$~m$^{3}$/As for the Co$_{2}$MnSn
phase. Combining with the electrical conductivity and assuming parabolic
conduction bands we estimate an electron mean free path $l_{e}$ at
4~K of $l_{e}=1.3$~nm for the Co$_{2}$MnSi phase and 2.2~nm for
the Co$_{2}$MnSn phase. Assuming that for the Co$_{2}$MnGe layer
we have a similar value for $R_{0}$ we estimate $l_{e}=5.5$~nm
for this thin film in figure \ref{fig:4-1}. These remarkably low
values for $l_{e}$ stresses the strong defect scattering in these
compounds.

\begin{figure}
\centering \includegraphics[height=8cm]{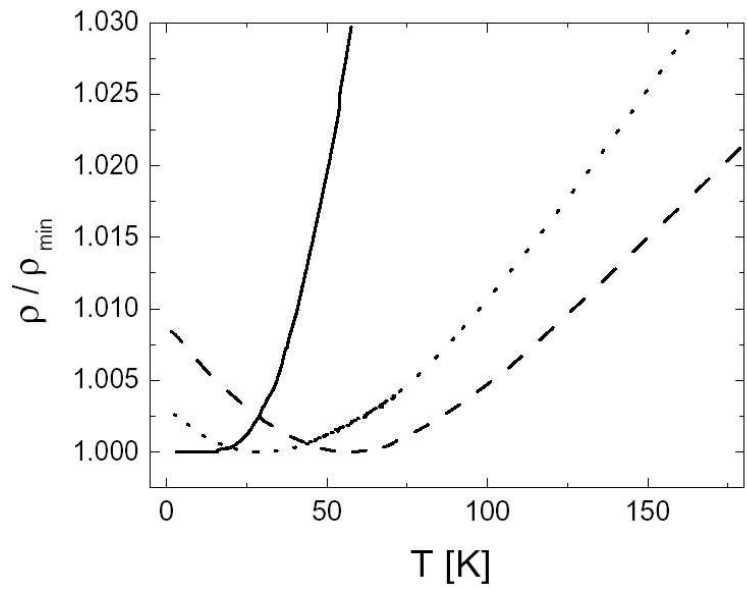}

\caption{Low temperature resistivity of Co$_{2}$MnSi (dashed line), Co$_{2}$MnSn
(dotted line) and Co$_{2}$MnGe (straight line) films. The Co$_{2}$MnSi
and the Co$_{2}$MnSn phase show an upturn of the resistivity towards
lower temperatures.}

\label{fig:4-3} 
\end{figure}

In figure \ref{fig:4-2} we show resistivity measurements of several
multilayers from section \ref{sec:multi}. We find high absolute values
for the resistivity, the resistivity ratio is below $RRR=1.1$ for
Co$_{2}$MnGe and Co$_{2}$MnSn with V and Au interlayers. In these
multilayers $l_{e}$ is very short namely of the order of 1~nm. If
one inspects the resistivity curves in figures \ref{fig:4-1} and
\ref{fig:4-2} in more detail, one observes a shallow minimum in $\rho$(T)
at temperatures between 20~K and 50~K for all samples with an $RRR$
below about $RRR=1.2$. This is a common feature for the Co based
Heusler thin films. The low temperature resistivity $\rho$(T) is
plotted on an enlarged scale in figure \ref{fig:4-3}. The amplitudes
of the resistivity upturn towards low temperatures and the temperature
of the resistivity minimum increases with increasing defect scattering,
thus the low temperature anomaly is clearly related to disorder. A
low temperature upturn in $\rho$(T) is well known for metallic thin
films and is usually associated with the Kondo effect \cite{melvin}
or weak localization \cite{lee85}. But these effects seem not very
plausible here, since in a strong ferromagnet Kondo scattering is
hardly possible and measurable weak localization effects are only
expected in two dimensional metals. There is a third classical effect
leading to a low temperature upturn in $\rho$(T) which is less well
known. This is a renormalization of the electronic density of states
at the Fermi level in metals with strong disorder \cite{altshuler79}.
This essentially is an electronic correlation effect and becomes relevant
if the scattering of the conduction electrons is strong and $l_{e}$
small. This renormalization effect has experimentally been observed
in amorphous metals \cite{mcmillan81}and gives a plausible explanation
for the low temperature anomaly in $\rho$(T) of the Heusler films,
too.

\begin{figure}
\centering \includegraphics[height=8cm]{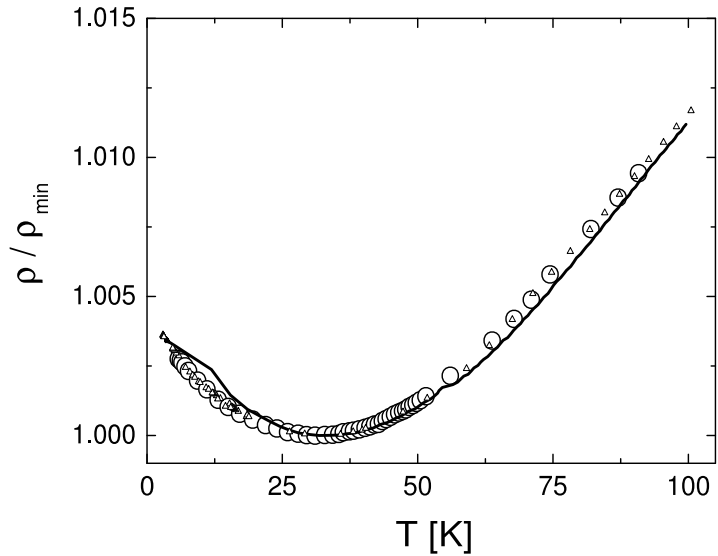}

\caption{Electrical resistivity versus temperature for the multilayer {[}Co$_{2}$MnGe(3nm)/Au(3nm)]$_{50}$
measured in zero field (straight line) and with an applied field of
4~T parallel to the film (open triangles) and perpendicular to it
(open dots).}

\label{fig:4-4} 
\end{figure}

A measurement which strongly supports this hypothesis is shown in
figure \ref{fig:4-4}, where for the example of a {[}Co$_{2}$MnSn/Au]
multilayer it is shown that a strong magnetic field of 4~T applied
in the direction parallel and perpendicular to the film does not change
the low temperature upturn in $\rho$(T). This is expected for the
renormalization effect, whereas for Kondo scattering and also in the
case of weak localization the upturn should be suppressed.

\begin{figure}
\centering \includegraphics[height=8cm]{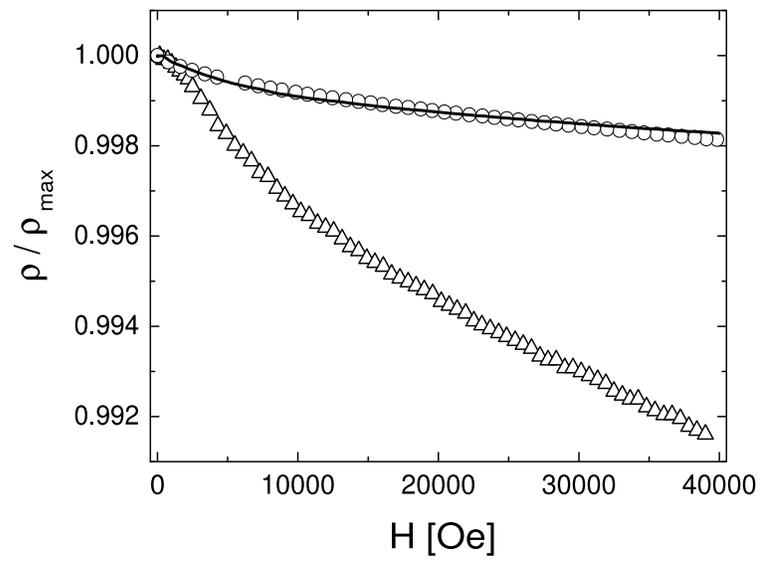}

\caption{High field magnetoresistance measured at 4~K for V(3nm)/Co$_{2}$MnSi(100nm)
(straight line), Au(3nm)/Co$_{2}$MnSn(100nm) (open dots) and V(3nm)/Co$_{2}$MnGe(100nm)
(open triangles).}

\label{fig:4-5} 
\end{figure}

\begin{figure}
\centering \includegraphics[height=8cm]{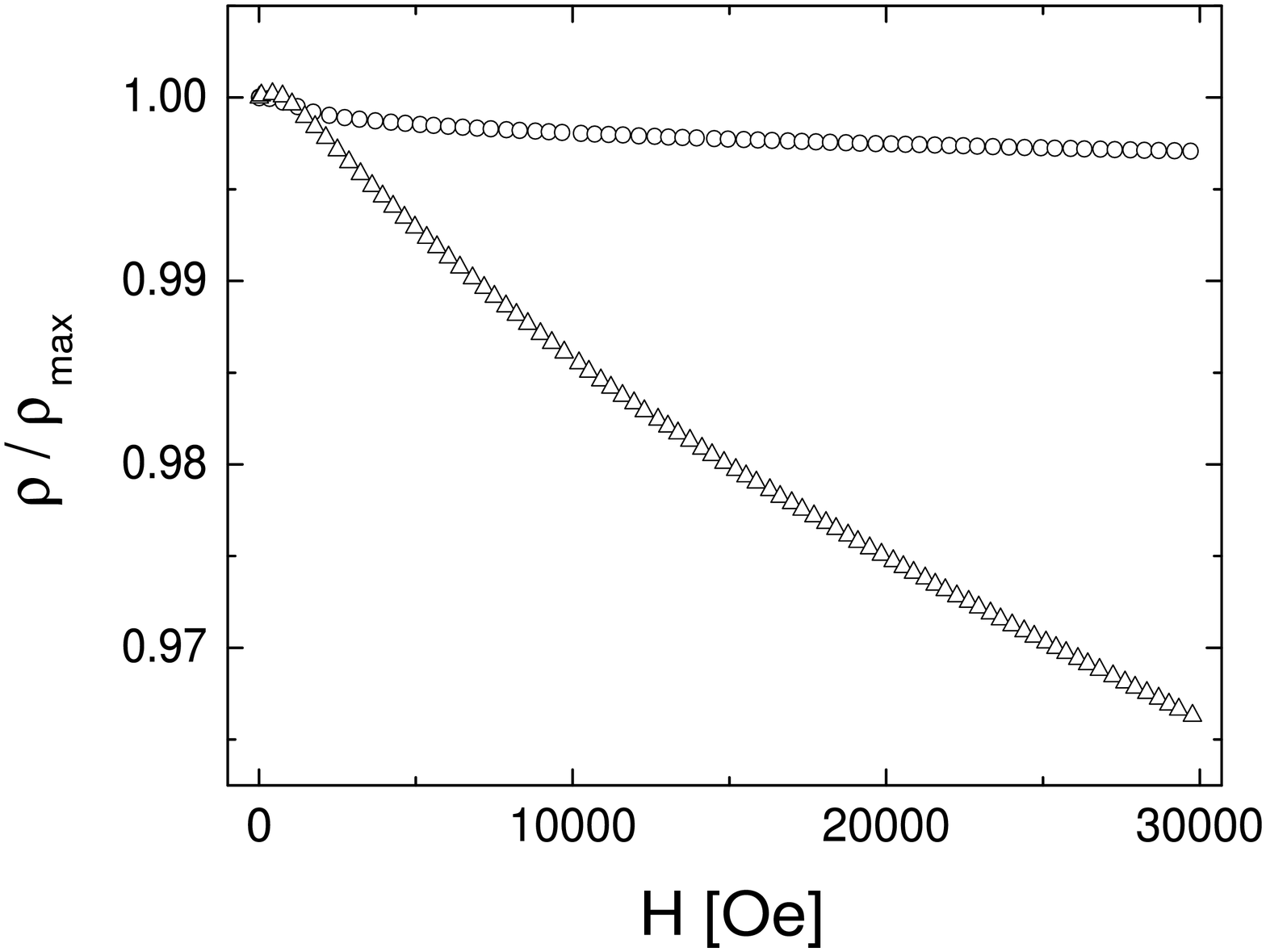}

\caption{High field magnetoresistance measured at 4~K for the multilayers
{[}Co$_{2}$MnSn(3nm)/V(3nm)]$_{30}$ (open dots) and {[}Co$_{2}$MnSn(3nm)/Au(3nm)]$_{30}$
(straight line).}

\label{fig:4-6} 
\end{figure}

Coming to the magnetoresistance (MR) of the pure thin films and the
multilayers we first discuss the high field MR i.e. the resistance
for fields above the ferromagnetic saturation field where the anisotropic
magnetoresitance (AMR) and the Giant magnetoresistance (GMR) play
no role. This high field magnetoresistance measured at 4~K is plotted
for the pure Heusler films in figure \ref{fig:4-5} and for several
multilayers in figure \ref{fig:4-6}. The high field magnetoresistance
for the pure Heusler films is negative and isotropic. The total decrease
of the resistance for fields up to 4 T is about 0.16\% for the Co$_{2}$MnSi
and the Co$_{2}$MnSn phase and definitely higher, namely 0.8\% for
the Co$_{2}$MnGe single crystalline film. We interpret the high field
MR as a reduction of spin disorder scattering in a ferromagnet with
non perfect ferromagnetic alignment of all spins. This interpretation
is supported by the results of the high field MR on the multilayers
in figure \ref{fig:4-6}. For the {[}Co$_{2}$MnSn/Au] multilayer
where the exchange bias effect discussed in section~\ref{sec:multi-2}
gives clear evidence for the existence of magnetic frustration and
canted spins at the interfaces, we find an unusually large MR value
with an amplitude of 3\% up to a field of 3~T, whereas for the {[}Co$_{2}$MnGe/V]
multilayer showing no indication of magnetic frustration the MR is
only 0.3\%.

In the low field MR measurements i.e. the MR for fields up to the
coercive force $H_{c}$ we observed the standard AMR of a ferromagnetic
film, however with a very small relative difference of the resistance
for the field applied parallel and perpendicular to the current of
the order $\Delta\rho/\rho=4\times10^{-4}$. For the multilayers the
AMR is even smaller i.e. below the experimental resolution limit for
$\Delta\rho/\rho$ of about 10$^{-5}$.

\begin{figure}
\centering \includegraphics[height=8cm]{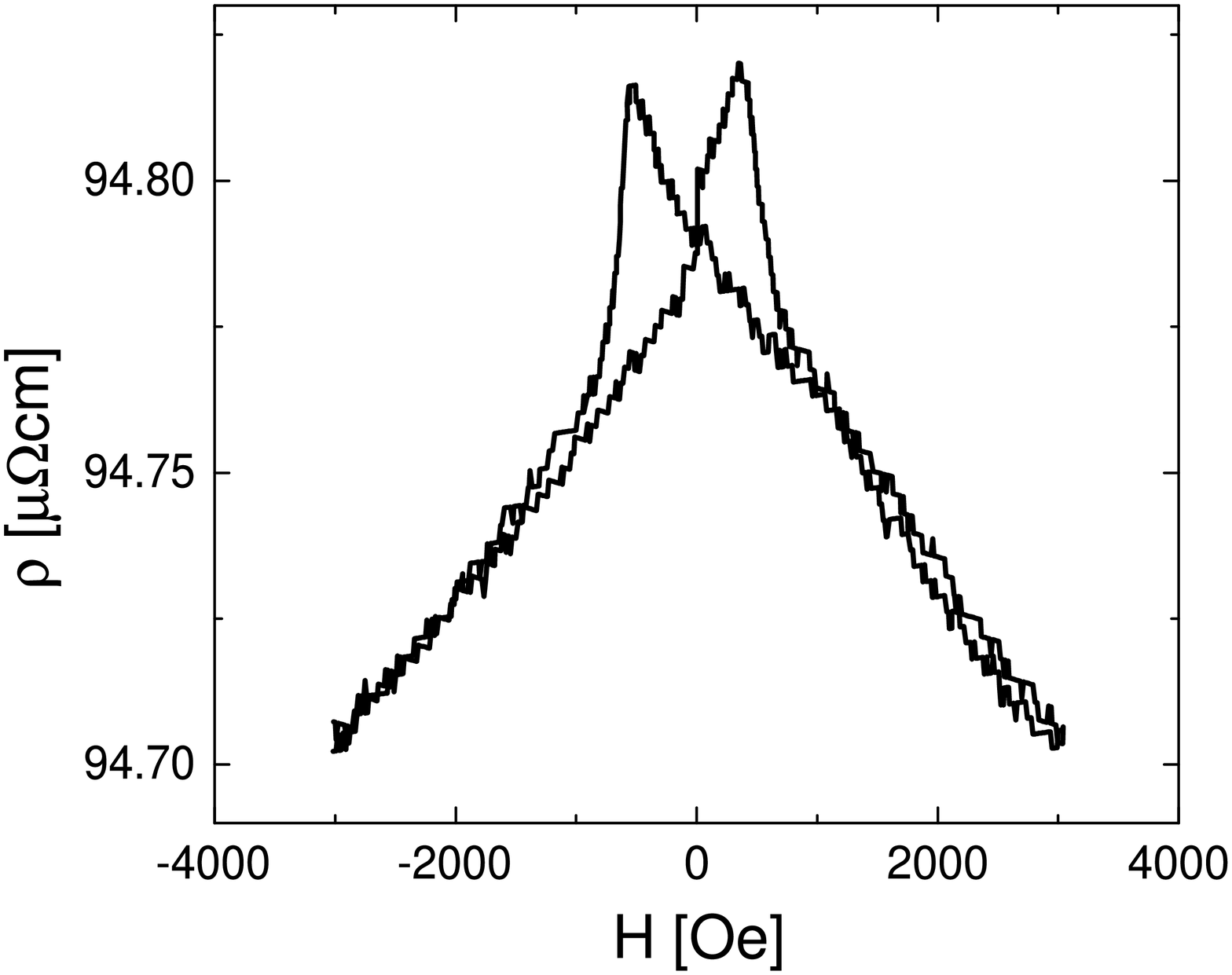}

\caption{Low field magnetoresistance measured at 4~K of a Co$_{2}$MnGe(3nm)/V(3nm)/Co$_{2}$MnGe(3nm)/CoO(2nm)
trilayer.}

\label{fig:4-7} 
\end{figure}

\begin{figure}
\centering \includegraphics[height=8cm]{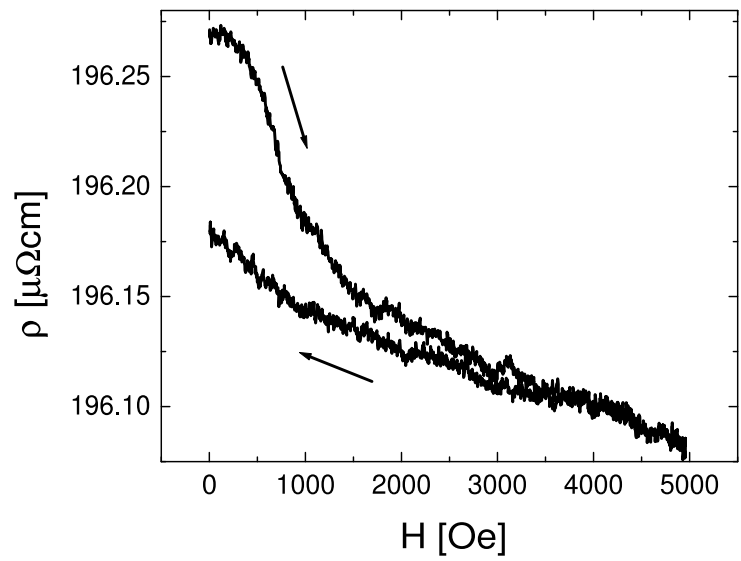}

\caption{Low field magnetoresistance measured at 4~K of a {[}Co$_{2}$MnGe(3nm)/V(3nm)]$_{50}$
multilayer, starting at the antiferromagnetic state after zero field
cooling (arrows indicate the field cycle).}

\label{fig:4-8} 
\end{figure}

In figures \ref{fig:4-7} and \ref{fig:4-8} we present low field
MR measurements related to the GMR effect. Figure \ref{fig:4-7} displays
the MR at 4~K of a Co$_{2}$MnGe/V/Co$_{2}$MnGe/CoO trilayer, where
an exchange bias with CoO is used to pin the magnetization direction
of the upper Co$_{2}$MnGe layer \cite{nogues99}. One finds a small
jump in the MR at $H_{c}$ with an amplitude $\Delta\rho/\rho=5\times10^{-4}$
which could be associated with a GMR effect. In figure \ref{fig:4-8}
we show the low field MR of a {[}Co$_{2}$MnGe/V] multilayer measured
at 4~K and starting from the antiferromagnetically ordered state
of the multilayer obtained after zero field cooling (see section \ref{sec:multi-3}).
Driving the antiferromagnetically coupled Co$_{2}$MnGe layers to
ferromagnetic saturation and returning to zero field, gives an irreversible
change of the MR $\Delta\rho/\rho=5\times10^{-4}$, which is the upper
limit for the GMR effect.

This chapter should demonstrate the difficulties one encounters when
trying to use the fully spin polarized Co based Heusler alloys in
magnetotransport applications such as GMR or AMR. The electrical conductivity
and the electron mean free path are small, pointing towards the existence
of lattice defects with a high conduction electron scattering efficiency.
Our results suggest that these are the same antisite defects in the
L2$_{1}$ structure which have the tendency to destroy the half metalicity.
The rather low values for the GMR in the spin valve device and the
antiferromagnetically coupled multilayers are not really surprising
since there are three detrimental factors which acting together are
able to nearly eliminate the GMR effect: First,non ferromagnetic interlayers
are present introducing strong spin independent scattering thereby
reducing the GMR. Second, for the observation of a sizable GMR effect,
the conduction electron mean free path should be larger than the double
layer thickness \cite{dieny94}, which is certainly not fulfilled
in our samples. Last but not least, the high spin polarization expected
for the ideal Co based Heusler compounds is most certainly lost in
very thin Co based Heusler layers with a saturation magnetization
of typically only 50\% of bulk samples.

\section{Summary and Conclusions}

\label{sec:summary}

We have shown that thick single films of the Heusler compounds Co$_{2}$MnGe
and Co$_{2}$MnSn can be grown with a saturation magnetization and
structural quality comparable to bulk samples. However, when decreasing
the thickness to the order of 3~nm the films show a definitely lower
value of the saturation magnetization and exhibit properties reminiscent
of small particle ferromagnets. We interpret this behaviour by a magnetic
decoupling of the Heusler films at the grain boundaries of the very
small grains forming the film. Structural disorder and a high concentration
of antisite defects at the grain boundaries weaken the ferromagnetism
drastically.

As we have shown by the XMCD measurements on the Mn and the Co $L_{2,3}$
absorption edges, it is only the Mn atom in the Heusler compounds
which reduces the saturation magnetization. Theoretical model calculations
suggest that this originates from an antiparallel spin orientation
of antisite Mn spin rather than from a loss of the atomic magnetic
moment of Mn. In the multilayers a similar phenomenon occurs at the
interfaces of the Heusler alloys with other metals. For a thickness
range of about 0.5 to 0.7~nm at the interfaces the Heusler alloys
are only weakly ferromagnetic or exhibit a spin glass type of order.
The exchange bias shift of the hysteresis loops observed at low temperatures
gives clear evidence for this.

In {[}Co$_{2}$MnGe/V] multilayers we have detected a peculiar antiferromagnetic
interlayer magnetic ordering with a well defined N\'{e}el temperature
far below the ferromagnetic Curie temperature of a single Heusler
layer of the same thickness. The antiferromagnetic ordering exists
in the thickness range of the V interlayer between 1~nm and 3~nm,
very unlike the antiferromagnetic interlayer order in multilayer systems
coupled by interlayer exchange mechanism. The antiferromagnetic order
is directly related to the granular ferromagnetic structure of very
thin Heusler layers. The small ferromagnetic particles defined by
the weak magnetic coupling at the grain boundaries exhibit superparamagnetic
behaviour above the N\'{e}el temperature. The interlayer dipolar
interactions between the superparamagnetic particles cause a reversible
magnetic phase transition with antiferromagnetic order between the
layers and ferromagnetic order within the layers at a well defined
N\'{e}el temperature.

Finally, the magnetotransport properties revealed that the electron
mean free path in the single layers and in the multilayers is very
small, indicative of the presence of large amounts of very effective
electron scattering centers. Simultaneously there exists an isotropic
negative spin disorder magnetoresistance up to very high fields, which
shows that the magnetic ground state is not a perfect ferromagnet
but possesses abundant magnetic defects. We think that the main source
of defects responsible for the small mean free path and the magnetic
disorder scattering are the antisite defects in the L2$_{1}$ structure.

Coming back to the issue discussed in the introduction, namely the
perspective of the Heusler compounds as full spin polarized ferromagnetic
layers in magnetoelectronic devices, our results indicate the severe
difficulties one inevitably encounters. The constraints in the preparation
of thin film heterostuctures in many cases prohibit the optimum preparation
conditions for the Heusler phase thus rendering it difficult to get
the L2$_{1}$ structure with a high degree of atomic order. At the
interfaces additional problems arise since they tend to be strongly
disordered and weakly ferromagnetic. What one ideally would need to
overcome these problems is a Heusler phase with a more robust spin
polarization at the Fermi level which principally might already exist
among the many new Heusler half metals which have been predicted theoretically.
In any case, the evolving field of magnetoelectronics motivated the
magnetism community to extend the experimental activities in thin
film preparation towards complex intermetallic ternary alloys, which
is certainly much more difficult than the preparation of simple films
but offers new and interesting perspectives for basic research and
applications.

\subparagraph{Acknowledgement}

The authors want to thank the DFG for financial support of this work
within the SFB 491 Bochum/Duisburg


\end{document}